\documentclass[useAMS,usenatbib]{mn2e}

\usepackage{graphicx}
\usepackage{paralist}
\usepackage{multirow}
\usepackage{amsmath}
\usepackage{color}
\usepackage[T1]{fontenc} 
\usepackage{aecompl}

\newcommand{\aj}{AJ}
\newcommand{\apj}{ApJ}
\newcommand{\apjl}{ApJL}
\newcommand{\apjs}{ApJS}

\newcommand{\aap}{A\&A}

\newcommand{\araa}{ARA\&A}
\newcommand{\jcap}{JCAP}
\newcommand{\mnras}{MNRAS}

\newcommand{\pasp}{PASP}

\newcommand{\procspie}{Proc. SPIE}

\title[X-ray selected AGN in the XMM-XXL]{A Spectroscopic Survey of X-ray Selected AGN in the Northern XMM-XXL Field}
\author[M.-L. Menzel et al.]
{M.-L. Menzel$^{1}$\thanks{E-mail: mlmenzel@mpe.mpg.de}, 
A. Merloni$^{1}$\thanks{E-mail: am@mpe.mpg.de}, 
A. Georgakakis$^{1}$,
M. Salvato$^{1}$,
E. Aubourg$^{2}$, \newauthor
W.N. Brandt$^{3}$$^{,4}$$^{,5}$,
M. Brusa$^{6}$$^{,7}$,
J. Buchner$^{1}$,
T. Dwelly$^{1}$,  
K. Nandra$^{1}$, 
I. P\^{a}ris$^{8}$,  \newauthor
P. Petitjean$^{9}$,
A. Schwope$^{10}$\\
$^{1}$ Max-Planck-Institut f\"ur Extraterrestrische Physik, Giessenbachstrasse 1, D-85741 Garching, Germany\\
$^{2}$ Laboratoire AstroParticule et Cosmologie, 12, rue Alice Domon et L\'eonie Duquet, F-75205 Paris Cedex 13, France \\
$^{3}$ Department of Astronomy \& Astrophysics, 525 Davey Lab, The Pennsylvania State University, University Park, PA 16802, USA\\
$^{4}$ Institute for Gravitation and the Cosmos, The Pennsylvania State University, University Park, PA 16802, USA\\
$^{5}$ Department of Physics, 104 Davey Lab, The Pennsylvania State University, University Park, PA 16802, USA\\
$^{6}$ Dipartimento di Fisica e Astronomia, Universit\`a di Bologna, viale Berti Pichat 6/2, 40127 Bologna, Italy\\
$^{7}$ INAF - Osservatorio Astronomico di Bologna, via Ranzani 1, 40127 Bologna, Italy \\
$^{8}$ Osservatorio Astronomico di Trieste, via G.B. Tiepolo 11, I-34143 Trieste, Italy \\
$^{9}$ Institut d'Astrophysique de Paris, 98 bis boulevard Arago, F-75014 Paris, France \\
$^{10}$ Leibniz-Institut f\"ur Astrophysik Potsdam (AIP), An der Sternwarte 16, D-14482 Potsdam, Germany \\
}
\begin{document}

\date{}

\pagerange{\pageref{firstpage}--\pageref{lastpage}} \pubyear{2002}

\maketitle

\label{firstpage}

\begin{abstract}
This paper presents a survey of X-ray selected active galactic nuclei (AGN) with optical spectroscopic follow-up in a $\sim 18\, \rm{deg^2}$ area of the equatorial XMM-XXL north field. A sample of 8445 point-like X-ray sources detected by \textit{XMM-Newton}  above a limiting flux of $F_{\rm 0.5-10\, keV} > 10^{-15} \rm\,erg\, cm^{-2}\, s^{-1}$ was matched to optical (\textit{SDSS}) and infrared (\textit{WISE}) counterparts. We followed up 3042 sources brighter than $r=22.5$ mag with the SDSS \textit{BOSS} spectrograph. The spectra yielded a reliable redshift measurement for 2578 AGN in the redshift range $z=0.02-5.0$, with $0.5-2\rm\, keV$ luminosities ranging from $10^{39}-10^{46}\rm\,erg\,s^{-1}$. This is currently the largest published spectroscopic sample of X-ray selected AGN in a contiguous area. The \textit{BOSS} spectra of AGN candidates show a distribution of optical line widths which is clearly bimodal, allowing an efficient separation between broad- and narrow-emission line AGN. The former dominate our sample (70 per cent) due to the relatively bright X-ray flux limit and the optical \textit{BOSS} magnitude limit. We classify the narrow emission line objects (22 per cent of the full sample) using standard BPT diagnostics: the majority have line ratios indicating the dominant source of ionization is the AGN. A small number (8 per cent of the full sample) exhibit the typical narrow line ratios of star-forming galaxies, or only have absorption lines in their spectra. We term the latter two classes ``elusive'' AGN, which would not be easy to identify correctly without their X-ray emission. We also compare X-ray (\textit{XMM-Newton}), optical colour (\textit{SDSS}), and IR (\textit{WISE}) AGN selections in this field. X-ray observations reveal, by far, the largest number of AGN. The overlap between the selections, which is a strong function of the imaging depth in a given band, is also remarkably small. We show using spectral stacking that a large fraction of the X-ray AGN would not be selectable via optical or IR colours due to host galaxy contamination. A substantial fraction of AGN may therefore be missed by these longer-wavelength selection methods. 

\end{abstract}


\section{Introduction}
\label{sec:intro}

A fundamental question in current astrophysics research on active galactic nuclei (AGN) is the origin and evolution of supermassive black holes (SMBH) as well as their interaction with their host galaxy. A large body of evidence suggests the presence of SMBH in nearly all local spheroids (e.g. \citealt{Magorrian98, Kormendy13} and references therein) and argues that their growth occurs predominantly through accretion processes  (e.g. \citealt{Soltan82, Marconi04}). In order to study the evolution and properties of SMBH, one should approach them by a preferably unbiased data set of AGN. The main limiting factor in AGN selection is the diversity of their spectral energy distributions (SED). This depends on the presence of dust and gas clouds along the line-of-sight, the contrast between accretion-luminosity and stellar emission from the host galaxy, and the detailed physics  of the  accretion process itself. Many methods have been developed to account for these factors using different wavelength bands and both photometric as well as spectroscopic selection criteria, but they are always, by construction, subject to selection effects. 

The accretion processes in SMBHs are highly complex and emit at different wavelengths. In the optical/UV spectra of AGN, there is the characteristic `blue bump' which is supposed to originate from a thermally radiating accretion disc. This feature allows AGN to be selected via their blue colours, and has been extensively used (see \citealt{Boyle00, Richards09, Ross12} and references therein). This selection has its limitations because the galaxy dilution as well as the optical obscuration suppress the blue bump. It is therefore only applied to optically point-like sources.

X-rays directly trace highly energetic processes occurring in the neighbourhood of the supermassive black hole  \citep{Brandt05, Brandt15}. Thus, X-ray selection is highly efficient and provides a clean separation from the host galaxies which typically emit at lower fluxes. However, the most heavily obscured, Compton-thick AGN are opaque even to X-rays. In this case, infrared AGN selection methods, tracing the reprocessed nuclear emission, benefit from the clearly distinct SED shapes of AGN and galaxy components in this wavelength range. Therefore, infrared is able to detect heavily obscured AGN \citep{Lacy07, Stern05,  Stern12, Assef13, Messias14}.

Another very well-known AGN selection criterion is the optical narrow emission line diagnostic \citep{Baldwin81, Ho97, Kauffmann03, Kewley06, Lamareille10}. It uses the fact that ionizing radiation from the AGN is harder than that of stars, which has an effect on the measured emission line ratios. However, this method is claimed to suffer from (i) contamination from the host galaxy which can potential swamp AGN signatures and (ii) the fact that the method selects a large fraction of Low Ionization Narrow Emission Region galaxies (LINERS) whose nature is still debated (e.g. \citealt{Yan12}).

Broad emission lines ($\rm FWHM>1000\,\rm km\,s^{-1}$) are unique spectral features which identify optical unobscured AGN, where the observer has a direct view on the fast moving clouds close to the black hole. Purely board line selected samples of AGN have been used for example in the VVDS and zCOSMOS surveys (e.g. \citealt{Gavignaud06, Trump09}). 

It is essential to understand  the selection function of the different methods for an unbiased census of AGN across cosmic time. Different selection methods may not only depend on obscuration or host galaxy contamination, but on other fundamental parameters of the active black hole, such as  Eddington ratio (i.e. the ratio between observed accretion rate and the Eddington rate), instantaneous accretion luminosity, black hole mass, etc. 

Several works \citep{Yan11, Hickox09,  Donley12, Mendez13} combined and compared multiple selection methods, such as X-ray, UV, optical, and infrared. These studies have been performed in areas benefiting from synergies of deep X-ray coverage, multi-wavelength observations and spectroscopic follow-up \citep{Brandt15}, e.g. the COSMOS and CDFS field (\citealt{Alexander03, Luo10, Brusa10, Civano12}) and Lockman Hole (\citealt{Mateos05}). However, most area surveys are limited to few $\deg^{2}$ or less. Larger area surveys, such as Bo\"otes (\citealt{Hickox09}), achieve better statistics for rare objects, e.g. luminous AGN, and improve clustering studies, but have been difficult to carry out, mainly due to the small field of view of X-ray imaging telescopes. 

In this work, we present one of the largest spectroscopic surveys of X-ray selected AGN (\textit{XMM-Newton}) in a homogenous and contiguous area with the same instrument (\textit{BOSS} spectrograph of the Baryonic Oscillation Spectroscopic Survey). This dataset allows for a more detailed analysis of properties of X-ray selected AGN, enables a comparison of selection methods for AGN based on different wavelengths, and provides a forecast for the AGN population in the \textit{eROSITA} survey and its follow-up programme {SPIDERS} (\citealt{MerlonieROSITA, Predehl14}, Merloni et al. 2015, in prep., Dwelly et al. 2015, in prep., Clerc et al. 2015, in prep.).
We chose the XMM-XXL north field (PI M. Pierre), which is an extension of the former XMM-LSS survey  \citep{Pierre04, Chiappetti13, Clerc14} and was conceived to study the large scale structure in the Universe by mapping  a well-defined statistical sample of galaxy clusters. It provides access to large subsamples of AGN when selected by luminosity, morphology or obscuration. Furthermore, many multi-wavelength datasets in optical (e.g. \textit{CFHTLS},  \textit{SDSS}), infrared (e.g.  \textit{UKIDSS},  \textit{WISE},  \textit{VIDEO}), UV (e.g.  \textit{GALEX}) and radio (e.g.  \textit{VLA}) bands are available in this field. Our reduction of the XMM-XXL area ($\sim 22\,\rm deg^2$) yields in a flux limit of $F_{0.5-2 \,\rm{keV}} \sim 3.5\times 10^{-15} \,\rm{erg\,cm^{-2}\,s^{-1}}$ for 50 per cent of the area.

The paper is organized as follows: Section \ref{sec:data sets} introduces the multi-wavelength data sets from \textit{XMM-Newton}, \textit{SDSS}, \textit{BOSS} and \textit{WISE}, along with the cross-matching and the spectroscopic target selection. Section \ref{sec:BOSSspectra} presents the redshift determination, as well as the classification of the \textit{BOSS} spectroscopic follow-up data. An overview of the spectroscopic redshift, and luminosity properties of the X-ray selected AGN is given in Section \ref{sec:Classproperties}. In Section \ref{sec:Colorproperties} we compare optical and infrared AGN selection criteria and evaluate them using our sample of X-ray selected AGN. We discuss the survey analysis in Section \ref{sec:Discussion} and provide a forecast for \textit{eROSITA} and \textit{SPIDERS} in Section \ref{sec:eROSITAforecast}. Finally, we summarize the results in Section \ref{sec:conclusion}. In this paper, we use the J2000 standard epoch and adopt a cosmology with $H{_0} = 70\, \,\rm{km\, s^{-1}\, Mpc^{-1}}$, $\Omega_M = 0.27$, and $\Omega_{\Lambda} = 0.73$ . The optical magnitudes from \textit{SDSS} are given in the AB system and infrared magnitudes from \textit{WISE} are given in the Vega system.

\section{Data sets and spectroscopic target selection in XMM-XXL north }
\label{sec:data sets} 

In this section, we introduce the multi-wavelength data sets in the XMM-XXL north field, the cross-matching of the optical and infrared counterparts, and the spectroscopic target selection of our X-ray selected AGN.  The full catalogue is presented in Appendix \ref{sec:Catalogue}.

\subsection{Introduction of imaging data sets}
\subsubsection[]{\textit{XMM-XXL} survey and X-ray source catalogue}

The \textit{XMM-Newton}  XXL survey (XMM-XXL,  PI Pierre) is a  medium-depth (10\,ks
per pointing) X-ray survey that covers a total area of $\rm 50\,deg^2$
split into two fields equal in size. In this paper,  we focus on
the  equatorial  sub-region  of  the XMM-XXL  (XMM-XXL  north),  which
overlaps with the  SDSS-DR8 imaging survey \citep{Aihara11}. The
XMM-XXL  north observations were  distributed around  the area  of the
original  $\rm 11\,deg^2$  XMM-LSS  survey \citep{Clerc14}  and
therefore build upon and extend that sample.

The X-ray  data used  in this paper  are primarily  from the
XMM-XXL and  XMM-LSS surveys. We also include  however, any additional
 \textit{XMM-Newton}  pointings that are contiguous  to the area covered by those
programmes,   such    as   the   \textit{XMM-Newton}    observations   of   the
\textit{Subaru}/\textit{XMM-Newton}  Deep Survey (SXDS; \citealt{Ueda08}).   The data
reduction, source  detection and sensitivity map  construction for the
X-ray catalogue follow the  methods described by \citet{Georgakakis11}.   
Specific   detail on  the  reduction   of  the   \textit{XMM-Newton} 
observations  in the  XMM-XXL  north  field and X-ray spectroscopic properties are  presented  in Liu  et
al. (2015, submitted).  In  brief, the X-ray  data reduction is  carried out
using the XMM Science Analysis  System (SAS) version 12.  Our survey 
comprises the XMM-XXL data observed  prior to  2012 January 23. 
At  that  date the XMM-XXL  programme   was  partially  complete, which   results  in  the
inhomogeneous X-ray coverage shown in Fig. 1. In the following, we always refer to this 
coverage as `XMM-XXL north' area. The catalogue contains in total 8445 sources including 8016 sources with detections in the soft band ($F_{0.5-2\rm keV}$), 4802 sources with detections in the hard band ($F_{2-10\rm keV}$) and 8309 sources with detections in the full band ($F_{0.5-10\rm keV}$). Within these three bands, $\sim 47$ per cent, $\sim 50$ per cent and $\sim 45$ per cent of the sources have an optical counterpart in the r-band ($r<22.5\,\rm mag$), respectively. For the total area of our survey with spectroscopic follow-up (18 $\rm deg^2$), the observations reach a flux limit of  $F_{0.5-10\rm keV} > 7.50 \times 10^{-15} \rm\,erg\, cm^{-2}\, s^{-1}$ for 10 per cent of the area and a flux limit of $F_{0.5-10\rm keV}> 1.27 \times 10^{-14} \rm\,erg\, cm^{-2}\, s^{-1}$ for 50 per cent of the area.

\subsubsection[]{Optical source catalogue of \textit{SDSS}}

The XMM-XXL north area is covered by the optical imaging of the third programme of the \textit{Sloan} Digital Sky Survey (\textit{SDSS-III}, \citealt{Eisenstein11}). This is an optical survey extending over $\sim 14,555 \,\rm deg^2$ at the ground based 2.5-meter telescope at the Apache Point Observatory, New Mexico \citep{Gunn06}. 
The five broad bands (average wavelength indicated) \textit{u} [$3551\,$\AA], \textit{g} [$4686\,$\AA], \textit{r} [$6165\,$\AA], \textit{i} [$7481\,$\AA], \textit{z} [$8931\,$\AA] \citep{Fukugita96} have the following (AB) magnitude limits: $22.0\,\rm mag$, $22.2\,\rm mag$, $22.2\,\rm mag$,  $21.3\,\rm mag$ and $20.5\,\rm mag$ (corresponding to $95$ per cent completeness for point sources). We retrieved the \textit{SDSS} imaging data from the DR8 \citep{Aihara11} and obtained $538\,508$ sources in the XMM-XXL north. 
In this work, we use the flux measurements from \textit{SDSS}: \texttt{psfmag} (for point-like sources, $\texttt{type}=6$) and \texttt{modelmag} (for extended objects, $\texttt{type}=3$). Both magnitudes are not corrected for extinction. Furthermore, we do not apply any cuts to correct for e.g. blending and moving objects.

\subsubsection[]{Infrared source catalogue of \textit{WISE}}
\label{sec:WISE}

The Wide-field Infrared Survey Explorer (\textit{WISE}) observed the entire sky using a four channel imager (mean wavelength indicated): \textit{W}1 [$3.4\,\rm\mu m$], \textit{W}2 [$4.6\,\rm\mu m$], \textit{W}3 [$12\,\rm\mu m$] and \textit{W}4 [$22\,\rm\mu m$] \citep{Wright10}. 
For the infrared AGN selection criteria in Section \ref{sec:WISEAGN}, we used the \texttt{mpro} magnitudes for both point-like and extended objects from the allWISE data release \citep{Wright10, Mainzer11, Cutri13}. We excluded diffraction spikes, persistent sources, halos, optical ghosts and blended sources ($\texttt {cc\_flag}=0$ in the \textit{W}1 and \textit{W}2 band, $\texttt{NB}\leq2$) following \citet{Stern12}. 
In the XMM-XXL north, we obtain in total $334\,697$ infrared sources which yielded $321\,352$ sources after cleaning for photometric failures. 

\subsection{Cross-matching of data sets}
\label{sec:crossmatch}

\begin{figure}
 \centering
\includegraphics[width=85mm]{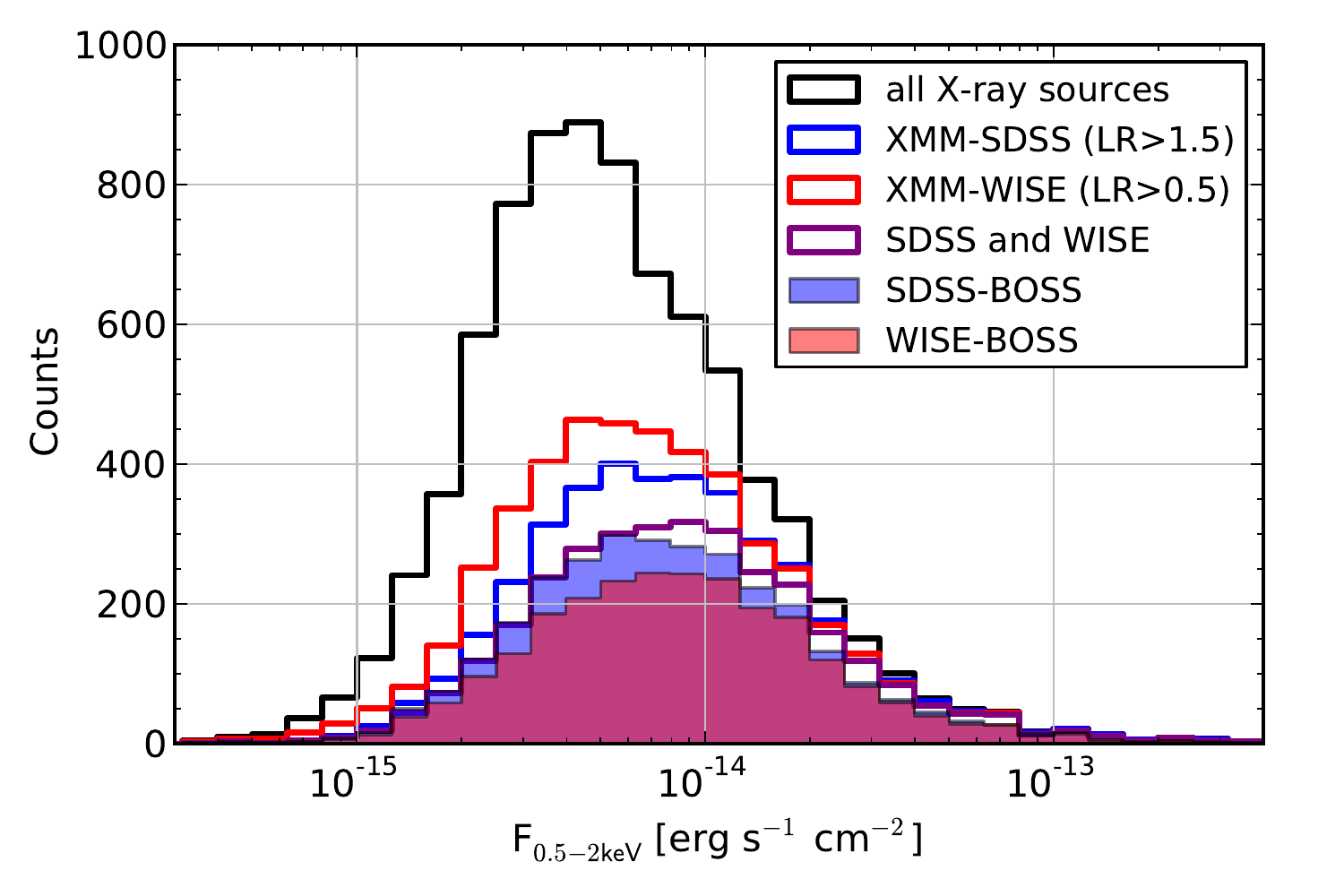}
 \caption{Histogram of the soft X-ray flux $F_{0.5-2\,\rm keV}$  for point-like X-ray sources in the northern XMM-XXL area: The plot shows all X-ray sources, the cross-matched sources to \textit{SDSS} with Likelihood-Ratio $\rm LR_{XMM,SDSS}>1.5$, the cross-matched sources to \textit{WISE} with $\rm LR_{XMM,WISE}>0.5$, all \textit{BOSS} spectra and the spectra with \textit{WISE} counterparts.}
 \label{fig:Xflux}
\end{figure}

The X-ray sources from the XMM-XXL north survey have been separately cross-matched to the \textit{SDSS} catalogue and to the allWISE catalogue. This allows an independent comparison of AGN selection criteria with the full multi-wavelength data set. The cross-matching was performed using the Likelihood-Ratio method as presented in \citet{Georgakakis11}. We apply a lower Likelihood Ratio limit for the matched catalogues, which allows for  an estimation of the spurious identification rate. Any cuts for e.g. photometric errors or flux have been applied after the matching process. Fig. \ref{fig:Xflux} demonstrates the full band X-ray flux histogram for the point-like X-ray sources, their associated counterparts and the followed-up spectra (as described in Section \ref{sec:targsel}).
 \begin{figure}
 \centering
\includegraphics[width=85mm]{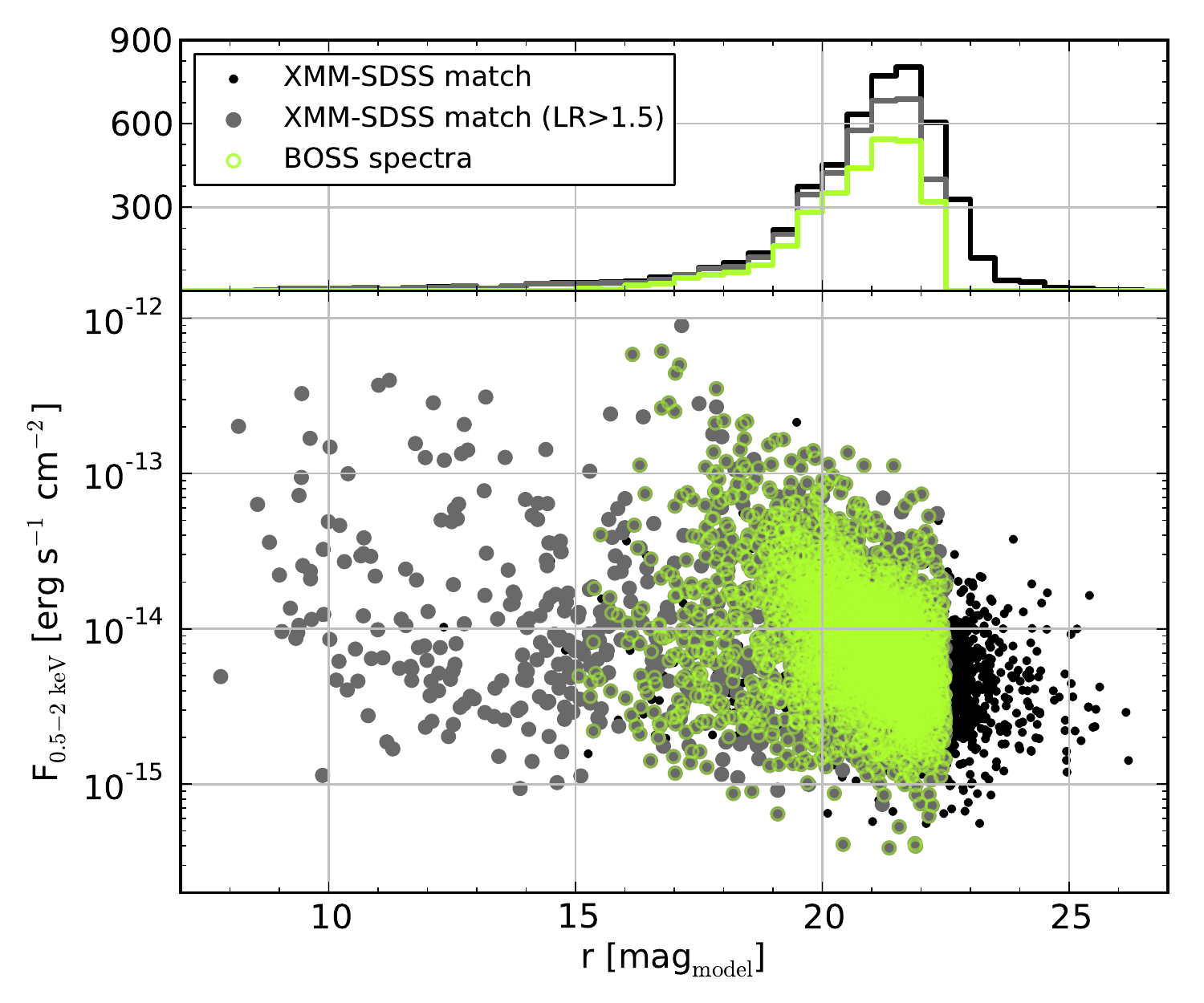}
 \caption{Distribution and histogram of \textit{r}-band magnitude (\textit{SDSS}: \texttt{modelmag}, AB) for X-ray source counterparts in the northern XMM-XXL area: The plot shows all cross-matched \textit{XMM-SDSS} sources (black), the cross-matched sources above the Likelihood-Ratio of $\rm LR_{\rm XMM,SDSS}> 1.5$ (grey) and the sources with \textit{BOSS} spectra (green). }
 \label{fig:rband}
\end{figure}

5294 X-ray sources have an optical counterpart in the \textit{SDSS} DR8 \citep{Aihara11} inside a maximal search radius of $4\,\rm arcsec$. Out of those, 4075 sources with a Likelihood Ratio $\rm LR_{\rm XMM,SDSS}> 1.5$ are selected as targets for the spectroscopic follow-up with \textit{BOSS}. The spurious identification rate for these objects is estimated to be about 7 per cent. Fig. \ref{fig:rband} shows the \textit{r$_{\rm model}$} magnitude distribution of both optically extended and point-like \textit{XMM-SDSS} cross-matched sources. The distribution has its maximum at $\textit{r$_{\rm model}$}\sim 22\,\rm mag$. The plot also shows the 3042 sources which have been targeted by \textit{BOSS} in our observation programme.  

\begin{figure}
 \centering
\includegraphics[width=85mm]{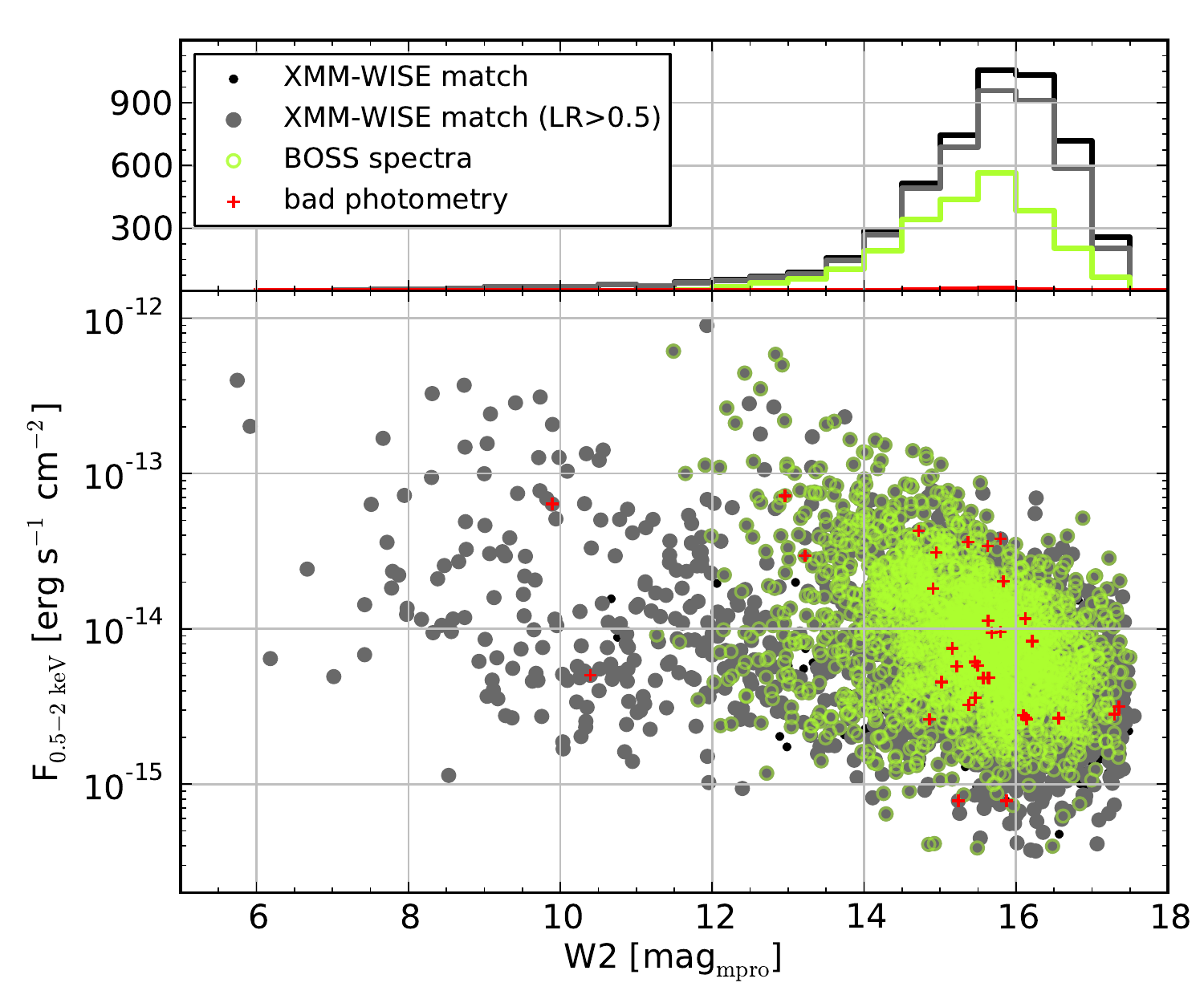}
 \caption{Distribution and histogram of \textit{W}2-band magnitude (\textit{WISE}: \texttt{mpro}, Vega) for X-ray source counterparts in the northern XMM-XXL area: The plot shows all cross-matched \textit{XMM-WISE} sources (black), the cross-matched sources above the Likelihood-Ratio of $\rm LR_{\rm XMM,WISE}> 0.5$  (grey) and the sources with \textit{BOSS} spectra (green). The red markers are sources with bad \textit{WISE} photometry and are excluded from the dataset.}
 \label{fig:w1band}
\end{figure}

5414 X-ray sources have an infrared counterpart in the \textit{WISE} imaging database within an maximal matching distance of  $5\,\rm arcsec$. We apply a Likelihood Ratio threshold of $\rm LR_{XMM,WISE}>0.5$, corresponding to a spurious identification rate  5 per cent, and retrieve 4844 sources, including 4811 with good \textit{WISE} photometry. In Fig. \ref{fig:w1band}, we show the \textit{W}2-band distribution and the histogram of all \textit{XMM-WISE} cross-matched objects. The distribution has its maximum at $\textit{W}2\sim16\,\rm mag$. In addition, we indicate the 2474 sources, which have spectroscopic \textit{BOSS} follow-up, too. 

The angular distance between the optical and infrared counterpart of X-ray sources gives the probability to be the counterpart of the same source. We assume an upper bound distance of 1 arcsec, which corresponds to the radius of a \textit{BOSS} fibre and is a very conservative limit for non-blended \textit{WISE} sources. Out of the 3305 X-ray sources with both \textit{SDSS} and \textit{WISE} cross-matches, 13 per cent have counterparts which are separated by more than 1 arcsec. Such a fraction decreases to 4 per cent for distances larger than 2 arcsec. This number is consistent with expectations, given the predicted numbers of spurious associations we estimate from the Likelihood-Ratio thresholds for the IR and optical matches to the X-ray sources (see above).

\subsection[]{Spectroscopic data set}
\label{sec:BOSSspec}

 \begin{figure*}
\includegraphics[width=180mm]{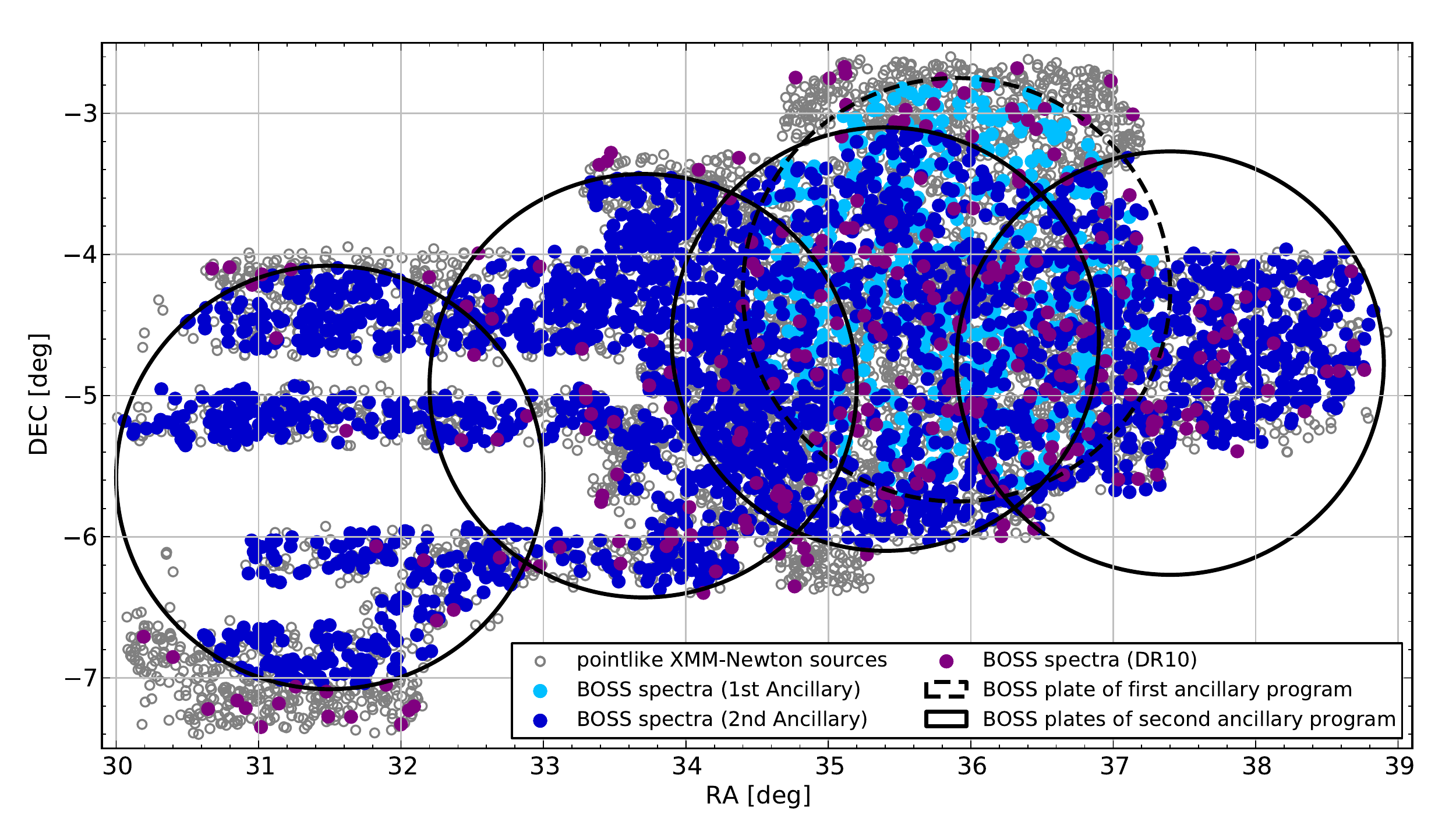}
 \caption{XMM-XXL north area with sky coordinates of \textit{XMM-Newton}  sources and associated \textit{BOSS} observed targets: We extracted 8445 point-like X-ray sources (grey) over a $\sim 18\, \rm{deg^2}$ area of the XMM-XXL north. By the time of our reduction (January 2012), the pointing of \textit{XMM-Newton}   in the XMM-XXL north area was not yet completed. The spectroscopic follow-up of the  \textit{XMM-Newton}  sources with \textit{BOSS} has been performed during two ancillary programmes (1st: dashed line circle / light blue markers, 2nd: solid black line circle / dark blue markers) and completed by former targets from \textit{BOSS}-DR10 (purple markers). }
 \label{fig:XXLarea}
\end{figure*}

\subsubsection[]{BOSS spectroscopic survey}

We performed optical spectroscopic follow-up of the \textit{XMM}-\textit{SDSS} matched sources with the \textit{BOSS} spectrograph of \textit{SDSS} \citep{Smee13}. It is a multiple fibre spectrograph using a standard plate of  $7 \rm \,deg^2$ which can host 1000 optical fibres of 2 arcseconds diameter and is typically exposed for $4500\,\rm s$. The covered wavelength range spans $\lambda=360 - 1040\,\rm{ nm}$ with an average resolution of $R = 2000$ and a redshift accuracy of $65 \rm\, km\,s^{-1}$. The \textit{BOSS} spectroscopic targets are limited in their \textit{r}-band magnitude range. We apply a magnitude cut of $15.0 < r < 22.5\rm\, mag$. The bright limit is necessary to avoid cross-talk effects in the optical spectrograph, while the faint one is introduced to ensure realistic chances of successfully obtaining a spectrum for strong emission line objects in a typical \textit{BOSS} exposure. 

The fibre allocation of one plate is split into fibres for the dedicated science programme and fibres which are assigned to standard calibration stars and sky calibration targets ($n_{\rm star}\approx 20$ and $n_{\rm calib}\approx80$) as well as repeated observations of other \textit{BOSS} programmes. The fibre tiling procedure of the \textit{BOSS} plates starts with the determination of the largest possible subset of targets which do not collide with each other. In a second step, the remaining fibres are optimally distributed to minimize fibre collisions \citep{Blanton03}. We can conclude that deselection from the initial target list was only due to random fibre collisions.

\subsubsection{Spectroscopic target selection}

\label{sec:targsel}
The spectroscopic follow-up of the 4075 matched \textit{XMM-SDSS} AGN comes from three different programmes, including two dedicated \textit{BOSS} ancillary programmes (published as a part of DR12, \citealt{Alam15}) and former \textit{BOSS} DR10 observations \citep{Ahn12} in the same region. In the following, we explain the target selection and refer for technical details to Appendix \ref{app:spectargsel}.
\\\\
\textit{First \textit{BOSS} Ancillary Programme}
\\
The first group of spectra has been retrieved from the \textit{BOSS} ancillary programme by PI P. Green and A. Merloni. It was devised to test target selection algorithms and strategies for both the \textit{SPIDERS} and \textit{TDSS} survey components of the \textit{SDSS-IV}. \textit{TDSS} (Time Domain Spectroscopic Survey, \citealt{Morganson15}) is a follow-up programme of time variable objects (e.g. from PAN-STARRS1), with either single as well as multi-epoch spectroscopy. The dedicated \textit{BOSS} targeting plate covered $7\rm\,deg^2$ including both the XMM-LSS as well as \textit{Pan-STARRS} Medium Deep Field MD01. In the area, we selected 1159 point-like X-ray sources at a flux limit of $F_{0.5-2 \,\rm{keV}} > 6\times10^{-15} \,\rm{erg\,cm^{-2}\,s^{-1}}$ and 11 point-like X-ray sources at a flux limit of $F_{2-10 \,\rm{keV}} > 6 \times 10^{-14} \,\rm{erg\,cm^{-2}\,s^{-1}}$. These flux limits were chosen to correspond to the planned \textit{eROSITA} limits in the deep exposed ecliptic pole regions. Applying an \textit{r}-band cut of $17.0 \,\rm{mag} <  r_{\rm psf} < 22.5 \,\rm{mag}$ and Likelihood-Ratio of $\rm LR_{\rm XMM,SDSS}> 1.5$, there were 795 \textit{BOSS} targets, out of which 499 sources have been observed. In Fig. \ref{fig:XXLarea}, we show the position of the \textit{BOSS} spectra and the ancillary plate in the northern XMM-XXL area.
\\\\
\textit{Second \textit{BOSS} Ancillary Programme}
\\
The second group of spectra comes from the larger ancillary programme comprising four \textit{BOSS} plates led by A. Georgakakis. This programme was fully dedicated to the follow-up of point-like X-ray  sources covering nearly the entire XMM-XXL north area. Due to the larger amount of available fibres, we selected all 8445 extracted X-ray point-like sources in the $18\,\rm\,deg^2$ XMM-XXL north reaching to a flux of $F_{\rm 0.5-10\, keV} \approx 10^{-15} \rm\,erg\, cm^{-2}\, s^{-1}$ and we applied  $15.0 < r_{\rm psf/model} < 22.5 \,\rm{mag}$ (see Fig. \ref{fig:rband}) for optically unresolved sources ($\texttt{TYPE}=6$: psf-magnitude) and optically resolved sources ($\texttt{TYPE}=3$: model-magnitude). Within the $18\,\rm deg^2$, our dataset contains 3885 \textit{XMM-SDSS} matched sources within the corresponding \textit{r}-band threshold and with $\rm LR_{\rm XMM,SDSS}> 1.5$. The footprint of the four plates (see Fig. \ref{fig:XXLarea}) comprises 3461 \textit{XMM-SDSS} matched sources, out of which 2357 have been followed-up by \textit{BOSS}. 
\\\\
\textit{DR10 Spectra}
\\
Some of the X-ray sources from the second ancillary programme in the same field had already been independently targeted as part of the \textit{BOSS} large-scale structure survey programme as LRG (luminous red galaxies) or candidates for high-redshift QSOs. Thus, in order to increase the spectroscopic completeness of our sample within $15.0 < r_{\rm psf/model} < 22.5 \,\rm{mag}$, we included this third group of 401 public available spectra from 14 different plates within \textit{BOSS}-DR10 \citep{Ahn12}.
\\\\
Summarizing, we selected 8445 X-ray sources and matched 3885 \textit{SDSS} counterparts by applying the r-band criteria $15.0 < r_{\rm psf/model} < 22.5 \,\rm{mag}$ and Likelihood Ratio $\rm LR_{XMM,SDSS}>1.5$. From the \textit{BOSS} observations, we obtained 3302 spectra. This number reduces to 3042 unique objects after correction for 260 objects with multiple spectra (257 double observations, 3 triple observations) by prioritizing the spectra with the higher signal-to-noise ratio (S/N ratio). 2474 of the \textit{BOSS} spectra have a \textit{WISE} counterpart with good photometry.

\section[]{Spectroscopic Redshift and Classification of AGN}
\label{sec:BOSSspectra}
\label{sec:Classification}

In this section, we present the processing of optical spectra in the \textit{BOSS} pipeline and discuss the determination of spectroscopic redshifts by visual inspection, as well as objective classification of our X-ray sources based on their spectroscopic properties.

\subsection[]{BOSS pipeline products}
\label{sec:BOSSpipeline}

The \textit{BOSS} pipeline processes the observational raw data in two steps. In the first pass, the \texttt{spec2d} pipeline converts the 2-dimensional CCD data into 1-dimensional spectra, introduces calibrations in wavelength and flux, and combines the red and blue spectral halves. 
In the second pass, the \texttt{spec1d} pipeline automatically analyzes the features of the 1D-spectra and assigns a redshift and a classification with a $\chi^2$-minimization method  \citep{Stoughton02, Bolton12}. 
The relevant information about redshift and emission line features of the spectra are stored in the pipeline products  \texttt{spZall} file and \texttt{spZline} file (DR12, \texttt{v5\_7\_0}, \citealt{Bolton12}). 

For the determination of the redshift, the pipeline fits templates of galaxies ($-0.01<z<1.0$) and quasars ($0.0033<z<7.0$) with a linear combination of four principal components. The stars have dedicated standard templates. After the fitting process, the \textit{BOSS} pipeline assigns a quality flag (\texttt{ZWARNING}) to each redshift $\texttt{Z\_BOSS}$. The optimal flag is $\texttt{ZWARNING}=0$, which is assigned to 80 per cent of our spectra. Redshift identifications with $\texttt{ZWARNING}>0$ may be caused by two best fitting templates with similar $\chi^2$, outlying points from best-fit model or a minimal $\chi^2$ at the redshift edge. 

The line features of every spectrum are provided by the \texttt{spZline} data model of the \textit{BOSS} pipeline. We employ these data without any additional line fitting procedure. The line fit is performed for 31 emission lines with a single gaussian on top of the continuum subtracted spectrum. The redshift is re-fit nonlinearly after the initial guess of the main redshift analysis and fixed for all lines beside Ly$\alpha$. Blueshifts of individual lines are not accounted for. The line widths are calculated as a strength-weighted average of dedicated emission line groups, independent of the broad or narrow line character:
\begin{compactenum}[(i)]
\item Balmer lines: $\rm H\alpha$, $\rm H\beta$, $\rm H\delta$, $\rm H\gamma$, $\rm H\epsilon$ 
\item Lyman lines: $\rm Ly\alpha$ 
\item Nitrogen line: NV
\item other lines: [ArIII], [SII], [NII], [OI], [SIII], HeI, HeII, [OIII], [NeIII], [OII], MgII, CIII], CIV.
\end{compactenum}
The fluxes of [OIII] 5007 and [OIII] 4959 as well as [NII] 6583 and [NII] 6548 are imposed with a ratio of 3:1.  
The average width of group (iv) lines is less representative of the true width. It includes both narrow and broad lines, and therefore underestimates the individual broad lines widths and overestimates the individual narrow lines widths. In our work, we focus on 9 emission lines of interest ($\lambda$ in vacuum wavelength): 
CIV ($1549.48\,$\AA), CIII] ($1908.73\,$\AA), MgII ($2800.32\,$\AA), [OII]-doublet ($3727.09\,$\AA) and ($3729.88\,$\AA), H$\beta$ ($4862.68\,$\AA), [OIII] ($5008.24\,$\AA), H$\alpha$ ($6564.61\,$\AA), and NII ($6585.27\,$\AA). 
There are three emission line parameters from \texttt{spZline}, which are important for the classification: 
\begin{compactenum}[(i)]
\item the emission line flux $A_{\rm gauss}$ (\texttt{LINEAREA}) and the error $\Delta A_{\rm gauss}$ (\texttt{LINEAREA\_ERR}), 
\item the Full-Width-Half-Maximum $\rm FWHM$ derived from $\rm FWHM = \sigma \times (2 \sqrt{2\ln{2}})$, where $\sigma$ is the gaussian width in $\rm km\,s^{-1}$ (\texttt{LINESIGMA}), and 
\item the equivalent width $\rm EW$ (\texttt{LINEEW}). 
\end{compactenum}
We impose a significance threshold of  $A_{\rm gauss}/\Delta A_{\rm gauss}>3$ for all emission lines.

\subsection[]{Redshift determination}
\label{sec:reddet}
The redshifts for the \textit{BOSS}-observed AGN are taken from the \texttt{spZall} file. Typically, all QSO observed by \textit{BOSS} are candidates for baryonic oscillations studies and therefore pass a visual inspection \citep{Paris12, Paris14}. We expect the sources of our pilot study to be different from the majority of the standard \textit{BOSS}-LRG and -QSO targets which are selected based on optical criteria. They will include e.g. narrow emission line AGN, host galaxy dominated AGN and AGN with less steep powerlaws.  For this reason, we started a visual screening of our data set. 
 We evaluate the redshift provided by the pipeline \texttt{Z\_BOSS} by visual inspection and assign both a new redshift \texttt{Z} and confidence parameter \texttt{Z\_CONF}, as presented in Appendix \ref{app:visred}. The following redshift variables are provided in the catalogue:
\begin{compactenum}[(i)]
\item $\texttt{Z\_BOSS}$: the redshift provided by the \textit{BOSS} pipeline before visual inspection;
\item $\texttt{Z}$: the adapted redshift after visual inspection; 
\item $\texttt{ZERR}$: the error of \texttt{Z};
\item $\texttt{Z\_CONF}$: the redshift confidence after visual inspection:
\begin{compactenum}[]
\item 3  - reliable pipeline redshift,
\item 2  - not robust pipeline redshift,
\item 1  - bad spectrum,
\item 30 - reliable visual redshift and pipeline failure,
\item 20 - not robust visual redshift and pipeline failure,
\end{compactenum}
\item $\texttt{STAR/BLLAC}$: star or BL Lac flag.
\end{compactenum}

After the visual inspection, 2525 sources have a reliable initial redshift assigned by \textit{BOSS} pipeline, which corresponds to 83 per cent of redshift success. Additional 53 sources get a different reliable redshift, out of which 8 have only visual redshifts because they cannot not be correctly fitted by the pipeline ($\texttt{Z\_CONF} = 30$). For the classification of the spectra in the following sections, we only use the 2570 \texttt{BOSS} spectra with $\texttt{Z\_CONF}= 3$, because they have the complete data products from the \textit{BOSS} pipeline.

\subsection[]{Spectroscopic Classification}

\begin{figure*}
 \includegraphics[width=160mm]{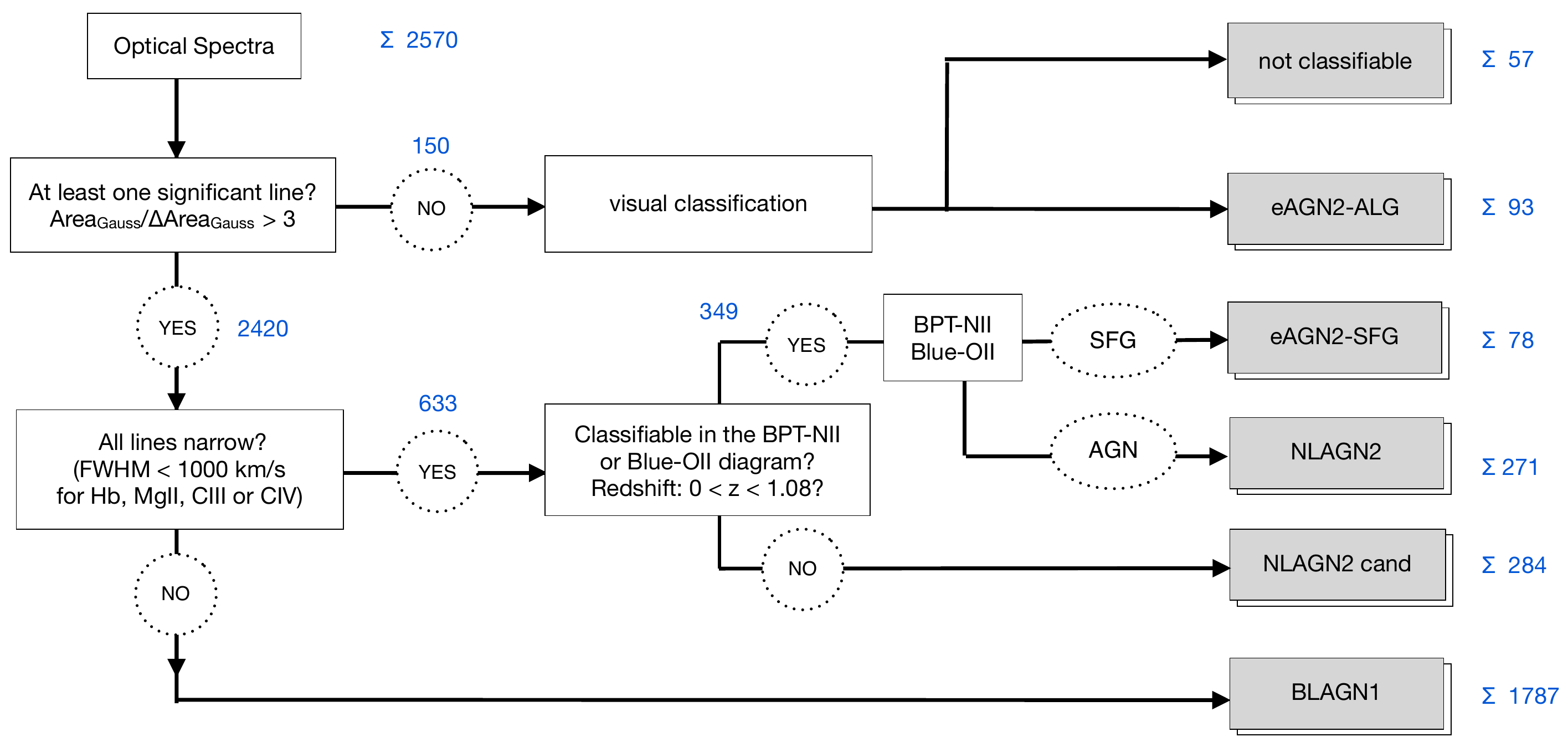}
 \caption{Classification flow-chart of \textit{BOSS} observed sources in the XMM-XXL north area: We use the emission line information provided by \texttt{spZline} from the \textit{BOSS} pipeline. For the classification, we refer to common optical AGN selection criteria and assign the classes: BLAGN1, NLAGN2/NLAGN2cand, eAGN(-ALG/-SFG) or not classifiable spectra.}
 \label{fig:flowchart02}
\end{figure*}

In this section, we present the classification rules for our X-ray selected sources based on spectroscopic properties. We first introduce the bimodal FWHM distribution of AGN emission lines to separate the population of broad line and narrow line emitters. In a second step, we determine the ionization source of the narrow line emitters with the help of optical emission line diagnostic diagrams. The flow chart in Fig. \ref{fig:flowchart02} visualizes the steps of the classification process.

\subsubsection{Line width bimodality of broad and narrow emission line emitters}
\label{sec:bimodal}

Our large sample of uniformly selected X-ray AGN allows for a systematic analysis of the FWHM of emission lines originating in the different regions of the AGN. For the classification process, we evaluate the properties of potentially broad lines H$\beta$, MgII, CIII] and CIV. As described in \ref{sec:BOSSpipeline}, the widths for H$\beta$ and MgII correspond to the strength-weighted averaged of their groups (i) and (iv). 

\begin{figure}
\includegraphics[width=85mm]{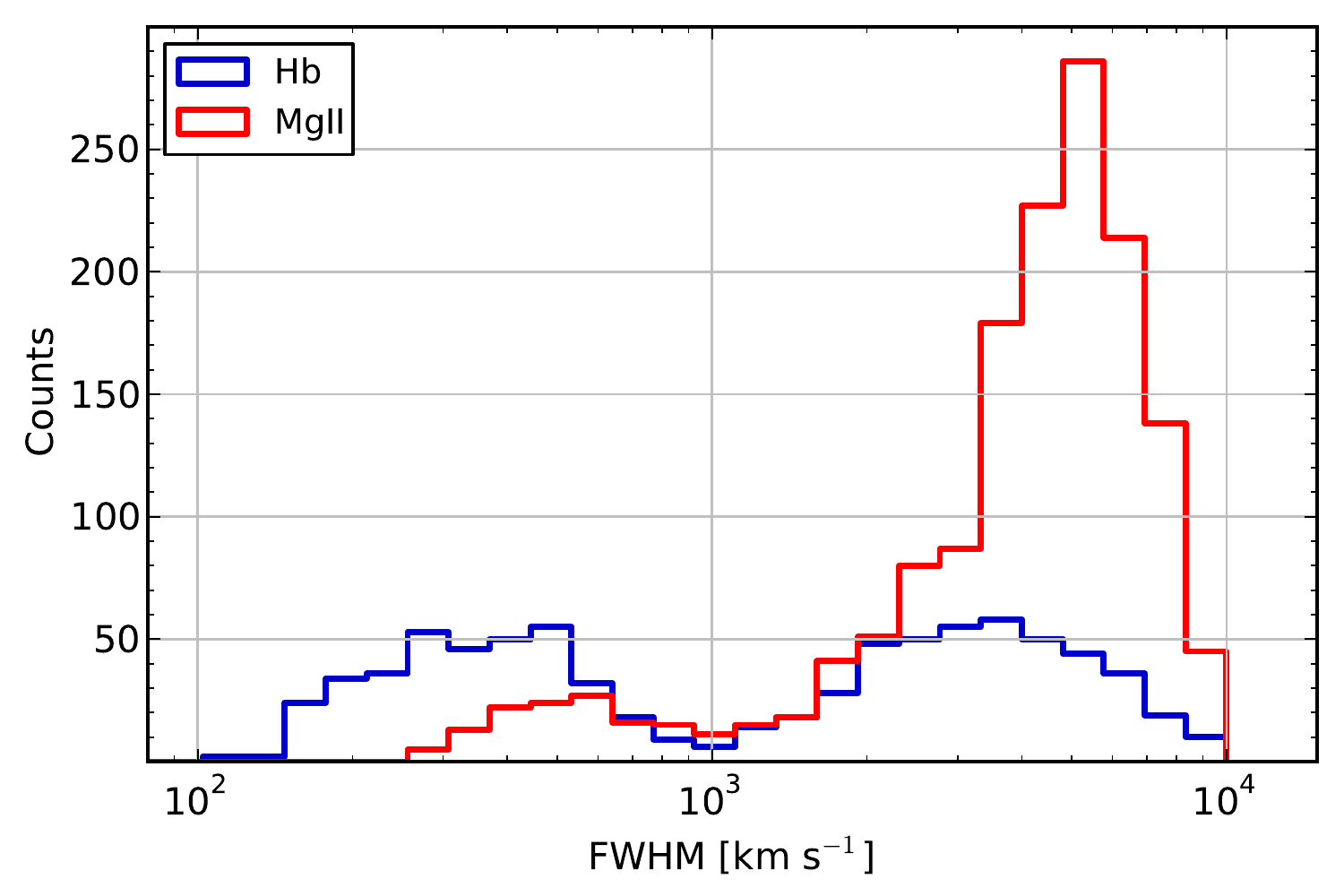}
 \caption{Average FWHM distribution of emission lines: We show the distribution of all significant H$\beta$ and MgII emission lines. The bimodal FWHM distribution implies the presence of two species: narrow line emitters ($\rm FHWM<1000\,\rm km\,s^{-1}$) and broad line emitters ($\rm FHWM\geq1000\,\rm km\,s^{-1}$).}
  \label{fig:hb fwhm}
\end{figure}

In Fig. \ref{fig:hb fwhm}, we show the FWHM-distribution of all significantly detected H$\beta$ and MgII emission lines. Both distributions show a clear bimodal shape, which suggests the presence of two physically distinct populations. In detail, the minima of the distributions are at $\rm FWHM_{(\rm H\beta)}=985\,\rm km\,s^{-1}$ and $\rm FWHM_{\rm MgII}=957\,\rm km\,s^{-1}$.  The high-FWHM sources are associated with broad line region (BLR) emission and the low-FWHM sources are associated with narrow line region (NLR) emission. In the following, we will use the FWHM threshold of $\rm FWHM=1000 \,\rm km\,s^{-1}$ to separate BLR and NLR emitters. 

For our classification, we analyze the FHWM of the H$\beta$, MgII, CIII] and CIV lines in sequence. The first significant emission line, which has a FWHM larger than $1000 \,\rm km\,s^{-1}$ defines an object to be a BLR emitter. For $z>1.14$, H$\beta$ is redshifted out of the \textit{BOSS} wavelength range. In this case, the width of MgII, CIII] or CIV are solely used for the classification. We find that at these redshifts, no more NLR emitters with significant emission lines are present in the sample. Therefore, the averaged FWHM for the group (iv) lines is not underestimated by strong narrow emission line contribution. 
\\\\
In the following sections, we will refer to the BLR emitters as `unobscured' (referring to type 1 AGN) from the optical point of view and call them `Broad Line AGN [of type] 1': \textbf{BLAGN1}. Their optical spectra show the blue/UV continuum from the accretion disc and the characteristic H$\beta$, MgII, CIV and CIII] broad emission lines from the high velocity clouds close to the black hole. 
\\\\
The NLR emitters with $\rm FWHM_{(\rm H\beta)}<1000\,\rm km\,s^{-1}$ can have different ionization sources, which will be discussed in the following section.

\subsubsection{Ionization source diagnostics of narrow emission line emitters}
\label{sec:BPT}
 
The ionizing radiation that excites the NLR ($\rm FWHM < 1000 \rm\,km\,s^{-1}$) of X-ray detected AGN candidates can originate from different regions in the galaxy: the accretion disc in the nuclear region (AGN), star formation in the host galaxy (SFG) or both. The narrow line emitters powered by an AGN typically have an orientation which prevents a direct look into the accretion disc  \citep{Antonucci93, Urry95}. The BLR clouds close to the black hole are obscured and only narrow lines from distant clouds in low density environments  are visible in the spectrum. The excitation of narrow emission lines by star formation typically happens through photoionization, e.g. in OB stars.

As is customary, we derive the dominating source of ionization for our sample of narrow line emitters between  $0<z<1.08$ by comparing the emission line ratios [OIII]/H$\beta$ vs. [NII]/H$\alpha$ (BPT-NII-diagram, \cite{Melendez14}) or [OIII]/H$\beta$ vs. [OII]/H$\beta$ (Blue-OII-diagram, \cite{Lamareille10}), as presented in Appendix \ref{app:linediag}. AGN have a stronger ionization continuum than SFG and typically reside in the top right corner of the diagrams, whereas SFG reside in the bottom left corner. In total, our dataset contains 633 narrow line emitters out of which 349 sources can be classified by the diagnostic lines diagrams. There are 271 `AGN' (BPT-NII-diagram: 244, Blue-OII-diagram: 27) which form the group of optically `obscured' AGN (referring to type 2 AGN) and, in this work, are named `Narrow Line AGN [of type] 2: \textbf{NLAGN2}. 
Furthermore, there are 78 `SFG' (BPT-NII-diagram: 40, Blue-OII-diagram: 38) which will be referred to as `elusive' AGN with star-formation: \textbf{eAGN-SFG}. They do not show any AGN-driven emission lines in their optical spectra, but their X-ray luminosities are characteristic of an active accretion process in the centre.

The ionization origin of the remaining 284 narrow line emitters cannot be reliably determined, because the required narrow emission lines are below the significance limit or the redshift of the objects is too high ($z>1.08$). The low significance of the lines  is caused by very strong host galaxy continuum contribution or very low S/N ratio spectra.  We classify this group as NLAGN2 candidates: \textbf{NLAGN2cand} and refer to their specific properties in Section \ref{sec:NLAGN2prop}.

Our dataset also comprises objects whose spectra do not have any significant emission lines at all. Similarly, to the eAGN-SFG, they indeed have characteristic X-ray luminosities for AGN, but only show the features of passive galaxies in their optical spectra. We classify them as `elusive' absorption line galaxies: \textbf{eAGN-ALG}. For their selection, we apply a significance threshold of $A_{\rm gauss}/\Delta A_{\rm gauss}=3$ to every emission line in the spectrum. 
Below this significance threshold, there is a large group of sources which are not eAGN-ALG. These are sources whose spectra mainly suffer from very low S/N ratio and do not show any stellar continuum or absorption lines at all. It is possible to derive a secure redshift by visual inspection, but our classification method is not able to automatically separate this group of objects. Therefore, we apply an additional conservative visual inspection to all of these objects. After the inspection, we sort them into two groups: 93 `eAGN-ALG' and 57 `not classifiable' spectra: \textbf{NOC}.

\subsection{Final classified sample}

Summarizing our final sample with 2570 spectra of optical counterparts to X-ray sources ($\texttt{Z\_CONF}=3$) with emission line information, we classified:
\begin{compactenum}[(i)]
\item 1787 BLAGN1 (70 per cent),
\item 271 NLAGN2  (11 per cent),
\item  284  NLAGN2cand (11 per cent), 
\item  78  eAGN-SFG (3 per cent),
\item  93 eAGN-ALG (4 per cent) and 
\item  57  not classifiable spectra  (2 per cent). 
\end{compactenum}
The percentages relate to the AGN selection only. We point out that our data set includes also 85 X-ray detected stars and 2 X-ray detected BL Lac.

\section{Class properties}
\label{sec:Classproperties}

  \begin{figure}
\includegraphics[width=85mm]{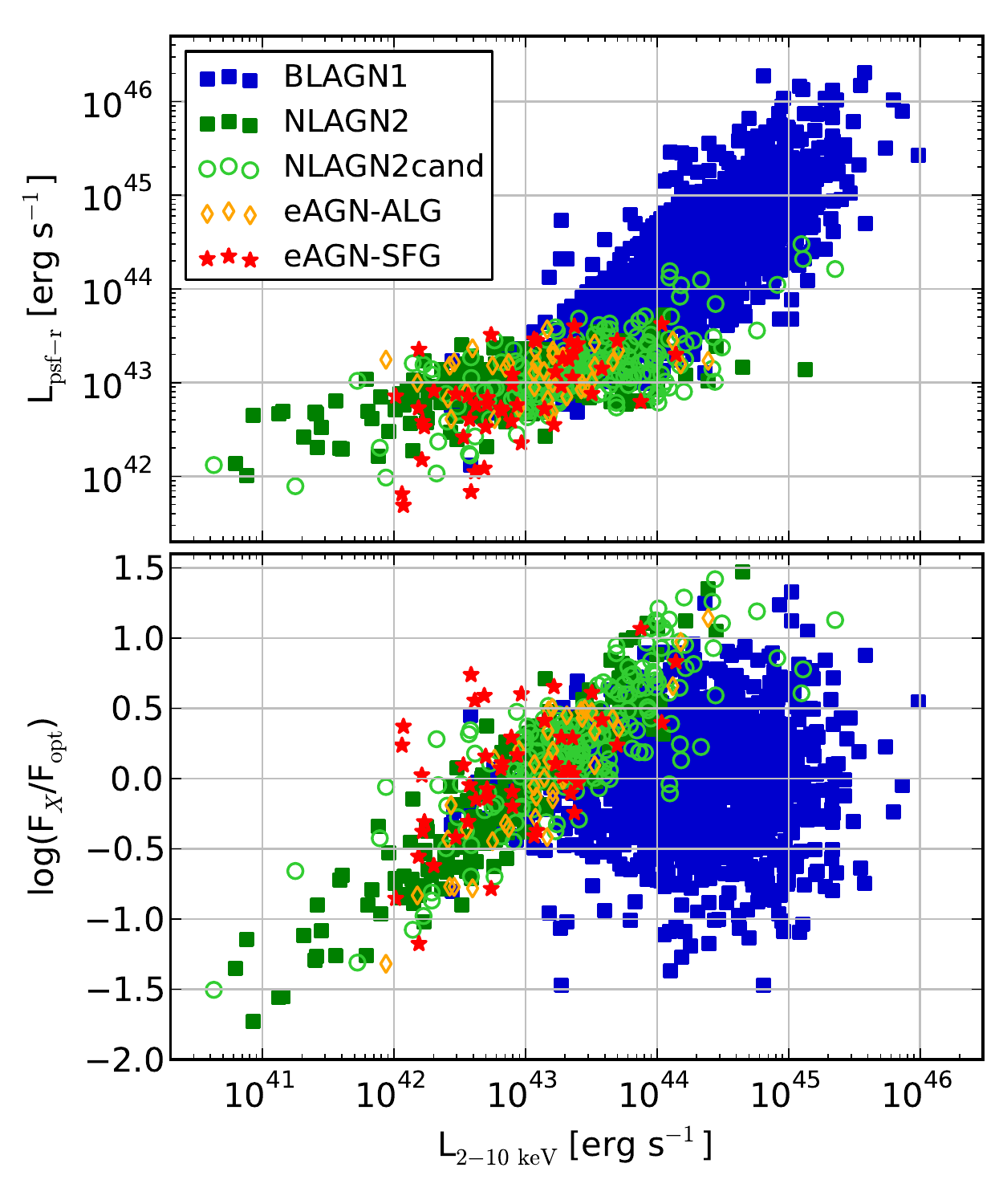}
 \caption{Top Panel: Optical luminosity ($r_{\rm psf}$-band) and hard
X-ray luminosity ($2-10\, \rm keV$). Bottom Panel: X-ray to optical flux
ratio and hard X-ray luminosity distribution of \textit{BOSS} observed
and classified AGN.}
  \label{fig:logxo}
\end{figure}

\begin{figure*}

\includegraphics[width=180mm]{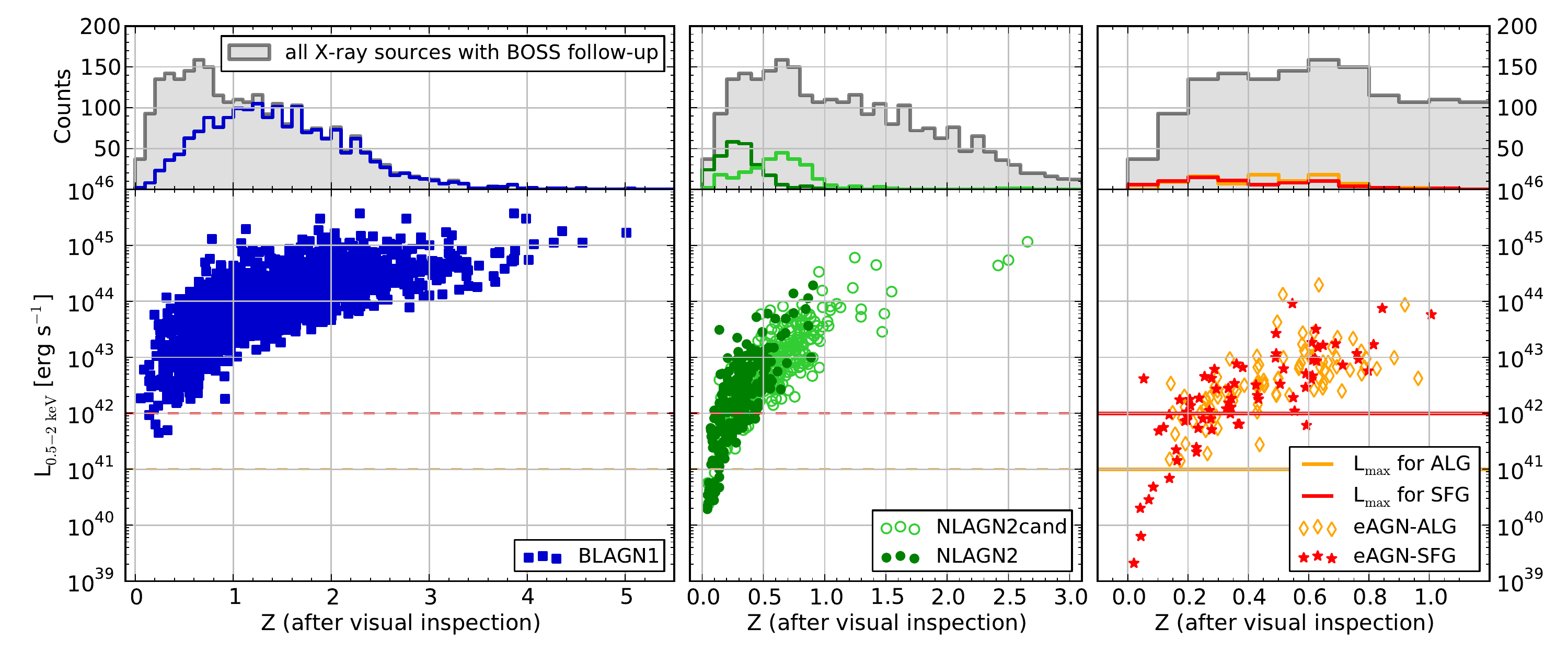}
 \caption{Luminosity and visual redshift distribution of classified \textit{BOSS} spectra. Left panel: BLAGN1. Central panel: NLAGN2 and NLAGN2cand. Right Panel: eAGN-SFG and eAGN-ALG. The luminosity thresholds indicate the upper limit of star forming galaxies (red) and absorption line galaxies (yellow) without AGN. In the histogram, we indicate the redshift distribution of all X-ray selected spectra (grey).}
 \label{fig:TY1}
 \end{figure*}

\begin{figure*}
 \includegraphics[width=175mm]{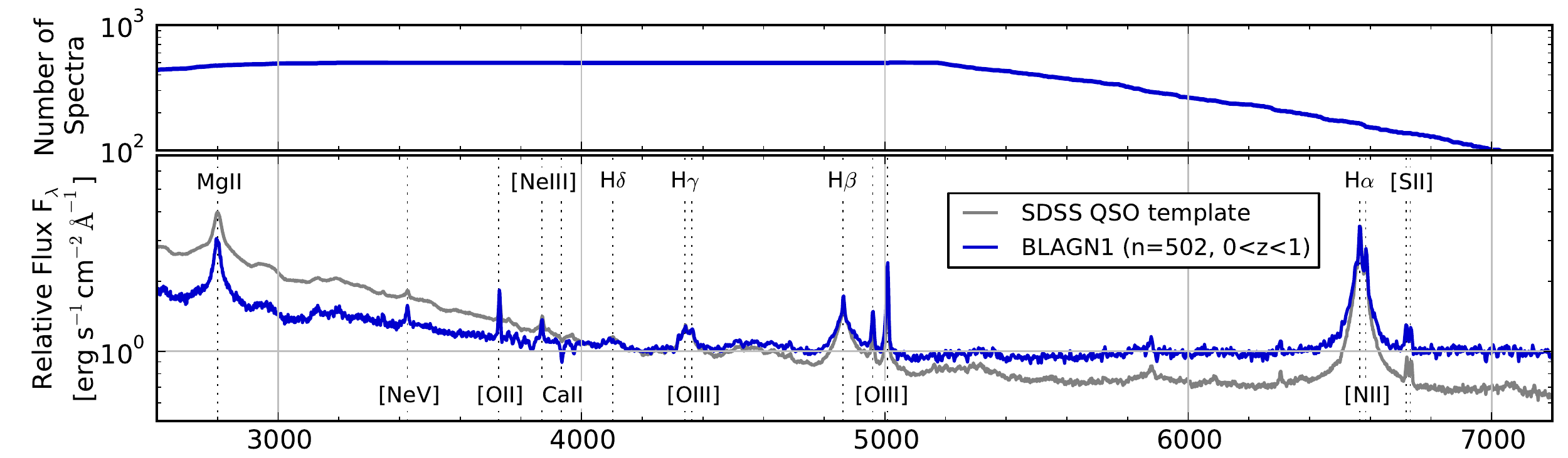}
 \includegraphics[width=175mm]{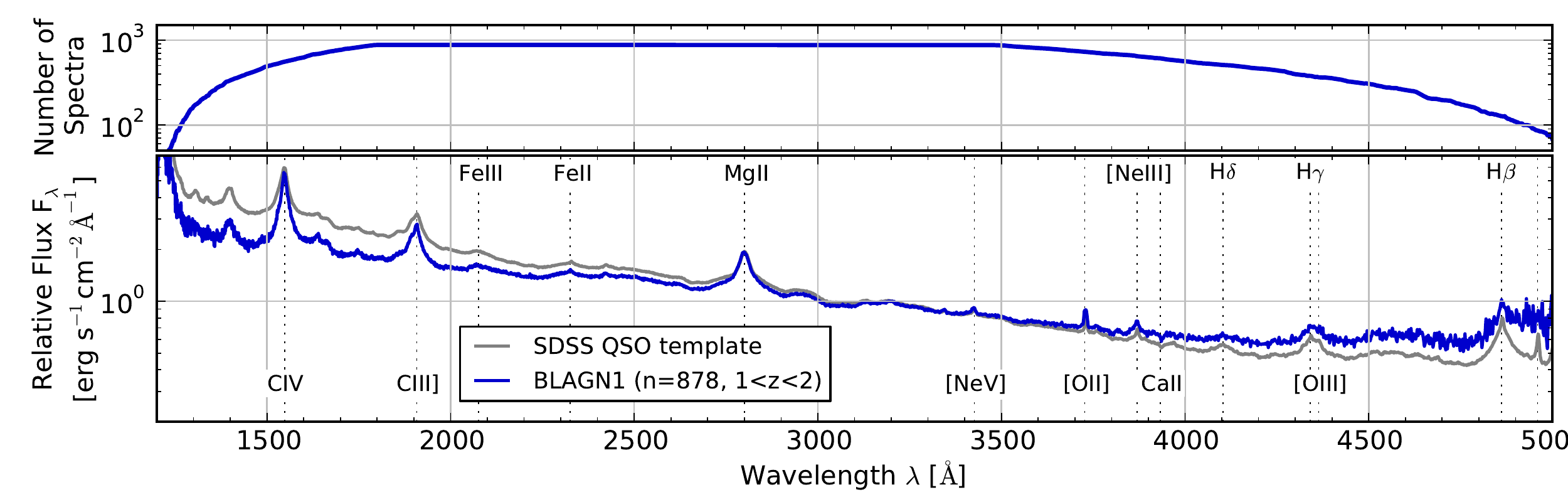}
 \includegraphics[width=175mm]{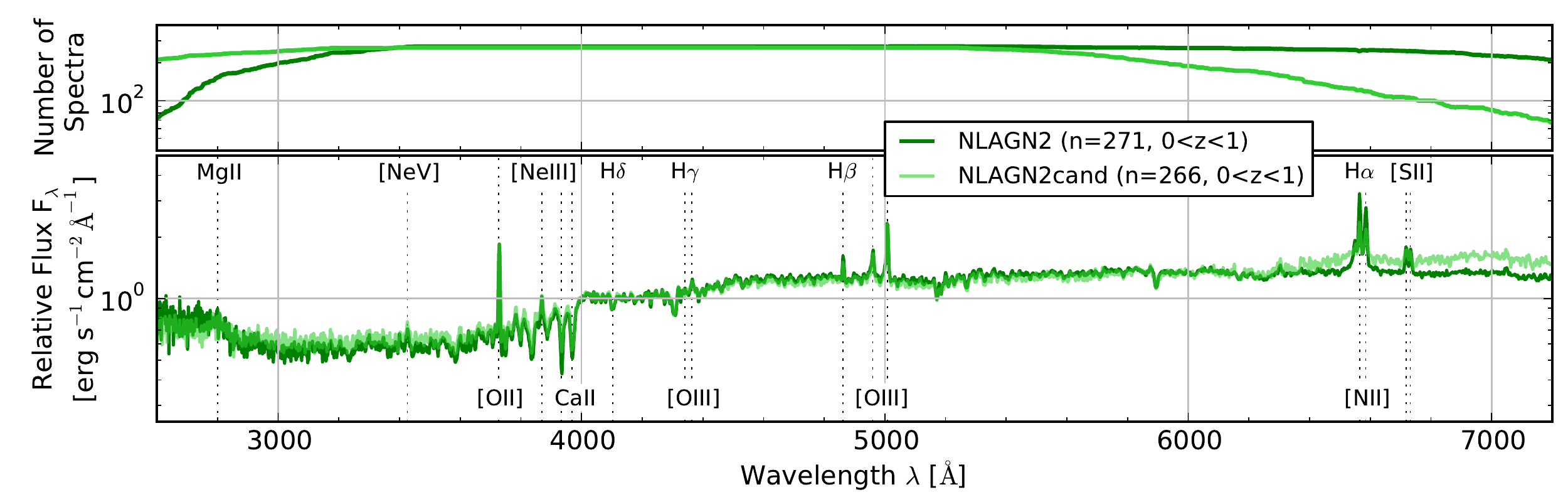}
 \includegraphics[width=175mm]{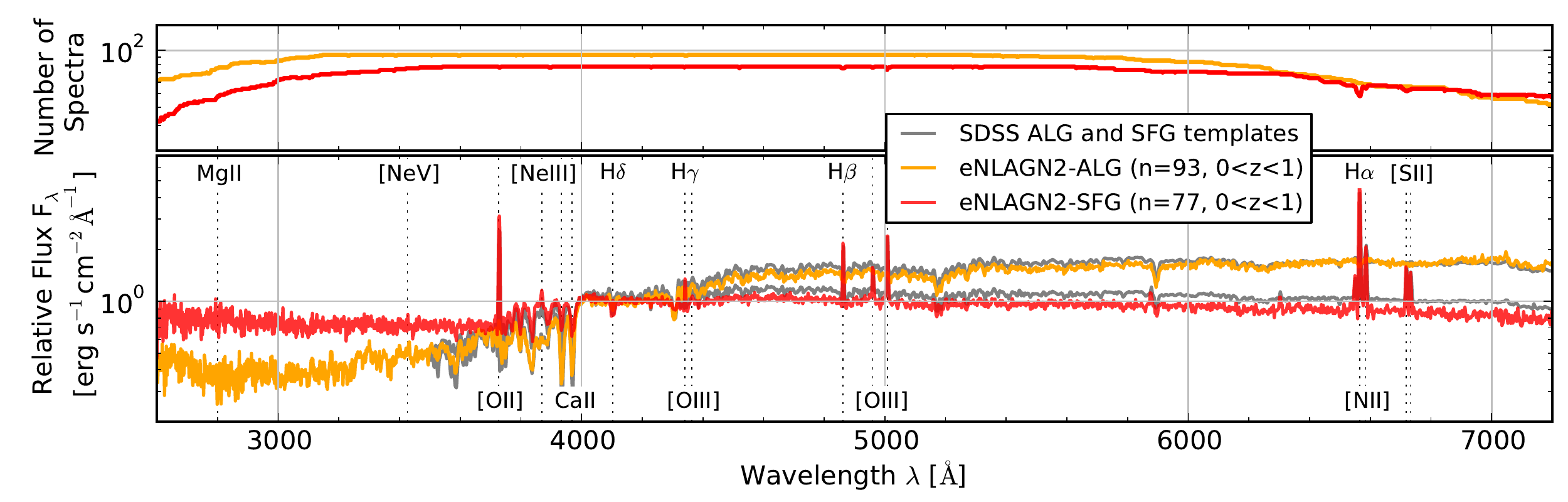}
 \caption{Top Panel: Median spectral stack and number of contributing spectra for BLAGN1 (blue) and the \textit{SDSS} template of a QSO (grey) \citep{Berk01} in the redshift range of $0.0<z<1.0$ (normalized at $4200\,$\AA) and $1<z<2$ (normalized at $3200\,$\AA).  Central Panel: Median spectral stack and number of contributing spectra for NLAGN2 and NLAGN2cand in the redshift range of $0.0<z<1.0$ (normalized at $4200\,$\AA). Bottom Panel: Median spectral stack and number of contributing spectra for eAGN-SFG, eAGN-ALG and the \textit{SDSS} template of a passive and active galaxy \citep{Yip04} in the redshift range of $0.0<z<1.0$ (normalized at $4200\,$\AA).}
    \label{fig:specstack}
 \end{figure*}

The optical classification of the X-ray selected AGN candidates from the previous section is also confirmed by their X-ray properties. The top panel of Fig. \ref{fig:logxo} presents the distribution of our classified \textit{BOSS} spectra in the plane of the optical \textit{r}$_{\rm psf}$ luminosity $L_{\rm opt}$ to hard X-ray luminosity $L_{x}$. The sources follow the distributions of \citet{Fiore03}. The BLAGN1 sequence is steep and shows a broad correlation between $L_{x}$ and $L_{\rm opt}$, both being tracers of accretion power. The NLAGN2/NLAGN2cand and eAGN-ALG/eAGN-SFG reside at lower optical luminosity ranges and have a flatter distribution. This is caused by the dominating host continuum in the optical bands, which only weakly correlates with X-ray luminosities. The bottom panel of Fig. \ref{fig:logxo} shows the X-ray to optical flux ratio which is calculated as follows:
\begin{equation}
\log\left(\frac{F_{2-10\,\rm keV}}{F_{\textit{r-band}}}\right) = \log F_{2-10\,\rm keV} - \left(-5.37 - \frac{\textit{r}_{\rm psf}}{2.5}\right).
\end{equation}
The populations of BLAGN1 and NLAGN2/eAGN have different locations in this diagram. For $L_{0.5-2\,\rm keV}\sim 10^{42} - 10^{44}\rm\, erg\,s^{-1}$, the NLAGN2/eAGN reside at higher X-ray to optical flux ratio, which is another clear indication for their optical nuclear obscuration. 
 
In the following paragraphs we describe in more detail the redshift, X-ray and optical spectra properties of each class.
We complement the analysis with spectral stacks, which are performed using the idl script \texttt{run\_composite} written by Min-Su Shin (January 2009). In this process, the contributing spectra are normalized to a chosen wavelength and corrected for Galactic extinction as well as the \textit{BOSS} spectroscopic resolution.

\subsection{BLAGN1}

The BLAGN1 span a large redshift range of $0.06<z<5.01$ and a luminosity range of $4.4 \times 10^{41} < L_{\rm 0.5-2\,keV} < 3.7\times 10^{45} \, \rm{erg\,s^{-1}}$ (see Fig. \ref{fig:TY1}). The upper redshift limit is strongly related to the faint \textit{SDSS} \textit{r}-band magnitude limit for the \textit{BOSS} targets. This limit hampers the observation of objects at $z>4.5$. At higher redshift, the Ly$\alpha$ absorption from the intergalactic medium strongly suppresses the spectroscopic features in the observed \textit{r}-band. Our sample contains one exception at $z=5.01$, which is caused by a Ly$\beta$-line in the \textit{r}-band ($ r_{\rm psf}=20.92\rm\, mag$). 

In the two top panels of Fig. \ref{fig:specstack}, we show the spectral stack of all BLAGN1 in the redshift ranges of $0<z<1$ and $1<z<2$, respectively. For comparison, we overplot the \textit{SDSS} DR4 QSO template \citep{Berk01}. The BLAGN1 show clear broad line features for the emission lines MgII, H$\beta$ and H$\alpha$. At low redshifts, the spectral stack of the BLAGN1 tends to be redder and flatter than the \textit{SDSS} QSO template. This is caused by the contribution of the host galaxy, indicated by the prominent Ca-doublet absorption and the continuum at $\lambda>4000\rm \,$\AA. At higher redshifts, the stack approaches the QSO template because the host contribution is outshone by the more luminous nuclear emission. The stack is still redder than the template, because our BLAGN1 have different optical colours than the QSO selection criteria and include mildly extinguished objects (see Section \ref{sec:XDQSO}).

\subsection{NLAGN2 and NLAGN2cand}
\label{sec:NLAGN2prop}

The NLAGN2 are detected at a relatively low redshift range of $0.04<z<0.91$ and a luminosity range of $ 1.9\times 10^{40} < L_{0.5-2\rm keV} < 1.9\times 10^{44}\,  \rm{erg\,s^{-1}}$ (see Fig. \ref{fig:TY1}). The upper redshift limit of NLAGN2  is hampered  by the optical line diagnostic diagrams. Within the redshift range of $0.5 < z < 0.9 $, the NLAGN2 magnitude distribution reaches $\textit{r}_{\rm model}=22.5\,\rm mag$ and the dominating host contribution becomes too faint for detection in SDSS images. Additionally, the fraction of obscured AGN is known to decrease at higher luminosities \citep{Ueda03, Hasinger08, Merloni14}. Thus, the high redshift and high luminosity objects in our sample tend to be more dominated by BLAGN1.

In the central panel of Fig. \ref{fig:specstack}, we show the spectral stack of all NLAGN2. We chose the redshift range of $0<z<1$ which corresponds to the upper redshift bound of the optical emission line diagrams. The stack displays the narrow AGN emission lines occurring for [OII],  H$\beta$, [OIII], H$\alpha$ and NII. Furthermore, there are strong host galaxy features, such as a clear Ca-doublet with the $4000\,$\AA-break and the stellar continuum in the red part of the spectrum. 

We now introduce the group of NLAGN2cands for narrow line objects whose emission origin cannot be determined by the BPT-NII/Blue-OII diagrams. We find that these sources either have spectra with strong host contribution, low S/N ratio or they are narrow line emitters at $z>1.08$. They span a redshift range of $0.04<z<2.66$ and luminosity range  of $2.4\times 10^{40} < L_{0.5-2\rm keV} < 1.2\times 10^{45}\,\rm{erg\,s^{-1}}$ (see Fig. \ref{fig:TY1}). Their spectral stack ($0<z<1$) in Fig. \ref{fig:specstack} reveals the continuum features of pure NLAGN2 but also the slightly broadened lines of BLAGN1. Furthermore, the optical \textit{SDSS} morphology of 94 per cent of NLAGN2cand is extended  and their optical colours differ from QSO (see Section \ref{sec:XDQSO}). Overall, it suggests that the majority of the sources are NLAGN2, which is reflected in their naming. Deeper exposed optical spectra or infrared spectra are needed to properly classify these objects. The three outstanding NLAGN2cand at $2.42<z<2.66$ have low S/N ratio spectra and show broad emission line features below the required significance threshold. They are probably candidates for faint BLAGN1.

\subsection{eAGN-ALG and eAGN-SFG}

The group of elusive AGN consists of sources whose spectra are consistent with either absorption line galaxies (ALG) or star-forming galaxies (SFG). 
 The eAGN-ALG reside within a redshift range of $0.14<z<0.96$ and a luminosity range of $ 1.4\times 10^{41} < L_{0.5-2\rm keV} < 2.0\times 10^{44}\,  \rm{erg\,s^{-1}}$ (see Fig. \ref{fig:TY1}). Their median soft X-ray luminosity is $L_{0.5-2\rm keV} = 3.43 \times 10^{42}\,  \rm{erg\,s^{-1}}$. The eAGN-SFG span a redshift range of $0.02<z<1.01$ and a luminosity range of $ 2.1\times 10^{39} < L_{0.5-2\rm keV} < 9.1\times 10^{43}\,  \rm{erg\,s^{-1}}$. Their median soft X-ray luminosity is at $L_{0.5-2\rm keV} = 2.89 \times 10^{41}\,  \rm{erg\,s^{-1}}$ (see Fig. \ref{fig:TY1}). The counterparts of the eAGN have the same distribution in \textit{r}-band, S/N ratio, Likelihood ratio, separation distance and number of counterparts as the entire population. Therefore, they should have the same probability distribution of counterpart association as the other objects.
 
The spectral stack of eAGN-ALG in the bottom panel of Fig. \ref{fig:specstack} only shows absorption line features and is very similar to the typical \textit{SDSS} DR4 early type galaxy templates \citep{Yip04}. On the other hand, the eAGN-SFG reveal the stellar emission lines, e.g. [OII] and [OIII], and resemble the \textit{SDSS} DR4 template for late type galaxies \citep{Yip04}. 
\\\\
\textit{What is the engine of elusive AGN?}
\\
In order to determine whether these objects really host an AGN we study their soft X-ray luminosities.
We can assume that ALG have a maximal stellar mass of $\sim 10^{12} M_{\odot}$, based on the massive \textit{BOSS} LRG sample of \citet{Maraston13}. According to \citet{Gilfanov04}, this corresponds to an X-ray luminosity of $L_{0.5-2\rm \, keV}\leq10^{41}\,\rm erg\,s^{-1}$ caused by low mass X-ray binaries in the galaxy. The soft X-ray luminosity of all 78 eAGN-ALG in our sample is above $10^{41}\,\rm erg\,s^{-1}$. Therefore, they can be considered as hosts of AGN and are not dominated by luminous massive X-ray binaries.
 
Focussing on the SFG population, their overall X-ray emission  - apart from the nuclear AGN - consists of the collective emission from stellar remnants in a galaxy. The upper limit on the X-ray luminosity can be associated with the maximum of the total Star Formation Rate \citep{Nandra02, Ranalli03, Grimm03}. Typically, the value of $L_{0.5-2\rm \,keV}\sim10^{42}\rm \, erg\,s^{-1}$ is considered as a threshold above which only AGN should be mainly responsible for the observed emission. Our sample of eAGN-SFG consists of 49 objects above this luminosity limit and 29 objects below this luminosity limit, corresponding to a ratio of $0.59\pm0.40$.
Comparing the number fraction to securely classified NLAGN2 in the same redshift and luminosity range, we obtain $0.72\pm 0.18$. The consistency of both ratios might be an indication that the majority of our eAGN-SFG with $L_{0.5-2\rm \,keV}<10^{42}\rm \, erg\,s^{-1}$ in our sample really host a comparably weak AGN.

\section[]{Assessment of optical and mid-IR AGN colour selection techniques }
\label{sec:Colorproperties}

One of the expected strengths of the  X-ray AGN selection is the ability to identify accreting black holes which 
are either obscured, or heavily contaminated by stellar emission processes at optical and NIR wavelengths. Indeed, our classification work presented in the previous section clearly demonstrates that among X-ray selected AGN, we find a variety of objects with a quite diverse set of optical properties. This significant fraction of AGN would otherwise not have been selected by other selection criteria.

In order to further evaluate how X-ray selection of AGN compares with photometric optical and mid-infrared colour selection, we analyze in this section the optical and infrared colour properties of our sample of X-ray selected AGN, and we compare it to the overall population of XDQSO (Section \ref{sec:XDQSO}) and \textit{WISE} (Section \ref{sec:WISEAGN}) colour-selected AGN in the XMM-XXL north.

\subsection[]{Optical: XDQSO targeting algorithm}
\label{sec:XDQSO}
For the optical selection of AGN, we refer to the XDQSO algorithm \citep{Bovy11}, which is a probabilistic selection technique developed
 for the efficient QSO selection on the basis of broad band optical imaging and point-like morphology. 
It has been employed as a target selection method for the measurement of baryon acoustic features with quasars within the \textit{eBOSS} programme (Myers et al. 2015, in prep.). The selection is limited to optically point-like objects ($\texttt{TYPE}=6$), with a dereddened \textit{i}-band magnitude of $17.75<\textit{i}<22.45\,\rm mag$ and at least one primary detection with $\textit{u}_{\rm psf}<22.5\,\rm mag$, $\textit{g}_{\rm psf}<22.5\,\rm mag$, $\textit{r}_{\rm psf}<22.5\,\rm mag$, $\textit{i}_{\rm psf}<22.0\,\rm mag$ or $\textit{z}_{\rm psf}<21.5\,\rm mag$. We only choose sources with a good \textit{BOSS} quasar targeting flag ($\texttt{GOOD}=0$, \citealt{Bovy11}) excluding the photometric blending, moving sources and interpolation effects. The algorithm uses the density estimation in flux space and assigns probabilities $P(\rm{QSO},z)$ 
of any \textit{SDSS} point source to be a QSO in dedicated redshift ranges. In this work, we refer to the quasar probability $P_{\rm sum}(\rm QSO)$  as the sum of the low-redshift ($z<2.2$), mid-redshift ($2.2\leq z\leq3.5$) and high-redshift ($z>3.5$) probabilities. As a threshold for the AGN selection, we define the arbitrary probability of $ P_{\rm sum}(\rm QSO)>0.5$. The algorithm has been trained using a star sample from \textit{SDSS} Stripe 82 and a quasar sample from the \textit{SDSS} DR7 Quasar catalogue.

  \begin{figure}
\includegraphics[width=82mm]{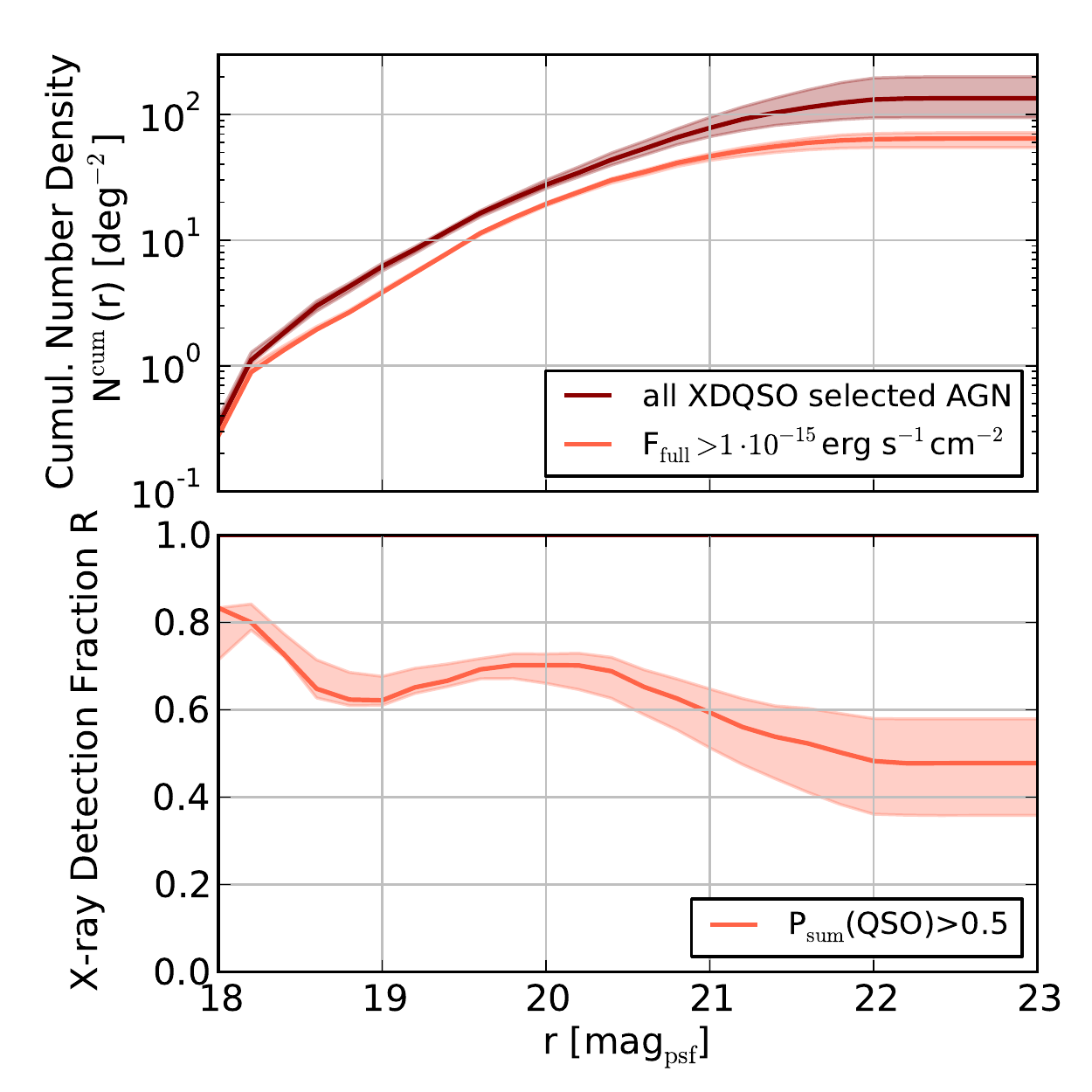} 
 \caption{Top Panel: Cumulative number density of all XDQSO selected AGN and of XDQSO AGN with X-ray counterparts in the XMM-XXL north field. We apply the optical selection threshold of $P_{\rm sum}(\rm  QSO)>0.5$ (central line) and $P_{\rm sum}(\rm  QSO)>0.8$ (lower border) as well as $>0.2$ (upper border).  Bottom Panel: X-ray detection fraction of XDQSO selected AGN. We divide the number of X-ray detected AGN with $P_{\rm sum}(\rm  QSO)>0.5$ over all XDQSO sources with $P_{\rm sum}(\rm  QSO)>0.5$ in the XMM-XXL north.}
 \label{fig:XDQSOfrac}
\end{figure}

\subsubsection{X-ray detection fraction}

In the $18\,\rm deg^2$ coverage area of \textit{XMM-Newton}  in the XMM-XXL north, the XDQSO catalogue contains $49\,172$ sources with assigned $P(\rm{QSO},z)$ and $\texttt{GOOD}=0$. There are 1617 XDQSO sources which can be associated to the \textit{XMM-SDSS} catalogue ($\rm LR_{\rm XMM,SDSS}> 1.5$). Considering a QSO probability of $ P_{\rm sum}(\rm QSO)>0.5$, the entire XMM-XXL north comprises 2408 XDQSO sources out of which 1159 are also X-ray selected. 
\\\\
In Fig.  \ref{fig:XDQSOfrac}, we show the detection fraction of X-ray selected XDQSO sources. We define the number density $N_{\rm XDQSO, X}$ of X-ray selected XDQSO sources and the number density $N_{\rm XDQSO}$ of all XDQSO selected sources in the XMM-XXL. The ratio $R$ of the cumulative number density (hereby X-ray detection fraction) is: 
\begin{equation*}
R=\frac{N^{cum}_{XDQSO,X}(\textit{r}, F_X,P_{\rm sum}(\rm QSO))}{N^{cum}_{XDQSO}(\textit{r}, P_{\rm sum}\rm (QSO))}.
\end{equation*}
and depends on the optical magnitude \textit{r}, the X-ray flux $F_X$ and the QSO probability 
$P_{\rm sum}(\rm QSO)$, as shown in Fig. \ref{fig:XDQSOfrac}.  We chose the \textit{r}-band magnitude to account for the magnitude threshold of the \textit{BOSS} observed sources. 
The individual QSO probabilities of the XDQSO sources are normalized and the sum of individual probabilities corresponds to the expected number of true QSO. For the group of 2425 XDQSO sources with $P_{\rm sum}(\rm QSO)>0.5$ in the XMM-XXL, we derive an averaged reliability of 86 per cent. The fractions in Fig. \ref{fig:XDQSOfrac} are not corrected for such estimated reliability.
\\\\
Assuming $P_{\rm sum}(\rm QSO)>0.5$ and $F_{0.5-10\rm keV}>10^{-15}\,\rm erg\, cm^{-2} \, s^{-1}$,  $\sim48$ per cent of all XDQSO selected sources will be picked up in X-ray at faint optical magnitudes ($r\leq22.5 \rm\, mag$) and up to $\sim 72$ per cent at bright \textit{r} magnitudes ($r\leq18.5\rm\, mag$).

\subsubsection{Morphological and spectroscopic properties of XDQSO-selected AGN}

\begin{figure*}
\includegraphics[width=180mm]{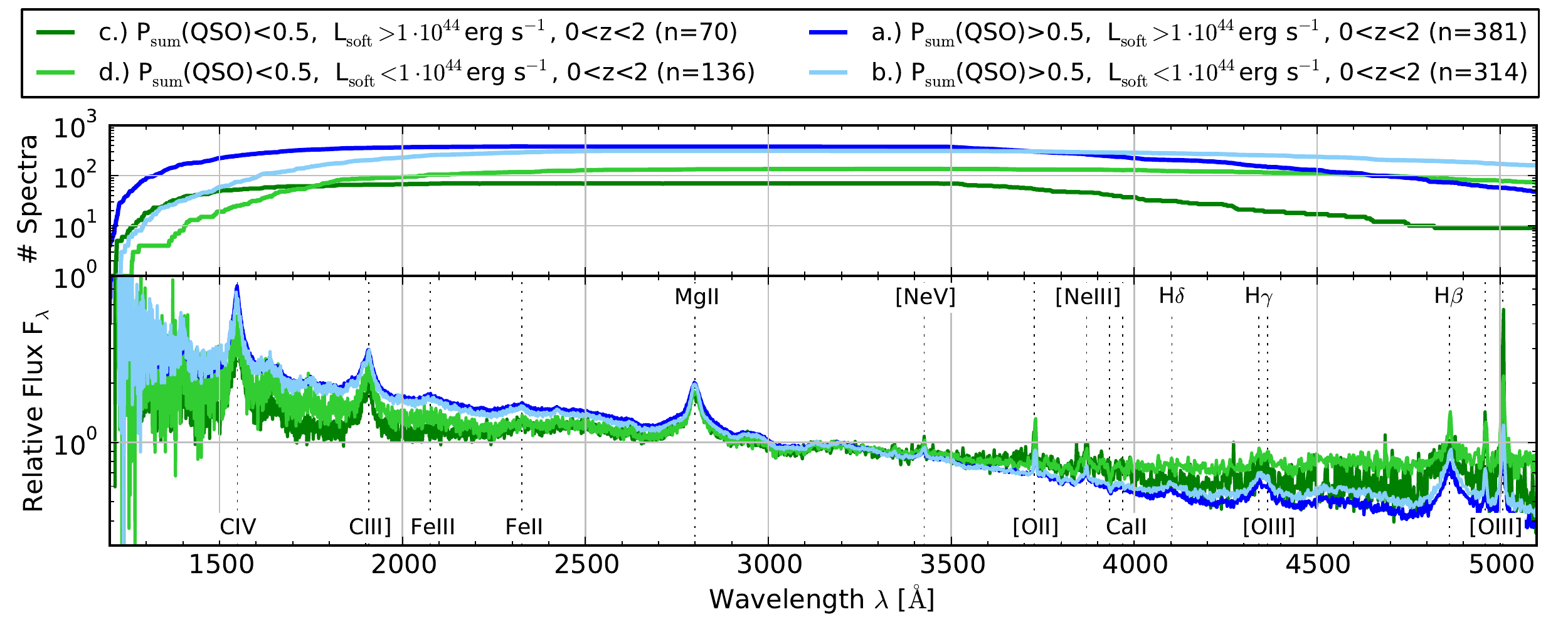}
 \caption{Median spectral stack and number of contributing spectra for optically point-like objects with a.) high soft X-ray luminosity objects and $\rm P(QSO)>0.5$, b.) low soft X-ray luminosity objects and $\rm P(QSO)>0.5$, b.) high soft X-ray luminosity objects and $\rm P(QSO)<0.5$, and b.) low soft X-ray luminosity objects and $\rm P(QSO)<0.5$. The spectra have a redshift of $0.0<z<2.0$ and are normalized to $\lambda=3200\,$\AA. }
 \label{fig:stackXDQSO}
   \end{figure*}

\begin{table}
    \caption{XDQSO properties \citep{Bovy11} of X-ray sources with spectroscopic information: We split the data set of X-ray selected AGN with reliable \textit{BOSS} follow-up into the different classes and optical morphology. We indicate number of XDQSO detection for each population.}
 \centering
\begin{tabular}[width=20mm]{lccccccccc} 
  \hline
spectroscopic & optical  & \multicolumn{3}{c}{ $P_{\rm sum}\rm (QSO)$} \\
classification & morph. &no  &$>0$&$>0.5$\\
\hline
BLAGN1 (n=1353)  & point-like & 196 & 1157 & 889 \\
BLAGN1 (n=434)& extended & 434 & - & - \\
NLAGN2 (n=7) & point-like & 1 & 6 & 1 \\
NLAGN2 (n=266) & extended & 266 & - & - \\
NLAGN2cand (n=32) & point-like & 16 & 16 & 3 & \\
NLAGN2cand (n=252) & extended & 252 & - & - & \\
eAGN (n=5) & point-like & 0 & 5 & 0\\
eAGN (n=166) & extended & 166 & - & - \\
not classified (n=57) & pl/ex & 38 & 19 & 9 \\
\hline
total & - & 1369 & 1203 & 902 \\ 
   \end{tabular}
   \label{tab:XDQSOclass} 
\end{table}

In the following section, we analyze the XDQSO properties of the X-ray sources with \textit{BOSS} follow-up: 1203 XDQSO sources have reliable \textit{BOSS} spectra, out of which 902 have $P_{\rm sum}(\rm QSO)>0.5$. As shown in Table \ref{tab:XDQSOclass}, these objects only comprise of optically point-like objects and are dominated by BLAGN1. Only few are classified as NLAGN2cand, NLAGN2 and eAGN. There are 196 point-like BLAGN1 which are not XDQSO selected due to their photometric properties ($\texttt{GOOD}>0$). 27 of them are too faint or bright for the initial magnitude thresholds of the XDQSO selection ($17.75<\textit{i}_{\rm dered}<22.45\,\rm mag$). By construction, the selection misses all optically extended objects. 
\\\\
In Fig.  \ref{fig:stackXDQSO}, we perform spectral stacks of X-ray selected XDQSO sources with different QSO probabilities and soft X-ray luminosities. We chose the redshift range of $0<z<2$ which covers a representative sample of the entire population. The stack includes the four following groups of X-ray and XDQSO selected targets:
\begin{compactenum}[a)]
\item $P_{\rm sum}(\rm QSO)>0.5$, $L_{0.5-2\rm keV}>10^{44}\rm\, erg\,s^{-1}$ ($n=420$),
\item $P_{\rm sum}(\rm QSO)>0.5$, $L_{0.5-2\rm keV}<10^{44}\rm\, erg\,s^{-1}$ ($n=336$),
\item $P_{\rm sum}(\rm QSO)<0.5$, $L_{0.5-2\rm keV}>10^{44}\rm\, erg\,s^{-1}$ ($n=95$),
\item $P_{\rm sum}(\rm QSO)<0.5$, $L_{0.5-2\rm keV}<10^{44}\rm\, erg\,s^{-1}$ ($n=197$).
\end{compactenum} 

The groups of high QSO probability spectra from a) and b) show clear broad line features. The low QSO probability spectra from c) and d) have a less steep power law and a more prominent galaxy contamination at redder wavelengths. Apparently, a more dominant host galaxy continuum affects the optical colour properties of the AGN which leads to a lower QSO probability. The X-ray luminosity for the same QSO probability range has only a very weak impact on the red part of the spectrum, because the host contribution mostly affects $P_{\rm sum}(\rm QSO)$.
\\\\
Summarizing, in the XMM-XXL, 48 per cent of the XDQSO selected AGN ($P_{\rm sum}(\rm QSO)>0.5$) are also X-ray selected and 13 per cent of all X-ray sources have XDQSO counterparts. In order to perform a full comparison of both AGN selections, we are missing a representative spectroscopic sample for only XDQSO selected objects without X-ray detection.  But we confirm that the XDQSO targeting algorithm selects optically point-like BLAGN1 showing an unobscured and blue optical spectrum with only weak host galaxy contribution. Among the XDQSO selected AGN with X-ray detection, the typical X-ray luminosity is $L_{0.5-2\rm keV}> 10^{42}\rm\, erg\,s^{-1}$ and redshift range is $0<z<5$.

\subsection[]{Infrared: \textit{WISE} colour selection}
\label{sec:WISEAGN}

The mid-infrared AGN selection benefits from the different spectral energy distributions (SED) of AGN and galaxies in the \textit{WISE} colour range: AGN show a steep power law-shaped spectrum and galaxies have a black body spectrum. This translates into a well-defined IR colour difference between the two classes, provided the nuclear AGN emission is the dominant component at these wavelengths. First, we evaluate the colour cut of \citet{Stern12}  using the two most sensitive \textit{WISE} bands \textit{W}1 [$3.4\, \mu\rm m$] and \textit{W}2 [$4.6\,\mu\rm m$]: 
\begin{equation}
 \textit{W}1-\textit{W}2>0.8\,\,\,\text{for}\,\, \textit{W}2\leq 15.05\rm\,mag
 \end{equation} The \textit{W}2 magnitude limit corresponds to $\texttt{w2snr}\geq10$. This threshold is established from a well studied AGN-sample in COSMOS. Taking the \textit{Spitzer} selected AGN sample of \citet{Stern05} as a truth sample, the \citet{Stern12} \textit{WISE} selection is 78 per cent complete and 95 per cent reliable.
  Secondly, we consider the AGN selection criteria of \cite{Assef13}: 
\begin{equation}
\textit{W}1-\textit{W}2 >
\begin{cases}
0.662\times e^{(0.232\times(W2 -13.97)^2)}\newline \,\, \\* \,\,\,\,\,\,\,\,\,\,\,\,\,\,\,\,\text{for}\,\, 13.97\leq\textit{W}2\leq17.11\rm\,mag\\
0.662 \,\,\text{for}\,\, \textit{W}2<13.97\rm\,mag
\end{cases}
\end{equation} 
The $\rm \textit{W}2$ -threshold corresponds to $\texttt{w2snr}\geq3$. The selection is established with a truth sample of UV and mid-IR photometry selected AGN from the deeper NOAO Deep Wide-Field Survey Bo\"otes field and has a reliability of 90 per cent.

  \begin{figure}
\includegraphics[width=85mm]{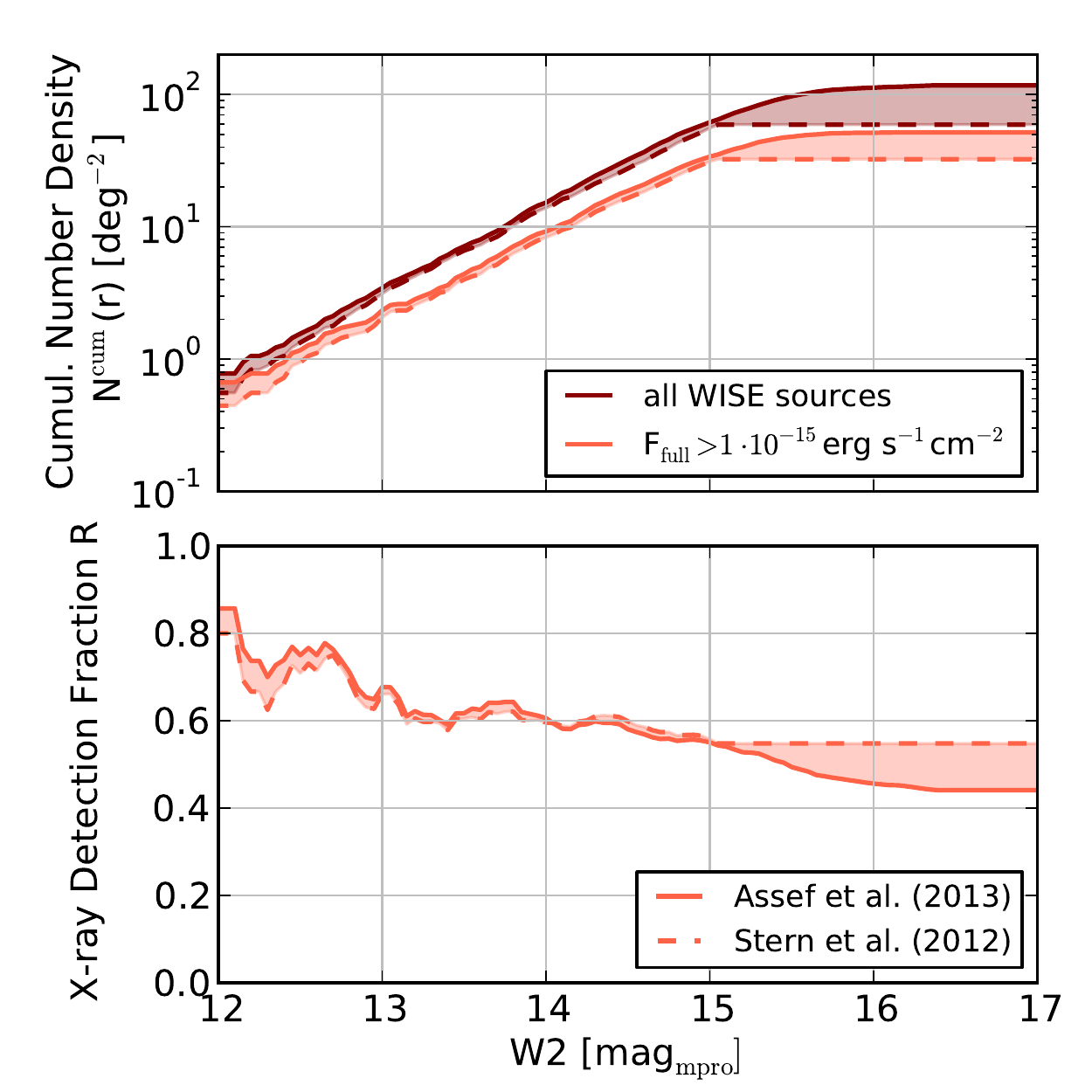}
 \caption{Top Panel: Cumulative number density of infrared selected AGN and of those with X-ray counterpart in the XMM-XXL north. We apply the infrared selection threshold of  \citet{Stern12} and \citet{Assef13}.  Bottom Panel: X-ray detection fraction of infrared selected AGN: We divide the number of X-ray detected \textit{WISE} AGN by all \textit{WISE} selected AGN in the XMM-XXL north.}
 \label{fig:WISEfracs}
 \end{figure}

\subsubsection{X-ray detection fraction}

In the $18\,\rm deg^2$ area of the XMM-XXL are $321\,325$ \textit{WISE} sources and 4811 of them can be matched as counterparts to the X-ray selected AGN considering $LR_{\rm XMM,WISE}>0.5$ with good photometry. In Fig. \ref{fig:WISEfracs}, we compare both infrared AGN selection criteria from \citet{Stern12} and \cite{Assef13}, and the X-ray selection for AGN in the XMM-XXL north. The \cite{Stern12} criterion selects 1063 infrared sources, containing 584 sources which are counterparts for the X-ray selected AGN. The \cite{Assef13} criterion selects 2109 infrared sources, including 930 X-ray selected sources.  We define the number density $N_{\rm WISE,X}$ for X-ray selected \textit{WISE} AGN at different soft X-ray flux limits, and the number density $N_{\rm WISE}$ for all \textit{WISE} selected AGN \citep{Stern12, Assef13} in the XMM-XXL north. The ratio $R$ of the cumulative number densities (hereby X-ray detection fraction) is:
\begin{equation*}
R=\frac{N^{cum}_{\textit{W}2, \rm X,WISE}(\textit{W}2, F_X)}{N^{cum}_{\rm WISE}(\textit{W}2)}.
\end{equation*}
and depends on the infrared magnitude \textit{W}2 as well as the X-ray flux $F_X$, as shown in Fig. \ref{fig:WISEfracs}. The datasets used for the calibration of the selection criteria in COSMOS and Bo\"otes have different depths than the XMM-XXL. Therefore, we cannot guarantee a similar reliability (by construction, typically around 90 per cent)  and do not correct the AGN detection fractions.

In Fig. \ref{fig:WISEfracs}, the \cite{Stern12} (dashed line) and the \cite{Assef13} (solid lines) selections do not differ much in their number densities up to $W2\leq15\,\rm mag$. At fainter magnitude, the Assef threshold allows the selection of further AGN. Assuming a soft X-ray flux of $F_{0.5-10\rm keV}> 10^{-15}\,\rm erg\, cm^{-2}\, s^{-1}$ and the faint \textit{W}2 limit of the \textit{WISE} selections, $\sim44$ per cent ($\textit{W}2\leq17\rm\, mag$) of all Assef and $\sim56$ per cent ($\textit{W}2\leq15\rm \,mag$) of all Stern selected AGN are detected in X-ray, too. For brighter magnitudes ($\textit{W}2\leq13\rm\,mag$), the fraction increases up to $\sim67$ per cent for the Assef and Stern selection.

\subsubsection{Infrared and spectroscopic properties of \textit{WISE} selected AGN}
\label{sec:Wisecomp}

\begin{table}
    \caption{\textit{WISE} properties  \citep{Stern12, Assef13} of X-ray selected AGN with spectroscopic information: We spilt the data set of X-ray selected AGN with reliable \textit{BOSS} follow-up into different classes and indicate the number of X-ray selected AGN as well as X-ray selected \textit{WISE} AGN for each population.}
 \centering
\begin{tabular}[width=20mm]{lccccccccc} 
  \hline
spectroscopic  & X-ray& X-ray \& & X-ray \& \\
classification    & selection & Stern et al. & Assef et al.\\
\hline
BLAGN1 & 1787 & 347 & 520 \\
NLAGN2 & 271 & 17 & 29\\
NLAGN2cand & 284 & 17 & 23 \\
eAGN & 171 & 5 & 6\\
not classified & 57 & 4 & 7 \\
\hline
total & 2570 & 390 & 585 \\
   \label{tab:WISE} 
   \end{tabular}
\end{table}

 \begin{figure*}
\includegraphics[width=180mm]{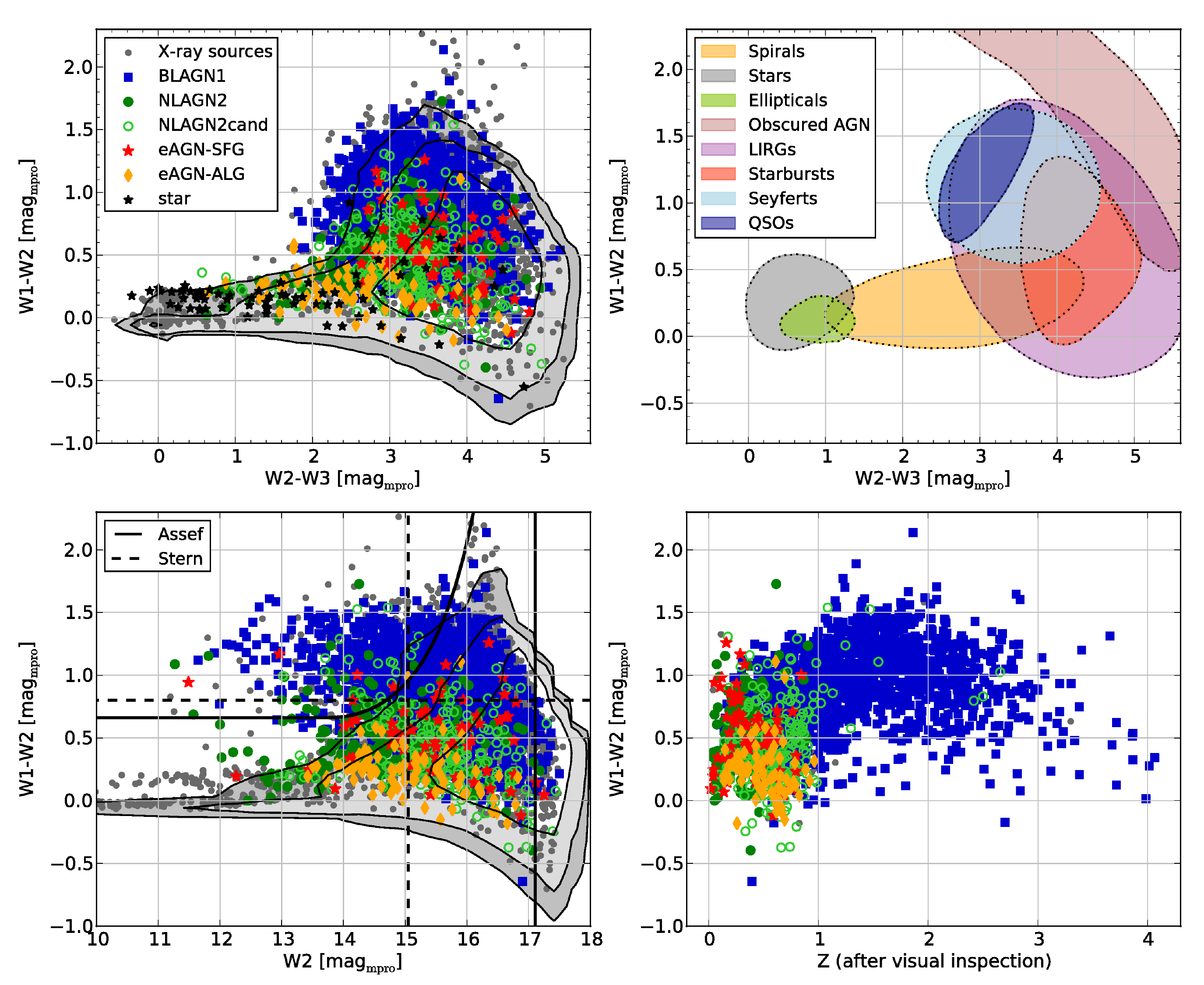}
 \caption{Top Panels: \textit{WISE} colour diagram: In the left panel, we plot the 68, 95 and 99 per cent contours of all \textit{WISE} sources,  the classified \textit{BOSS} spectra (coloured) and the unclassified X-ray sources (black) in the XMM-XXL north. In the right panel, we plot the class regions of \citet{Wright10}. Bottom Left Panel: \textit{WISE} colour magnitude diagram and AGN selections: We plot the contours of all \textit{WISE} sources and over plot the classified \textit{BOSS} spectra and unclassified X-ray sources. The solid line and the dashed horizontal line mark the infrared AGN selection from \citet{Assef13} and \citet{Stern12}, respectively. The vertical lines indicate the lower magnitude limits for this two selections. Bottom Right Panel: \textit{WISE} colour redshift diagram of \textit{BOSS} spectra with \textit{WISE} counterpart.}
 \label{fig:WISEbands}
   \end{figure*}

\begin{figure*}
\includegraphics[width=180mm]{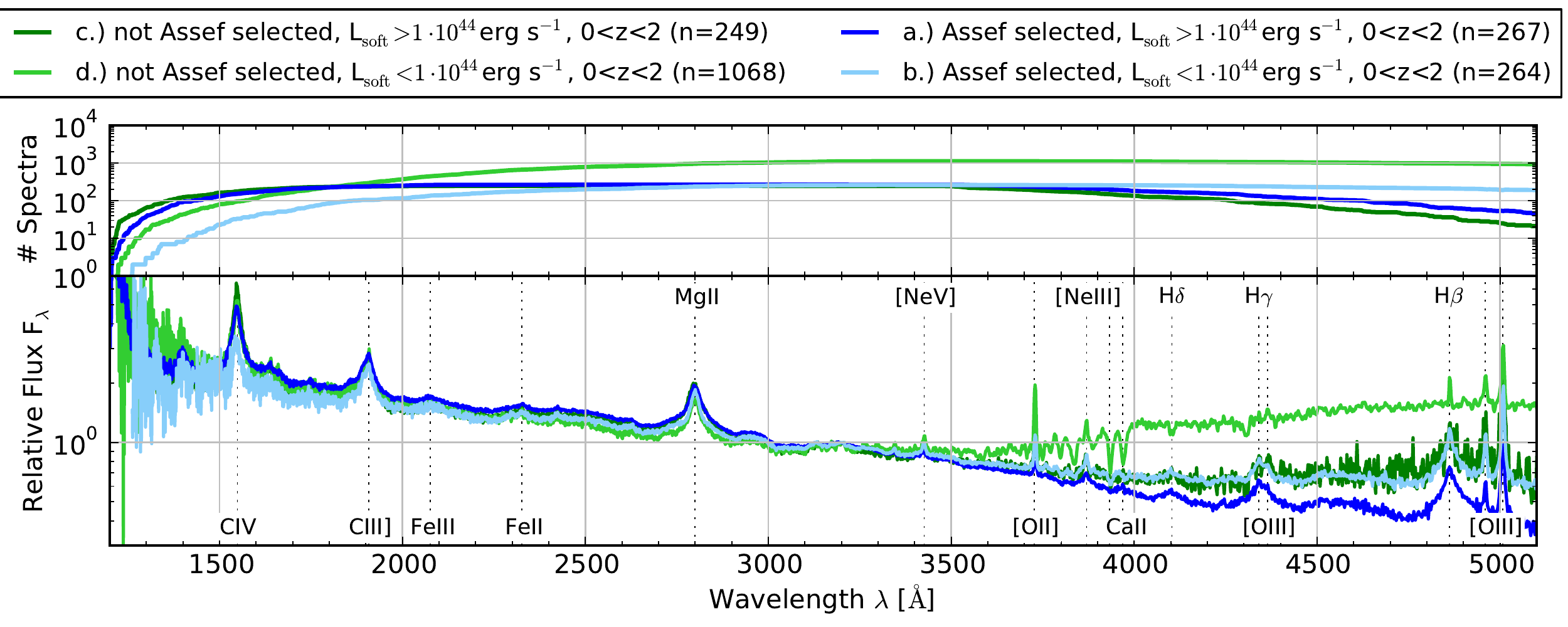}
 \caption{Median spectral stack and number of contributing spectra for \citet{Assef13} selected sources with a.) high soft X-ray luminosity and b.) low X-ray luminosity, and not \citet{Assef13} selected AGN with c.) high soft X-ray luminosity and d.) low X-ray luminosity. The spectra have a redshift of $0.0<z<2.0$ and are normalized to $\lambda=3200\,$\AA. }
 \label{fig:stackWISE}
   \end{figure*}

The overall population of X-ray sources with infrared counterpart comprises 4811 sources, out of which 2117 also have reliable \textit{BOSS} classifications. In the top panels of Fig. \ref{fig:WISEbands}, we plot these sources in the $\textit{W}1-\textit{W}2$ colour and $\textit{W}2-\textit{W}3$ colour diagram and compare the source locations with characteristic source regions from \citet{Wright10}. The classifications of our \textit{BOSS} sources are consistent: the BLAGN1 mainly reside in the QSO and Seyfert area, the NLAGN2 extend from the spiral region in the Seyfert region. The eAGN-SFG live in the starburst region extending to the Seyfert region. The eAGN-ALG are in the spiral region, which might be caused by the contribution of the AGN emission in the \textit{W}2 and \textit{W}3 band. The stars selected via X-rays fall into the common \textit{WISE} star region but also spread beyond. 
\\\\
In the bottom left panel of Fig. \ref{fig:WISEbands}, we indicate the \textit{WISE} selections from \cite{Stern12} and \cite{Assef13} in the $\textit{W}1-\textit{W}2$ colour and $\textit{W}2$ magnitude diagram. The plot comprises the density contours of all \textit{WISE} sources in the XMM-XXL north area and we overplot the X-ray sources with \textit{WISE} counterparts and the \textit{BOSS} observed spectra.  The \cite{Stern12} criterion selects 390 \textit{BOSS} observed X-ray sources and the \cite{Assef13} criterion selects 585 targets. The number statistics and classifications of the \textit{WISE} and X-ray selected sources are listed in Table \ref{tab:WISE}. The selected sources mainly consist of BLAGN1 (89 per cent for both selections), NLAGN2/NLAGN2cand and few eAGN. The lower right panel demonstrates the $\textit{W}1-\textit{W}2$ colour and redshift dependency of X-ray selected \cite{Assef13} AGN with spectroscopic information. Both selections are limited to a redshift range of $z<3$ due to the overall SED shape of AGN in the infrared restframe (see \citealt{Stern12}).
\\\\
In Fig. \ref{fig:stackWISE}, we show the spectral stacks of X-ray selected \cite{Assef13} AGN at different luminosities in the redshift range  $0<z<2$. We chose the average luminosity of $L_{0.5-2\rm keV}= 10^{44}\rm\, erg\,s^{-1}$ as a threshold for the high and low luminosity populations. This redshift range comprises a representative population from the BLAGN1 and both NLAGN2 and eAGN classes.  The four groups of X-ray selected spectra are:
\begin{compactenum}[a.)]
\item Assef selected: $L_{0.5-2\rm keV}>10^{44}\rm\, erg\,s^{-1}$ ($n=267$),
\item Assef selected: $L_{0.5-2\rm keV}<10^{44}\rm\, erg\,s^{-1}$ ($n=264$),
\item not Assef selected: $L_{0.5-2\rm keV}>10^{44}\rm\, erg\,s^{-1}$ ($n=249$),
\item not Assef selected: $L_{0.5-2\rm keV}<10^{44}\rm\, erg\,s^{-1}$ ($n=1068$).
\end{compactenum}
The high luminousity X-ray and \citet{Assef13} selected spectra from a.) comprise only BLAGN1 which show clear broad line features and nearly no host galaxy contribution. The stack of low luminosity sources from b.) shows emission lines which are less broad and a redder continuum indicating host galaxy contribution from NLAGN2 and host galaxy dominated BLAGN1. The stacks from c.) and d.) are X-ray sources not selected by \citet{Assef13}. They show stronger host galaxy contribution than the \citet{Assef13} selected sources in the same luminosity range. The stack from d.) includes the large majority of host dominated AGN.  
\\\\
Summarizing, 44 per cent of the \citet{Assef13} selected AGN are also selected by X-ray and 11 per cent of all X-ray selected AGN are also infrared AGN. Our comparison of X-ray selection criteria with \citet{Assef13} and \citet{Stern12} is biased because the spectroscopic sample does not comprise only infrared selected AGN and is limited in the \textit{r}-band. But applying the two infrared AGN selection criteria  to our X-ray selected sample, they access a population of AGN with very weak host contribution and red $\textit{W}1-\textit{W}2$ colours. This group includes mainly BLAGN1 and few NLAGN2 in the redshift range of $0<z<3$ and has a large luminosity range of $4\times10^{40}<L_{0.5-2\rm \,keV}<2\times10^{45}\rm \, erg\,s^{-1}$.

\section{Discussion}
\label{sec:Discussion}
\subsection{Understanding the AGN population in the XMM-XXL north}

The identification and classification of X-ray sources has been already performed in many subfields of the XMM-XXL north region. \citet{Tajer07} and \citet{Polletta07} used a sample of 136 X-ray point 
like sources at a flux of $F_{2-10 \,\rm{keV}} >10^{-14} \,\rm{erg\,cm^{-2}\,s^{-1}}$ in a $1\, \rm{deg^2}$ area and obtained a reliable photometric redshift and classification for 107 sources with optical and infrared photometry. 
\citet{Garcet07} used 612 point-like X-ray sources at a limiting flux of 
$F_{2-10 \,\rm{keV}} > 8\times 10^{-15} \,\rm{erg\,cm^{-2}\,s^{-1}}$  in an area of $3 \, \rm{deg^2}$ and 
associated 99 objects followed up by optical spectroscopy 
(\textit{2dF}, \textit{VIMOS}, \textit{SALT}) with secure redshift and classification ($\textit{R} < 22 \,\rm{mag}$). 
In the same field, \citet{Stalin10} selected 829 objects at $F_{0.5-2 \,\rm{keV}} > 10^{-15} \,\rm{erg\,cm^{-2}\,s^{-1}}$ and obtained a secure identification for 487 sources with optical spectroscopy at \textit{AAT} ($g' < 22 \,\rm{mag}$).

Compared to all these programmes, our campaign stands out for the sheer number of spectra available from the same instrument, under stable, uniform observing conditions, and for the use of a well tested, well understood and robust data analysis pipeline. In the XMM-XXL north, we obtained 3042 unique \textit{BOSS} spectra and reached a spectroscopic completeness of 32 per cent for all X-ray selected point sources at $F_{0.5-10 \,\rm{keV}} > 10^{-15} \,\rm{erg\,cm^{-2}\,s^{-1}}$ and 78 per cent for all optical cross-matched sources with $17<\textit{r}<22.5 \,\rm mag$. In order to increase the number of reliable spectra, we performed a visual inspection for the spectra with critical redshift estimates of the \textit{BOSS} pipeline. We find that 85 per cent of the X-ray selected and \textit{BOSS} observed targets have a reliable redshift identification after the visual inspection. Our classification of X-ray selected and optically followed-up spectra is based on the optical emission line features of the sources. The algorithm presented in this work classifies 2513. For 2 per cent of all spectra with reliable redshift, the emission line information are not sufficient for a classification.

The principal element of our classification algorithm is the bimodal distribution of the emission line FWHM, separating narrow line region (NLR) emitters and broad line region (BLR) emitters at a FHWM-threshold of $\rm FWHM\sim1000\rm \,km\,s^{-1}$.  The bimodality can be explained by the distinct locations of NLR and BLR. Similar distributions have also been shown in the paper of \cite{Hao05} with a threshold of $\rm FWHM\sim1200\rm\,km\,s^{-1}$ based on a sample of low redshift AGN with subtracted galaxy continuum. The authors fit each individual emission line with multiple gaussians, whereas in this work we use averaged widths. The major group of X-ray selected objects in the XMM-XXL north are the BLR emitters: BLAGN1. The group of NLR emitters includes objects whose X-ray emission originates either from an obscured AGN, the star forming region or both. They are selected based on their position in the BPT-NII and Blue-OII diagram. The NLAGN2 are powered by a central AGN and form the second large group with 11 per cent of all classified X-ray selected objects. Our data set comprises a group of equal size with NLAGN2 candidates (11 per cent) which do not have enough significant emission lines or lie at a redshift of $z>1.08$ and cannot be projected in the BPT-NII and Blue-OII diagram. Their spectral stack as well as optical colours and morphology indicate that the majority of the sources are NLAGN2. Deeper exposed optical spectra or infrared spectra are needed to properly classify these objects. For the majority of BLAGN1 and NLAGN2, the optical classification is also confirmed by their X-ray spectra. But there are known exceptions of e.g. optically type 1 AGN with obscured X-ray spectra (e.g. \citealt{Brusa03, Merloni14}) or optically type 2 AGN with un-obscured X-ray spectra. These objects require more detailed X-ray spectral analysis (see Liu et al. 2015, submitted). 

Our sample contains 6 per cent elusive AGN (eAGN) living in star forming galaxies (eAGN-SFG) and absorption line galaxies (eAGN-ALG). They do not stand out in any cross-matching related variable compared to the entire population. The fraction of elusive AGN is comparable to the COSMOS sample (\citealt{Brusa10}, 4 per cent, considering $L_{0.5-2\rm keV} >  10^{42}\,  \rm{erg\,s^{-1}}$). Elusive AGN have been referred to as optically `dull' AGN or XBONGS by \citet{Comastri02, Maiolino03,  Caccianiga07, Trump09, Pons14}. The absence of AGN signatures in their optical spectra is still a matter of debate. According to \citet{Georgantopoulos05}, it is caused by dilution effects of a strong host galaxy component.  \citet{Comastri02} argues for a strong obscuration of the nuclear source preventing optical AGN emission. \citet{Yuan04} suggest that the XBONGS have truncated disc close to the black hole causing the absence of the characteristic AGN emission line. Extreme obscuration of the NLR by AGN-fueling dusty spirals \citep{Kraemer11}, or AGN flickering effects \citep{Schawinski15} have also been suggested. Based on their soft X-ray luminosity, we can confirm that the eAGN-ALG host an AGN. The group of eAGN-SFG partly falls below the upper luminosity threshold of SFG ($L_{0.5-2\,\rm keV}=10^{42}\,\rm erg\,s^{-1}$), but the number fraction is consistent with NLAGN2 in the same redshift and luminosity range. Liu et al. (2015, submitted) analyse the X-ray properties of the elusive AGN. 75 per cent of them have $N_H<21.5\,\rm cm^{-2}$ indicating the presence of a low luminosity type 1 AGN, whose optical emission is probably fully dominated by stellar emission processes \citep{Davies15}.

Narrow Line Seyfert 1 AGN (NLSey1) are also part of our X-ray selected AGN. 
As shown by e.g. \citet{Osterbrock85, Boller96}, this population of optically unobscured AGN typically has FWHM for H$\beta$ in the range of $500 < \rm FWHM_{\rm H\beta} < 1500\,\rm km\,s^{-1}$, very steep X-ray spectra, strong optical FeII emission, and strong variability.
The NLSey1 contribute to our BLAGN1 group, but probably also affect narrow emission line objects with $\rm FWHM_{\rm H\beta} < 1000\,\rm km\,s^{-1}$. \citet{Castello12} find that NLSey1 are classified as SFG by the optical line diagnostic diagram due to their smaller H$\beta$ EW. The NLAGN2 of our dataset are securely identified as AGN because of their projection in the AGN region of the optical line diagnostic diagrams. 
But the group of NLAGN2cand and eAGN-SFG could potentially be contaminated by NLSey1. According to \citet{Castello12}, the NLSey1 which are projected in the SFG region have high X-ray luminosities $L_{2-10\,\rm keV}>10^{42}\,\rm erg \,s^{-1}$ and large X-ray to optical flux ratios ($\log F_{\rm X}/F_{\rm opt} > 0.1$). This applies to a subset of our eAGN-SFG as shown in Fig. \ref{fig:logxo}.
From the stack of the NLAGN2 and eAGN-ALG, however, we see that the majority of contributing spectra have very strong host galaxy continua (see bottom panel of Fig. \ref{fig:specstack}). This indicates that our classification procedure for objects with $\rm FWHM_{\rm H\beta} < 1000\,\rm km\,s^{-1}$ mainly selects optically obscured AGN and is only weakly affected by NLSey1. A dedicated study of the NLSey1 contribution in our sample would require further FeII complex and X-ray spectra analysis, which is not in the subject of this paper.

The X-ray selected dataset also contains Galactic stars. We can estimate the star fraction of the entire X-ray data set, assuming a fraction of 3 per cent for all \textit{XMM}-\textit{SDSS} cross-matched objects with $\textit{r}>15$ and a fraction of 100 per cent for all objects with $\textit{r}<15$. This results in a maximal star fraction of $\sim7$ per cent.

\subsection{Uniqueness of X-ray AGN selection}
Considering current and future large area surveys in different wavelength bands, the number of observable AGN populations is going to increase dramatically. Many studies underline the strong impact of selections on the characteristic features of AGN. In this work, we have access to a very large area of $18\rm\,deg^2$, covered by multi-wavelength data, allowing for comparison of three AGN selection criteria in X-ray, optical and infrared, with high statistic reliablity. We find interesting trends regarding the optical morphology, the X-ray luminosity, the redshift, and host galaxy contribution.

  \begin{figure}
\includegraphics[width=85mm]{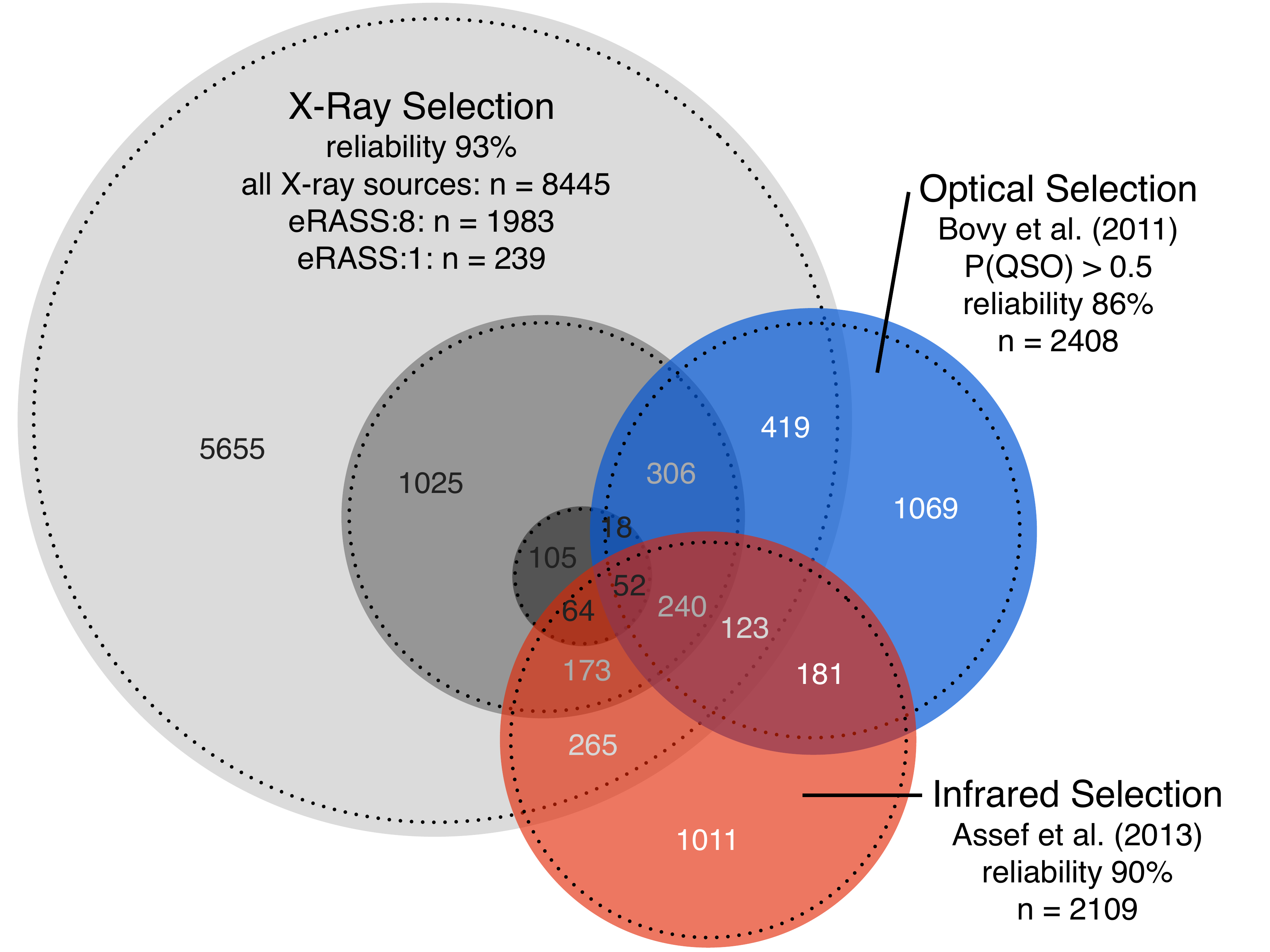}
 \caption{Venn diagram of X-ray, optical (XDQSO) and infrared \citep{Assef13} AGN selections in the XMM-XXL north: We indicate the flux depths for eRASS:8 and eRASS:1 of \textit{eROSITA}, the reliable fraction (dashed line) and number of AGN in each intersection.}
 \label{fig:venn}
   \end{figure}

In Fig. \ref{fig:venn}, we illustrate the relative sizes of the differently selected AGN samples and their reliability fractions in the XMM-XXL. In addition, we show the subsets of sources with X-ray fluxes above the expected limit of the first and final \textit{eROSITA} all sky coverage (eRASS:1 and eRASS:8, respectively), as explained in Section \ref{sec:eROSITAforecast} below.  The \textit{WISE} and \textit{SDSS} catalogues have been matched via their source positions to be within a distance of 1 arcsec. The associated X-ray sources belong to the Likelihood-Ratio match catalogue from Section \ref{sec:crossmatch}. 
 
Considering the limiting fluxes of \textit{SDSS}, \textit{WISE}, and \textit{XMM-Newton} in the XMM-XXL north, the combination of all AGN selection criteria picks up about $600$ AGN $\rm deg^{-2}$. Their common subgroup includes only $\sim 23 $ AGN $\rm deg^{-2}$ out of which 99 per cent are optically point-like BLAGN1 with very weak host galaxy components. Applying a less conservative limit for the positional distance between \textit{SDSS} and \textit{WISE} sources, the number of objects in the common subgroup with \textit{XMM-Newton} increases from 415 (1 arcsec), to 430 (2 arcsec) and finally 434 ($\sim$ 8.5 arcsec). The total intersection of only  \textit{SDSS} and \textit{WISE} comprises 596 sources for 1 arcsec and 623 sources for 2 arcsec.

Each of the three AGN selection methods also retrieves an exclusive population of AGN. At the full  depth of $F_{\rm 0.5-10\, keV} >  10^{-15} \rm\,erg\, cm^{-2}\, s^{-1}$, $\sim 380 $ AGN $\rm deg^{-2}$ are only selected in X-rays, corresponding to 80 per cent of the total X-ray selection. XDQSO and infrared selections have both access to nearly $\sim 60$ exclusive AGN $\rm deg^{-2}$ each, which corresponds to $\sim 50$ per cent of their full selections. 

The last two selections preferentially pick up AGN which are outshining their hosts at high optical or infrared luminosities, respectively \citep{Bahcall97}. The XDQSO selection \citep{Bovy11} is trained for typical optically point-like \textit{BOSS}-QSO and is highly sensitive within the available depth of optical \textit{SDSS} data. This is crucial e.g. to reach a high number of QSO in the baryonic oscillations programme of \textit{eBOSS} \citep{Dawson13}. Our spectroscopic sample does not include \textit{BOSS} spectra that are exclusively XDQSO selected, but these sources are most likely to be optically point-like low X-ray luminosity BLAGN1. The infrared selection from \citet{Assef13} is limited to the redshift range of $z<3$ and the majority of selected sources are classified as BLAGN1 having red $\textit{W}1-\textit{W}2$ colours. The exclusively infrared selected AGN are expected to be low X-ray luminosity AGN (both BLAGN1 and NLAGN2) with a weak host galaxy contribution, or heavily obscured AGN (both BLAGN1 and NLAGN2). 

Our study confirms that X-rays are less sensitive to dilution effects in the host galaxy, optical morphology, obscuring material, and star formation, as firstly shown by \citet{Nandra07b} and \citet{Bundy08}. In our sample, they pick up a broad variety of AGN, such as typical QSO, optically unobscured AGN with strong host galaxies, optically obscured AGN, and elusive AGN in a large luminosity and redshift range.

Based on our results, we expect that the selection method has a strong effect on statistical properties such as  clustering and dark halo mass of AGN populations. As also shown by \citet{Merloni15}, at any given AGN luminosity, X-rays select systems accreting at lower Eddington rates $\lambda$. These objects reside in more massive hosts with lower contrast of host galaxy to nuclear emission, and are missed by any other selection method. This is in line with the study of \citet{Mendez15} focussing on the clustering properties of X-ray, radio, and infrared selected AGN from the \textit{PRIMUS} and \textit{DEEP2} surveys. The authors find that X-ray selected AGN cluster more than infrared galaxies and reside in more massive dark matter halos. This is not the case if the stellar mass, specific star formation rate, and redshift distributions of the selected galaxies are matched to the same control sample.

\section{\textit{eROSITA} and \textit{SPIDERS} forecast}
\label{sec:eROSITAforecast}

In light of the upcoming all sky X-ray survey by \textit{eROSITA} on-board the \textit{SRG} mission \citep{MerlonieROSITA, Predehl14}, our study forms an outstanding data set to study the characteristics of the X-ray AGN population which will be discovered in large numbers. Comparing to the \textit{ROSAT} all-sky survey \citep{Voges99}, the X-ray selected AGN density over the whole sky is going to increase from to $\sim 2\,\rm deg^{-1}$ to $\sim90\,\rm deg^{-1}$. For an angular resolution averaged across the field of view of about 26" (half energy width, \citealt{Burwitz14}), the predicted flux limits of \textit{eROSITA} are:
\begin{compactenum}[-]
\item $F_{0.5-2 \,\rm{keV}}>4.0\times 10^{-14}\rm \,erg\,s^{-1}\,cm^{-2}$  (\textbf{eRASS:1}, 0.5 yr),
\item $F_{0.5-2 \,\rm{keV}}>2.5\times 10^{-14}\rm \,erg\,s^{-1}\,cm^{-2}$ (\textbf{eRASS:2}, 1 yr),
\item $F_{0.5-2 \,\rm{keV}}>1.5\times 10^{-14}\rm \,erg\,s^{-1}\,cm^{-2}$ (\textbf{eRASS:4}, 2 yr), 
\item $F_{0.5-2 \,\rm{keV}}>9.8\times 10^{-15}\rm \,erg\,s^{-1}\,cm^{-2}$ (\textbf{eRASS:8}, 4 yr). 
\end{compactenum}
The scanning strategy is still a subject of debate, but the XMM-XXL survey reaches a depth comparable to the deepest planned \textit{eROSITA} exposure ($F_{0.5-2 \,\rm{keV}}\sim4.0\times 10^{-15}\rm \,erg\,s^{-1}\,cm^{-2}$) near the ecliptic poles covering a solid angle of $\ga 500\,\rm deg^2$. 

The eROSITA X-ray catalogue of AGN is going to be the target of the spectroscopic follow-up programme \textit{SPIDERS}  (Spectroscopic Identification of \textit{eROSITA} Sources, Merloni et al. 2015, in prep.). This programme belongs to the \textit{eBOSS} survey (Dawson et al. 2015, in prep.) in the \textit{SDSS-IV} to follow-up \textit{ROSAT}, \textit{XMM-Newton}, and eventually \textit{eROSITA} sources with the optical \textit{BOSS} spectrograph (see \texttt{http://www.sdss.org}). It aims to be one of the largest optical spectroscopic follow-up surveys of X-ray selected AGN. In the following, we will provide both a scientific as well as technical forecast for \textit{eROSITA} and \textit{SPIDERS} based on our spectroscopic XMM-XXL dataset.

 \begin{figure}
\includegraphics[width=85mm]{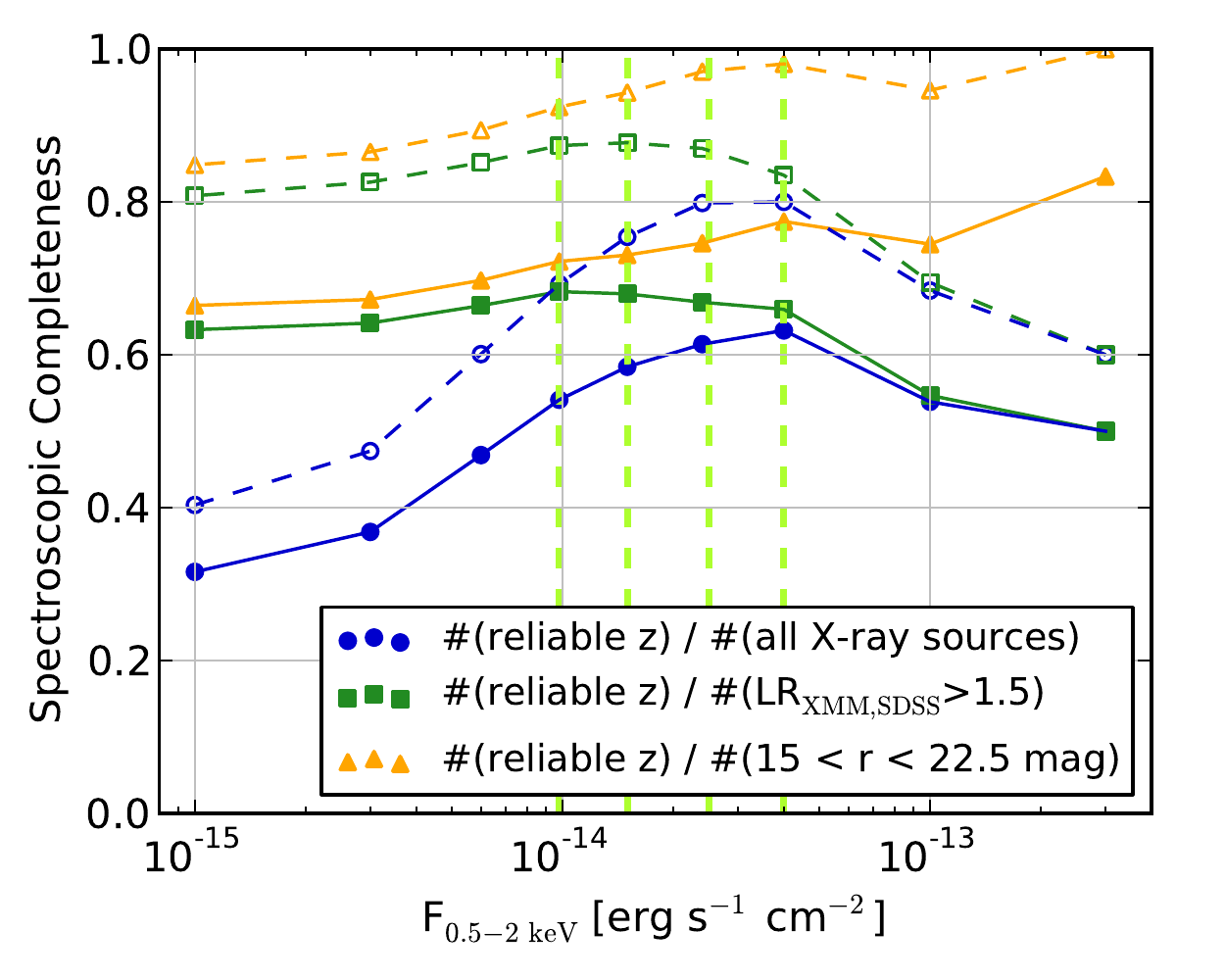}
 \caption{Spectroscopic completeness in the XMM-XXL north: We show the fractions of X-ray sources with reliable redshift ($\texttt{Z\_CONF}=3$ and $\texttt{Z\_CONF}=30$) over the number of  all X-ray sources (blue),   matched \textit{SDSS} counterparts ($LR_{\rm XMM,SDSS}>1.5$) (green), and \textit{r}-band within $15<\textit{r}<22.5\,\rm mag$ (yellow). In addition, we apply the fibre collision correction to the fraction of sources with reliable redshift (dashed lines with empty markers) to obtain the intrinsic spectroscopic completeness. The vertical green dashed lines indicate the flux depths of \textit{eROSITA} (from left: eRASS:8, eRASS:4, eRASS:2, eRASS:1). }
 \label{fig:speccomp}
   \end{figure}

\subsection{Spectroscopic completeness}

In order to predict the spectroscopic completeness of \textit{SPIDERS} (and other future \textit{eROSITA} follow-up programmes), we first calculate, for the XMM-XXL survey, the number densities of:
\begin{compactenum}
\item all X-ray sources, and of those with
\item matched \textit{SDSS} counterparts ($LR_{\rm XMM,SDSS}>1.5$),
\item \textit{r}-band within $15<\textit{r}<22.5\,\rm mag$,
\item \textit{BOSS} follow-up, or
\item reliable redshift ($\texttt{Z\_CONF}=3$ and $\texttt{Z\_CONF}=30$). 
\end{compactenum}
In addition, we define the fibre collision correction:
\begin{equation}
\mu = \frac{\#(LR_{\rm XMM,SDSS}>1.5,\,15<r<22.5\,\rm mag) }{ \#({\textit{BOSS}\,\,\rm spectra})}
\end{equation}
which has an influence on the number densities of \textit{BOSS} followed-up AGN. The number densities as well as the fibre collision correction at different soft X-ray fluxes is listed  in Table \ref{tab:eROSITA_I}.

In Fig. \ref{fig:speccomp}, we show the spectroscopic completeness of our dataset. The fractions correspond to the number of sources with reliable redshifts over the total number of X-ray sources (yellow line), sources with matched \textit{SDSS} counterparts (green line) and sources within the \textit{r}-band limits (blue line). We indicate the fractions at different X-ray fluxes and highlight the \textit{eROSITA} survey depths. After the correction for fibre collision (dashed lines), the spectroscopic completeness for all \textit{BOSS} targets ($15<\textit{r}<22.5\,\rm mag$) ranges from 85 per cent at XMM-XXL depth to 96 per cent for \textit{eROSITA} depth. The spectroscopic completeness for all X-ray sources with optical counterparts ($LR_{\rm XMM,SDSS}>1.5$) drops at $F_{0.5-2\,\rm keV}\ga 4\times 10^{-14}\,\rm erg\,s^{-1}$. This is due to the counterpart distribution, which shifts towards brighter optical magnitudes with shallow X-ray fluxes (see Fig. \ref{fig:rband}) and reduces the number of \textit{BOSS} targets within the \textit{r}-band threshold ($\textit{r}\leq 15-17 \,\rm mag$). The overall spectroscopic completeness for all X-ray sources at deep X-ray flux limits reflects both the sensitivity limits of SDSS for the optical counterparts and the faint \textit{r}-band threshold of the  \textit{BOSS} spectrograph. 
  
The optical classification of the AGN also correlates with the X-ray flux.  For the XMM-XXL depth, our AGN sample comprises 73 per cent BLAGN1, 20 per cent NLAGN2/NLAGN2cand and 6 per cent eAGN. For the \textit{eRASS:8}, these ratios change to 80 per cent BLAGN1, 16 per cent NLAGN2/NLAGN2cand and 4 per cent eAGN. 
 
  \begin{figure}
\includegraphics[width=85mm]{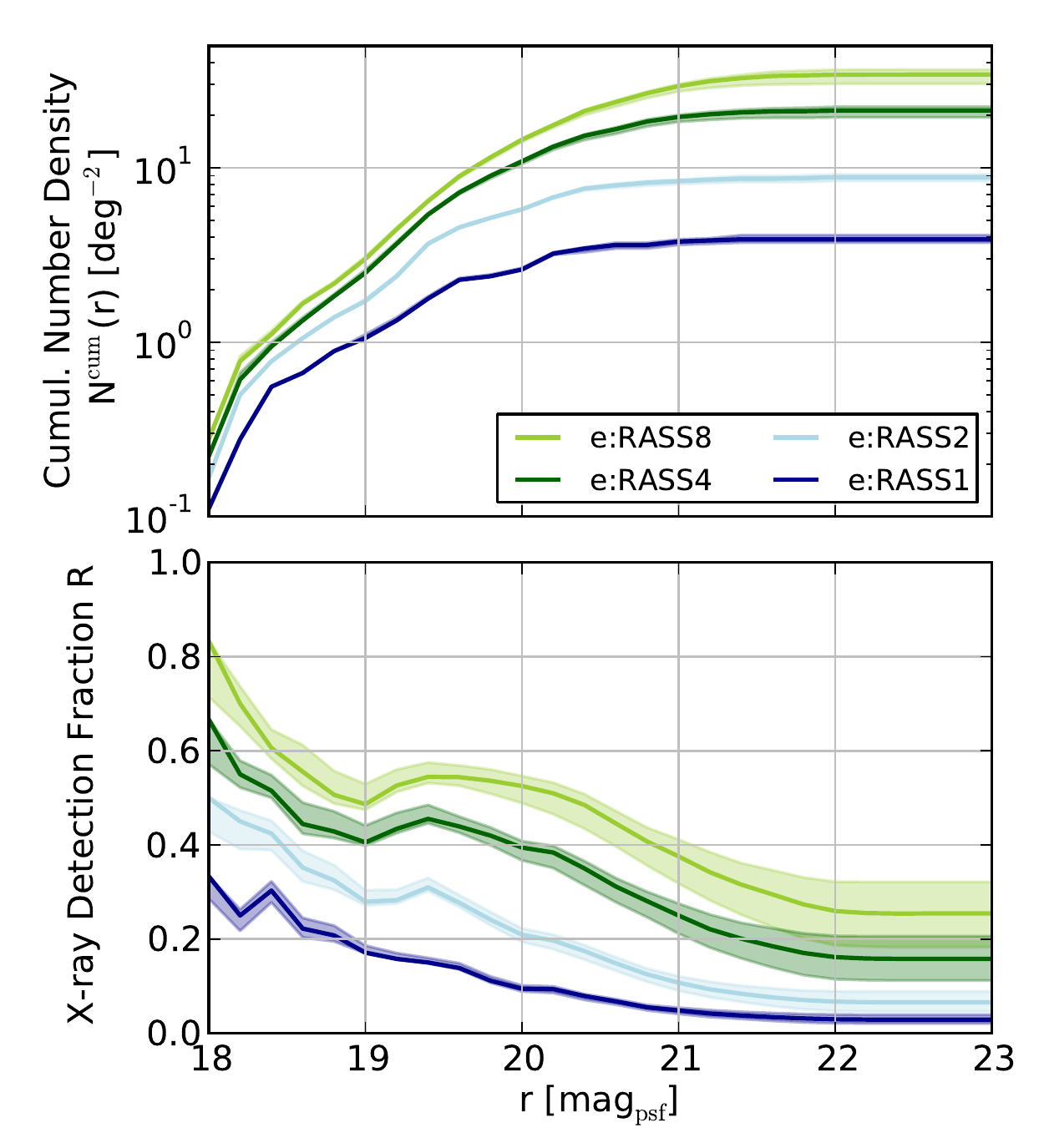} 
 \caption{Top Panel: Cumulative number density of XDQSO selected AGN with X-ray counterparts: We plot the number densities for the \textit{eROSITA} X-ray depths. We apply the optical selection threshold of $P_{\rm sum}(\rm  QSO)>0.5$ (central line) and $P_{\rm sum}(\rm  QSO)>0.8$ (lower border) as well as $>0.2$ (upper border).  Bottom Panel: X-ray detection fraction of XDQSO selected AGN at \textit{eROSITA} fluxes. We divide the number of X-ray detected AGN with $P_{\rm sum}(\rm  QSO)>0.5$ over all XDQSO sources with $P_{\rm sum}(\rm  QSO)>0.5$ in the XMM-XXL north.}
 \label{fig:XDQSOfracerass}
\end{figure}

   \begin{figure}
\includegraphics[width=85mm]{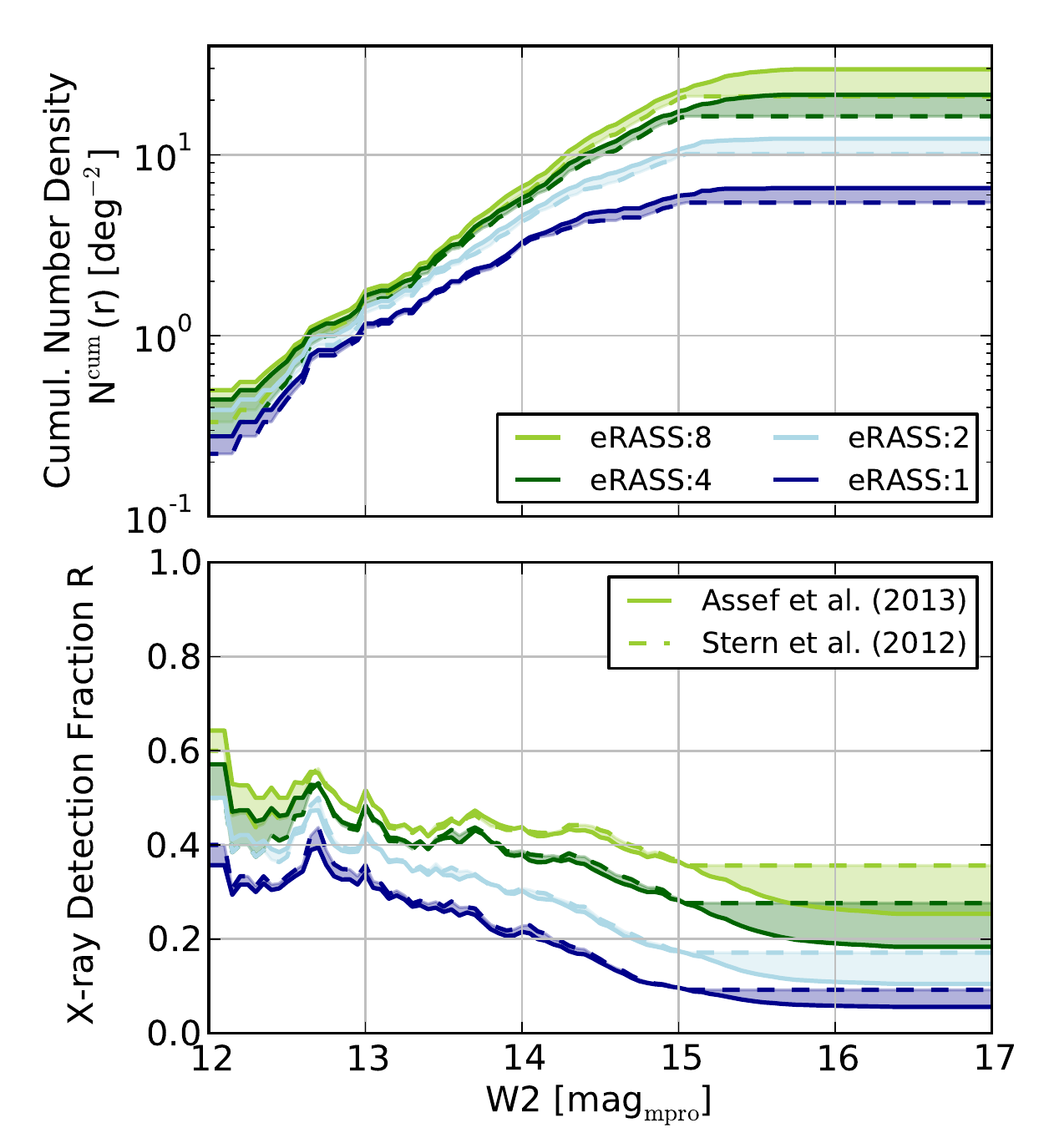}
 \caption{Top Panel: Cumulative number density of infrared selected AGN with X-ray counterpart. We apply the infrared selection threshold of  \citet{Stern12} and \citet{Assef13}, and plot the number densities for the\textit{eROSITA} X-ray depths. Bottom Panel: X-ray detection fraction of infrared selected AGN at \textit{eROSITA} fluxes: We divide the number of X-ray detected \textit{WISE} AGN by all \textit{WISE} selected AGN in the XMM-XXL north.}
 \label{fig:WISEfracserass}
 \end{figure}

\subsection{X-ray detection fraction of XDQSO and WISE selected AGN}

Following the explained procedures from Section \ref{sec:XDQSO} and Section \ref{sec:WISEAGN}, we derived the number densities and X-ray detection fraction of XDQSO (see Fig. \ref{fig:XDQSOfracerass}) and 
and \textit{WISE} (see Fig. \ref{fig:WISEfracserass}) for an X-ray selected sample at \textit{eROSITA} fluxes.

\section[]{Conclusions}
 \label{sec:conclusion}
  
We present and publicly release one of the largest contiguous catalogues of X-ray selected and spectroscopically observed AGN to date. It provides a unique data set to study the optical properties of X-ray selected AGN and also serves as a pilot study for the \textit{eROSITA} follow-up programme \textit{SPIDERS} in SDSS-IV. 
   
The survey contains 8445 point-like X-ray sources covering an area of $\sim 18 \,\rm deg^2$ in the XXM-XXL north area reaching down to a flux of $F_{0.5-10\,\rm keV}>10^{-15}\,\rm erg\,cm^{-2}\,s^{-1}$. They have been cross-matched to their \textit{SDSS} and \textit{WISE} counterparts via the Maximum-Likelihood-Ratio method. The \textit{BOSS} spectrograph followed up 3042 sources within $15<\textit{r}_{\rm SDSS\, psf/model}< 22.5\,\rm mag$ and after visual inspection, we obtained 2578 sources with reliable redshifts. The sample covers a redshift range of $0<z<5$ and a luminosity range of $2\times 10^{39}<L_{0.5-2\rm \,keV}<4\times10^{45}\rm \, erg\,s^{-1}$.  

For the study of the AGN properties of our X-ray selected sample, we introduce a spectral classification method which is based on optical emission line properties provided by the \textit{BOSS} pipeline. We analyzed properties, such as the line widths of AGN-induced  and star-formation-induced emission lines and finally derived the following classes for a subset of 2570 spectra: 1787 BLAGN1, 271 NLAGN2,  284 NLAGN2cand, 78 eAGN-SFG, 93 eAGN-ALG, 85 stars and 2 BL Lac. 57 sources cannot be classified, because of missing emission line information. 
We compared the X-ray AGN selection to common AGN selection using X-ray, optical \citep{Bovy11}, and infrared colours \citep{Stern12, Assef13}. 
\\
In the following, we conclude with the main scientific outcomes of the analysis of X-ray selected AGN in the XMM-XXL north:
\begin{compactenum}[(i)] 
\item The bimodal FWHM-distribution of optical AGN-induced emission lines widths (e.g. H$\beta$ and MgII) clearly separates the population of X-ray selected AGN into broad line region emitters and narrow line region emitters. The minimum of the FWHM-distribution is at $\rm FWHM\sim 1000\rm\,km/s $.
\item The X-ray selection probes a wide variety of AGN with respect to the obscuring material along the line of sight and the contribution of the passive or active host galaxy. It allows for a selection of particular classes, such as optically unobscured BLAGN1 with strong host galaxy contribution, optically obscured NLAGN2, and optically elusive AGN. Because of the \textit{r}-band magnitude limits for the optical spectroscopy, we are biased against optically faint sources, which mainly affects optically obscured NLAGN2. 
\item Applied to our X-ray selected sample, we find that the optical AGN selection via the XDQSO targeting algorithm \citep{Bovy11} is, by construction, biased towards optically point-like sources and selects BLAGN1 with weak host features in the entire redshift range of the sample.  The \textit{WISE} colour AGN selections from \citet{Assef13} and \citet{Stern12} applied to our sample preferentially selects BLAGN1 at $z<3$ with weak hosts and red $\textit{W}1-\textit{W}2$ colours. 
\item In the coming years, multi-object optical spectrographs with characteristics similar to \textit{BOSS} or with higher performance (e.g. \textit{4MOST}, \citealt{deJong14}; \textit{DESI}, \texttt{http://desi.lbl.gov}) will be able to provide highly complete, and very efficient follow-up programmes for the upcoming \textit{eROSITA} all-sky X-ray surveys, bringing the study of AGN populations to an unprecedented level of statistical accuracy.
\end{compactenum}

\section*{Acknowledgments}
Funding for SDSS-III has been provided by the Alfred P. Sloan Foundation, the Participating Institutions, the National Science Foundation, and the U.S. Department of Energy Office of Science. The SDSS-III web site is \texttt{http://www.sdss3.org}.

SDSS-III is managed by the Astrophysical Research Consortium for the Participating Institutions of the SDSS-III Collaboration including the University of Arizona, the Brazilian Participation Group, Brookhaven National Laboratory, Carnegie Mellon University, University of Florida, the French Participation Group, the German Participation Group, Harvard University, the Instituto de Astrofisica de Canarias, the Michigan State/Notre Dame/JINA Participation Group, Johns Hopkins University, Lawrence Berkeley National Laboratory, Max Planck Institute for Astrophysics, Max Planck Institute for Extraterrestrial Physics, New Mexico State University, New York University, Ohio State University, Pennsylvania State University, University of Portsmouth, Princeton University, the Spanish Participation Group, University of Tokyo, University of Utah, Vanderbilt University, University of Virginia, University of Washington, and Yale University. Marcella Brusa acknowledges support from the FP7 grant `eEASy' (CIG321913). 

We thank Adam Bolton (University of Utah), Damien Coeffey (MPE), Jeremy Sanders (MPE), Alina Streblyanska (IAC), Michael di Pompeo (Dartmouth College) and Adam Myers (University of Wyoming) for their support.
Furthermore, we express our gratitude towards anonymous reviewers from \textit{MNRAS} for their constructive comments.

\appendix

\section{Spectroscopic Target Selection}
\label{app:spectargsel}

\begin{figure*}
\includegraphics[width=170mm]{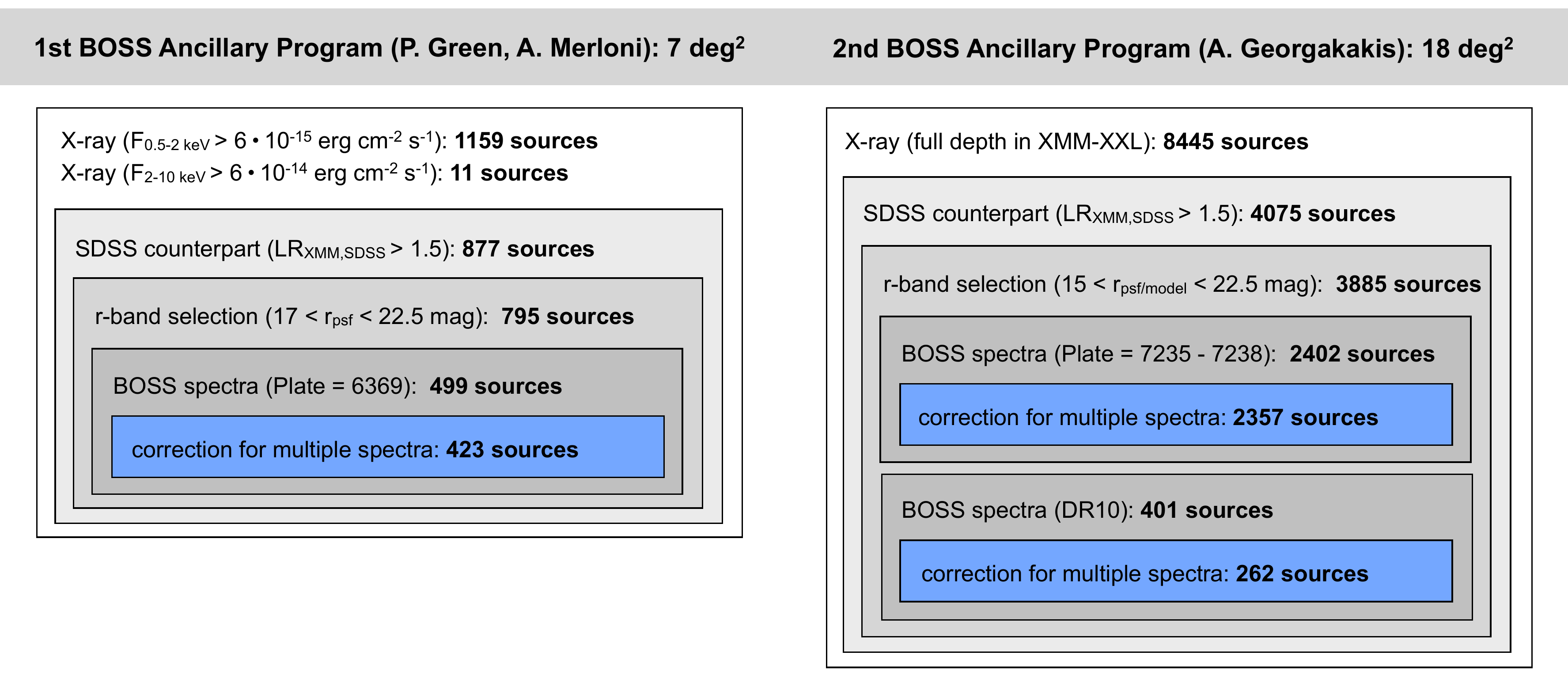}
 \caption{Spectroscopic target selection in the XMM-XXL north field: The targets for the follow-up of AGN come from three different groups, including two dedicated \textit{BOSS} ancillary programmes and former \textit{BOSS} DR10 observations \citep{Ahn12} in the same region. We indicate the number of point-like X-ray sources, the matched \textit{SDSS} counterparts, the applied \textit{r}-band cut, the \textit{BOSS} observed spectra, and the correction for multiple spectra for each programme.}
 \label{fig:targsel}
\end{figure*}

The spectroscopic follow-up data come from two \textit{BOSS} ancillary programmes comprising 7 \textit{BOSS} plates and furthermore the released DR10 catalogue. We indicate the source numbers of every spectroscopic targets selection step in Fig. \ref{fig:targsel}, provide additional technical information for the different programmes and list the plate centre coordinates of the dedicated ancillary programmes in the Table \ref{tab: summary BOSS plates}. 

The first SDSS Ancillary Project is named `TDSS/SPIDERS/eBOSS Pilot Survey', the \textit{BOSS} targeting primary programme is called `TDSS/SPIDERS/BOSS Pilot Survey' and the Ancillary Bit Numbers are 23, 24 and 25. The \textit{BOSS} spectrograph observed the dedicated plate 6369 in October 2012 without the use of washers. 

The second SDSS Ancillary Project is named `Follow-up spectroscopy of wide-area XMM fields'. The observations of the plates 7235, 7236, 7237, 7238 took place in November 2013 and January 2014 without the use of washers, resulting in 2357 spectra. The \textit{BOSS} targeting programme is called `Wide-Area XMM fields' and the Ancillary Bit Numbers are 32 and 33.

The added sources with \textit{BOSS} spectroscopy from the DR10 are located both outside and inside the ancillary plates footprint as indicated in Fig. \ref{fig:XXLarea}. There are 30 spectra from the plate 6369 of the \textit{TDSS} targets which did not meet the flux threshold from the first ancillary programme, but are within the flux criteria of the second ancillary programme.

 \begin{table}
 \caption{\textit{BOSS} Plate information of the \textit{XMM-SDSS} targets in the XMM-XXL north area: We list plate number and plate centre coordinates.}
 \label{tab: boss plates 01}
 \centering
\begin{tabular}[width=20mm]{ccccccccccc} 
  \hline
Plate & Observation &  RA [deg] of &Dec [deg] \\ 
Number & MJD &Plate Centre&Plate Centre \\ 
 \hline
 6369 & 56217  & 35.9000  & -4.2500  \\
 7235 & 56603 & 37.3965  & -4.7678  \\
 7236 & 56605 & 35.4477 & -4.5985  \\
7237 & 56662 & 33.7033 & -4.9263  \\
7238 & 56660 &  31.5008 & -5.5752  \\
   \end{tabular}
   \label{tab: summary BOSS plates} 
\end{table}

\section{Visual Redshift determination}
\label{app:visred}

In order to determine critical parameters which are correlated to redshift failures of the \textit{BOSS} pipeline, we first checked all spectra of the first ancillary programme (Plate Number: 6369). Assuming this dataset to be a good representative for all our targets, we applied this knowledge to the second ancillary programme and only inspected the subset of spectra most likely to be problematic. In total, we inspected $\sim1200$ spectra. The remaining spectra were part of the visual inspection process of the \textit{BOSS} QSO group and are published in Paris et al. (2015, in prep.).

We evaluate the redshift provided by the pipeline \texttt{Z\_BOSS} by visual inspection and assign both a new redshift \texttt{Z} and confidence parameter \texttt{Z\_CONF} (see Section \ref{sec:reddet}). The spectra, whose visual inspection redshift coincide with the \textit{BOSS} pipeline redshifts, keep their initial redshifts $\texttt{Z}=\texttt{Z\_BOSS}$. For the spectra with wrong  \textit{BOSS} pipeline redshifts, we rerun the \textit{BOSS} pipeline (\texttt{spreduce1d.pro}) in a small redshift range suggested by the visual inspection and assign a new redshift $\texttt{Z}\neq\texttt{Z\_BOSS}$. The redshift confidence parameter \texttt{Z\_CONF} contains information about the reliability of the visual redshift assignment. There are flags for `reliable' and `not robust' redshifts. Furthermore, we assign a flag for `bad spectra' where no redshift can be obtained by visual inspection. For a subsample of spectra, the pipeline fails to determine a correct redshift as suggested by the visual inspection. In this case, we assign the flags `reliable visual redshift and pipeline failure' or `not robust visual redshift and pipeline failure'. During the visual inspection, we mark all stars and BL Lac with the \texttt{STAR/BLLAC} parameter. Their precise redshift determination is not included in our visual evaluation, because stars and BL Lacs are not in the focus of this study. 
In Table \ref{tab:zstat}, we give an overview about the redshifts obtained from the \textit{BOSS} pipeline after visual inspection. 

In the following, we want to highlight the critical spectral characteristics which lead to a redshift failure by the \textit{BOSS} pipeline. 

These characteristics are:
\begin{compactenum}[(i)]
\item low redshift quality: $\texttt{ZWARNING}>0$ (583 spectra with 293 failures), 
\item large redshift error: $\texttt{ZERR\_BOSS}>0.01$ (56 spectra with 45 failures), 
\item low S/N ratio: \texttt{SN\_MEDIAN\_ALL} $<1.6$ (837 spectra with 272 failures), 
\item very high redshifts: $\texttt{Z\_BOSS}>4 $ (49 spectra with 39 failures), 
\item very low redshifts: $\texttt{Z\_BOSS}<0.05$ (131 spectra with 37 failures).
\end{compactenum}
The items (ii) to (v) are often associated with $\texttt{ZWARNING}>0$. The redshifts failures at high redshifts (iv) are often caused by emission line confusion of Ly$\alpha$, MgII, CIV, CIII] and H$\alpha$.
We have to point out that problematic sources can also be caused by external influences, such as:
\begin{compactenum}[(vi)]
\item blending of a neighbour object (1 out of 423 spectra from plate 6369)
\item instrumental fibre problems (4 out of 423 spectra from plate 6369). 
\end{compactenum} These features are not correlated with the 
S/N ratio or the redshift quality flag.  The influence of neighbour 
objects can be reduced by cross-checking with the optical images. 
There are some fibres with instrumental problems resulting in spectra which are missing correct spectroscopic information in parts of the observed wavelength range. These spectra provide redshift information for the objects, but classification is not possible. For our catalogue, we will exclude these objects.
\\\\
The group of spectra with correct and robust \textit{BOSS} redshifts is defined by $\texttt{ZWARNING}=0$, 
\texttt{SN\_MEDIAN\_ALL} $>1.6$, and $0.05<\texttt{Z\_BOSS}<4$, and we obtained an extremely high reliability of 99 per cent for the \textit{BOSS} pipeline results. Therefore, within this parameter space, we did not perform any visual inspection for the spectra of the secondary ancillary programme, nor for all added spectra from the DR10.

\begin{table}
 \caption{Final redshifts and redshift confidence of \textit{BOSS} spectra after the visual inspection: The data set is divided in reliable redshifts, not robust redshifts, bad redshifts and stars/BL Lac. We indicate the redshift confidence and the coincidence of the redshift of the \textit{BOSS} pipeline and the visual inspection. }
 \label{tab: boss plates 01}
 \centering
\begin{tabular}[width=20mm]{lllccccccc} 
 \multicolumn{2}{l}{\textbf{a.) reliable redshift }}& \textbf{2578 spectra}\\
$\texttt{Z\_CONF}=3$: & $\texttt{Z\_BOSS} = \texttt{Z}$ & 2525   \\ 
$\texttt{Z\_CONF}=3$: & $\texttt{Z\_BOSS} \neq \texttt{Z}$ & 45 \\ 
$\texttt{Z\_CONF}=30$:  & good visual redshift,  & 8 \\
& but pipeline failure & \\

\hline 

 \multicolumn{2}{l}{\textbf{b.) not robust redshift}} & \textbf{122 spectra}\\
  $\texttt{Z\_CONF}=2$:&  $\texttt{Z\_BOSS} = \texttt{Z}$& 82  \\
  $\texttt{Z\_CONF}=2$: & $\texttt{Z\_BOSS} \neq \texttt{Z}$ & 29  \\
 $\texttt{Z\_CONF}=20$: & not robust visual redshift & 11 \\
 & and pipeline failure & \\

\hline
 \multicolumn{2}{l}{\textbf{c.) bad redshift}} & \textbf{255 spectra}\\
 $\texttt{Z\_CONF}=1$:  & $\texttt{Z\_BOSS} \neq \texttt{Z}$ & 255 \\
 \hline
 \multicolumn{2}{l}{\textbf{d.) stars, BL Lac}}& \textbf{87 spectra} \\
  \multicolumn{2}{l}{$\texttt{STAR/BLLAC}= \texttt{star}\,\, \rm{or}\,\, \texttt{BL Lac}$:}  & 87   \\
   \end{tabular}
   \label{tab:zstat} 
\end{table}

\section{Optical Line Diagnostic Diagrams}
\label{app:linediag}

\subsubsection*{BPT-NII-diagram}
For spectra with significant H$\beta$, [OIII], H$\alpha$, and [NII], we follow the BPT-NII 
selection criteria \citep{Baldwin81, Kauffmann03, Kewley06, Melendez14} with the 
ratios [OIII]/H$\beta$ vs. [NII]/H$\alpha$. These ratios are less sensitive to 
reddening because of the small wavelength separation of the relevant emission lines. In addition, we use EW which removes the direct reddening dependence. The BPT-NII-diagram 
only classifies objects at a redshift of $0<z<0.54$, because H$\alpha$ and [NII] are shifted out of the \textit{BOSS} wavelength range for higher redshifts.

\begin{figure}
\includegraphics[width=85mm]{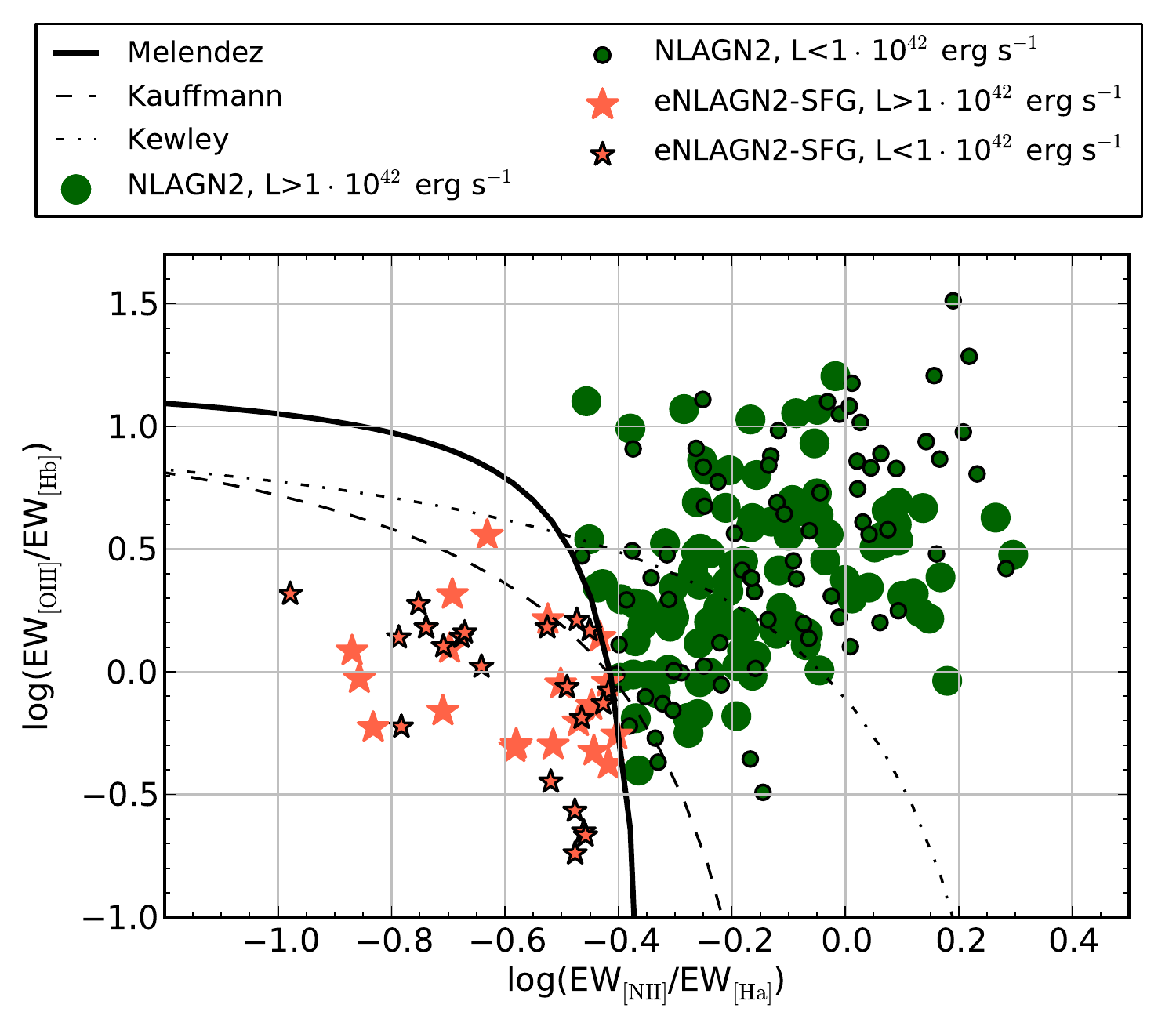}
\caption{BPT-NII-diagram of NLAGN2 and eAGN-SFG: The size of the marker indicates the luminosity below or above the threshold of  $L_{0.5-2\rm keV}\sim10^{42}\rm\, erg\,s^{-1}$. This luminosity threshold is the upper limit of X-ray emission caused by star formation in SFG.  This diagram applies for sources in the redshift range $0<z<0.54$.}
\label{fig:NII}
\end{figure}

According to \cite{Melendez14}, the demarcation of star-forming galaxies and AGN-dominated sources, including the effect of dust, is described by:
\begin{equation}
\log\left(\frac{\rm{EW([OIII])}}{\rm{EW(H\beta)}}\right)=\frac{0.13}{\log{ \left( \frac{\rm{EW([NII])}}{\rm{EW(H\alpha)}} \right)} + 0.31}+1.24
\end{equation}
and shown as solid line in Fig. \ref{fig:NII}.
This demarcation line is based on photoionization modeling for AGN and SFG galaxies. The authors derive a two-zone model including matter-bounded and radiation-bounded components which allows for a theoretical prediction for the BPT-NII diagram.  
In Fig. \ref{fig:NII}, we also add the demarcation line of \cite{Kauffmann03}, which separates between SFG and AGN as well as the demarcation line of \cite{Kewley06}, which separates AGN-SFG composite objects and pure AGN. 
For our classification, we only adopt the demarcation of \cite{Melendez14}.

\subsubsection*{Blue-OII-diagram}
\begin{figure}
\includegraphics[width=85mm]{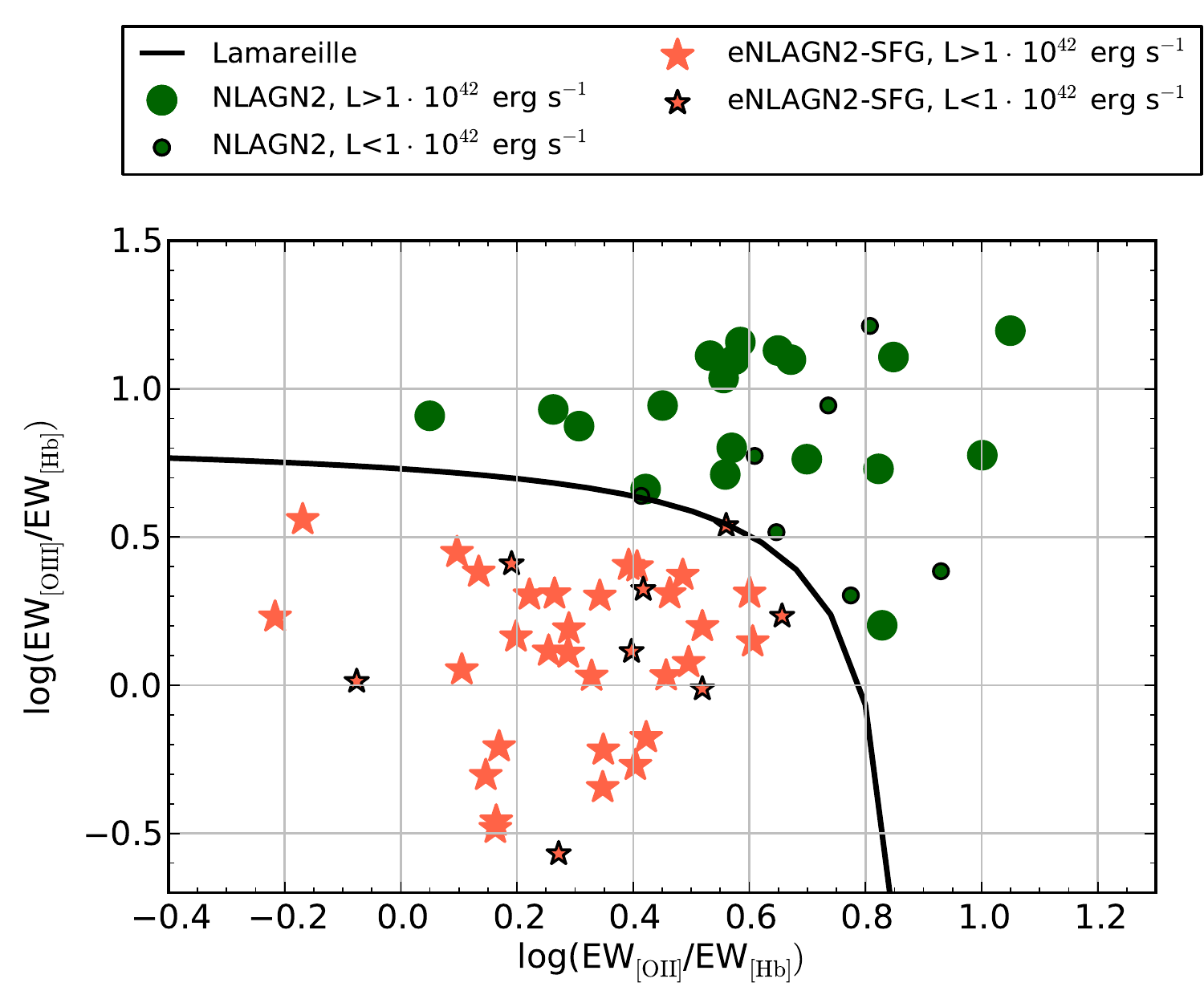}
\caption{Blue-OII-diagram of NLAGN2 and eAGN-SFG: The demarcation curve corresponds to \citet{Lamareille10}.   The size of the marker indicates the luminosity below or above the threshold of  $L_{0.5-2\rm keV}\sim 10^{42}\rm\, erg\,s^{-1}$. This luminosity threshold is the upper limit of X-ray emission caused by star formation in SFG. This diagram applies for sources in the redshift range $0<z<1.08$, and we only project objects which cannot be placed in the BPT-NII-diagram.}
\label{fig:OII}
\end{figure}

The [OII]/H$\beta$ emission line ratio diagram identifies objects with significant [OII]-doublet (EW of $\lambda=3726.032, 3728.815\,$\AA) at higher redshifts ($0<z<1.08$) \citep{Lamareille04, Lamareille10}. For the sake of simplicity, we refer to the [OII] doublet by the name of `[OII]'. The emission lines H$\beta$ and [OII] have a very large separation in wavelength space and the reddening influences the emission lines as well as the underlying continuum differently. Therefore, even the EW is effected by reddening. For the separation of AGN and SFG, we choose the demarcation line of \cite{Lamareille10}:
\begin{equation}
\log\left(\frac{\rm{EW([OIII])}}{\rm{EW(H\beta)}}\right)=\frac{0.11}{\log{\left(  \frac{\rm{EW([OII])}}{\rm{EW(H\beta)}}  \right)} -0.92}+0.85,
\end{equation} 
shown as the solid line in Fig. \ref{fig:OII}. This equation is the most recent demarcation line derived from the \textit{SDSS} DR7 release and allows to separate between regions of SFG, Seyfert 2/LINERS. This classification is designed to yield highly pure SFG samples, with a correct classification of 99.7 per cent of all SFG and a contamination from AGN/LINERS of 16 per cent. 
In this work, we only apply the Blue-OII diagnostics for objects which cannot be placed in the BPT-NII-diagram. 

\section{eROSITA}
\label{sec:eROSITA}

In Table \ref{tab:eROSITA_I}, we provide the number densities and fibre collision correction for the X-ray selected AGN in the XMM-XXL north at different flux depths, including \textit{eROSITA} depths.

\begin{table*}
\caption{Top Panel: List of X-ray sources with \textit{SDSS} counterparts and unique \textit{BOSS} spectra in the XMM-XXL north adapted to different soft X-ray depths. The fibre collision correction is calculated by the number of \textit{BOSS} followed-up spectra over \textit{BOSS} targets within the \textit{r}-band limit. Bottom Panel: List of X-ray sources with reliable redshifts ($\texttt{Z\_CONF}=3$ or $\texttt{Z\_CONF}=30$) and spectroscopic classification in the XMM-XXL north.}
\begin{scriptsize}
\begin{tabular}[width=20mm]{lcccccccccc} 
\hline
\textit{eROSITA} scan& $F_{0.5-2 \,\rm{keV}}$ & X-ray sources & \textit{SDSS} counterparts & \textit{SDSS} counterparts &\textit{BOSS} follow-up  & fibre collision\\ 
&  && $\rm LR_{\rm XMM,SDSS}>1.5$& $15<\textit{r}<22.5\,\rm mag$   &spectra  & correction \\
(Scanning time) &[$\rm erg\,s^{-1}\,cm^{-2}$] &  [$\rm deg^{-2}$] & [$\rm deg^{-2}$] & [$\rm deg^{-2}$] & [$\rm deg^{-2}$]  &$\mu$\\
\hline
	&	$3.0\times 10^{-13}$	&	0.56	&	0.56	&	0.33	&	0.28	&	1.20	\\
	&	$1.0\times10^{-13}$	&	3.61	&	3.56	&	2.61	&	2.06	&	1.27	\\
\textbf{eRASS:1} (0.5 yr) 	&	$4.0\times 10^{-14}$	&	13.2	&	12.7	&	10.8	&	8.56	&	1.27	\\
\textbf{eRASS:2} (1 yr) 	&	$2.5\times 10^{-14}$	&	27.5	&	25.2	&	24.5	&	17.1	&	1.30	\\
\textbf{eRASS:4} (2 yr)	&	$1.5\times 10^{-14}$	&	61.3	&	52.7	&	49.1	&	38.0	&	1.29	\\
\textbf{eRASS:8} (4 yr)	&	$9.8\times 10^{-15}$	&	110	&	87.3	&	82.6	&	64.5	&	1.28	\\
	&	$6.0\times 10^{-15}$	&	187	&	132	&	126	&	98.4	&	1.28\\
	&	$3.0\times 10^{-15}$	&	331	&	190	&	182	&	141	&	1.29	\\
	&	$ 1.0\times10^{-15}$	&	438	&	219	&	208	&	163	&	1.28	\\
	\\
\end{tabular}

\begin{tabular}[width=20mm]{lccccccccc} 
 \hline
\textit{eROSITA} scan & $F_{0.5-2 \,\rm{keV}}$  & reliable redshift  & BLAGN1 & NLAGN2 & NLAGN2cand & eAGN-ALG & eAGN-SFG  \\ 
(Scanning time) & $[\rm erg\,s^{-1}\,cm^{-2}]$ &  [$\rm deg^{-2}$] & [$\rm deg^{-2}$] & [$\rm deg^{-2}$] & [$\rm deg^{-2}$]  &[$\rm deg^{-2}$]  & [$\rm deg^{-2}$]   \\
\hline
	&	$3.0\times 10^{-13}$	&	0.28	&	0.17	&	0.06	&	0.00	&	0.00	&	0.06	\\
	&	$1.0\times10^{-13}$	&	1.94	&	1.72	&	0.06	&	0.00	&	0.11	&	0.06	\\
\textbf{eRASS:1} (0.5 yr) 	&	$4.0\times 10^{-14}$	&	8.39	&	6.89	&	0.78	&	0.33	&	0.22	&	0.11	\\
\textbf{eRASS:2} (1 yr) 	&	$2.5\times 10^{-14}$	&	16.6	&	13.3	&	1.94	&	0.83	&	0.22	&	0.17	\\
\textbf{eRASS:4} (2 yr)	&	$1.5\times 10^{-14}$	&	35.8 	&	28.7	&	3.33	&	2.00	&	0.67	&	0.72	\\
\textbf{eRASS:8} (4 yr)	&	$9.8\times 10^{-15}$	&	59.6 	&	47.3	&	5.28	&	4.22	&	0.94	&	1.17	\\
	&	$6.0\times 10^{-15}$	&	87.9	&	67.8	&	7.28	&	7.50	&	1.94	&	1.72	\\
	&	$3.0\times 10^{-15}$	&	122	&	91.1 	&	10.4	&	11.3	&	3.61	&	2.94	\\
	&	$1.0\times10^{-15}$	&	138	&	98.8	 &	13.2	&	14.3	&	4.78	&	3.94	\\
\label{tab:eROSITA_I}
\end{tabular}
\end{scriptsize}
\end{table*}

\section{Catalogue}
\label{sec:Catalogue}

  \begin{table*}
    \caption{Catalogue parameters of X-ray selected AGN in the northern XMM-XXL field}
\begin{scriptsize}
\begin{tabular}[width=20mm]{clllc|ccccccc} 

\hline
Column & Name & Description \\
\hline
1& \texttt{UXID} & UXID from Liu et al. (2015, in prep.) \\
2& \texttt{RA\_XMM} & RA-position from Liu et al. (2015, in prep.) in deg\\
3&\texttt{DEC\_XMM} & Dec-position from Liu et al. (2015, in prep.) in deg  \\
4&\texttt{RADECERR\_XMM} & Error RA- and Dec-position from Liu et al. (2015, in prep.)  in deg\\
5& \texttt{FLUX\_FULL} & Full X-ray flux $F_{0.5-10\,\rm keV}$ from Liu et al. (2015, in prep.)  in $\rm\, erg\,s^{-1}\rm cm^{-2}$\\
6& \texttt{FLUX\_SOFT} & Soft X-ray flux $F_{0.5-2\,\rm keV}$ from Liu et al. (2015, in prep.)  in $\rm\, erg\,s^{-1}\rm cm^{-2}$ \\
7& \texttt{FLUX\_HARD} & Hard X-ray flux $F_{2-10\,\rm keV}$ from Liu et al. (2015, in prep.)  in $\rm\, erg\,s^{-1}\rm cm^{-2}$ \\
8& \texttt{LUM\_FULL} & Full X-ray luminosity $F_{0.5-10\,\rm keV}$ from Liu et al. (2015, in prep.)  in $\rm\, erg\,s^{-1}\rm$\\
9& \texttt{LUM\_SOFT} & Soft X-ray luminosity $F_{0.5-2\,\rm keV}$ from Liu et al. (2015, in prep.)  in $\rm\, erg\,s^{-1}\rm$ \\
10& \texttt{LUM\_HARD} & Hard X-ray luminosity $F_{2-10\,\rm keV}$ from Liu et al. (2015, in prep.)  in $\rm\, erg\,s^{-1}\rm$ \\

 \hline
11&\texttt{OBJID} &\textit{SDSS}-ID from \textit{SDSS} DR-8 \citep{Aihara11}\\
12&\texttt{RA\_SDSS} &RA-position from \textit{SDSS} DR8 in deg\\
13&\texttt{DEC\_SDSS} &Dec-position from \textit{SDSS} DR8 in deg\\
14&\texttt{RAERR\_SDSS} &Error for RA-position from \textit{SDSS} DR8 in deg\\
15&\texttt{DECERR\_SDSS} &Error Dec-position from \textit{SDSS} DR8 in deg\\
16&\texttt{LR\_XMMSDSS} &Likelihood-Ratio of the matching with \textit{SDSS} counterpart \\
17-21&\texttt{PSF\_MAG}  &Psf-magnitudes for \textit{u}, \textit{g}, \textit{r}, \textit{i}, \textit{z} in AB-mag from \textit{SDSS} DR8 \\
22-26&\texttt{PSFERR\_MAG} &Error for psf-magnitudes for \textit{u}, \textit{g}, \textit{r}, \textit{i}, \textit{z} in AB-mag from \textit{SDSS} DR8 \\
27-31&\texttt{MODEL\_MAG}& Model-magnitudes for \textit{u}, \textit{g}, \textit{r}, \textit{i}, \textit{z} in AB-mag from \textit{SDSS} DR8 \\
32-36&\texttt{MODELERR\_MAG}  & Error for model-magnitudes for \textit{u}, \textit{g}, \textit{r}, \textit{i}, \textit{z} in AB-mag from \textit{SDSS} DR8 \\
37&\texttt{TYPE} & Optical morphology: \texttt{3} - galaxy like, \texttt{6} - star like \citep{Aihara11}\\
38& \texttt{PQSOSUM} & Summed XDQSO QSO probability \citep{Bovy11} \\
39 & \texttt{GOOD} & \textit{BOSS} good flag \citep{Bovy11}\\

\hline 
40&\texttt{WISE\_DESIGN}& Designation from all\textit{WISE}\\
41&\texttt{RA\_WISE} &RA-position from all\textit{WISE} in deg \\
42&\texttt{DEC\_WISE} &Dec-position from all\textit{WISE} in deg\\
43&\texttt{RASIG\_WISE} & One-sigma uncertainty  for RA-position from all\textit{WISE} in deg \\
44&\texttt{DECSIG\_WISE} &One-sigma uncertainty for Dec-position from all\textit{WISE} in deg\\
45&\texttt{LR\_XMMWISE} &  Likelihood-Ratio of the matching with \textit{WISE} counterpart (Georgakakis et al) \\
46-49&\texttt{MPRO\_MAG}  & Mpro-magnitudes of \textit{W}1, \textit{W}2, \textit{W}3 and \textit{W}4 in Vega-mag from all\textit{WISE}\\
50-53&\texttt{MPROSIG\_MAG} & Profile-fit photometric measurement uncertainty for \textit{W}1, \textit{W}2, \textit{W}3 and \textit{W}4 in Vega-mag from all\textit{WISE}\\
54&\texttt{CLEAN} & Exclusion of diffraction spikes, halos, optical ghosts and blended sources ($\texttt{ccflag}=0$ for \textit{W}1 and \textit{W}2, $
\texttt{NB}\leq2$): \\
&& \texttt{1} - photometric conditions are fulfilled, \texttt{0} - photometric conditions are not fullfilled\\

\hline
55&\texttt{PLATE} & \textit{BOSS} plate number \\
56&\texttt{FIBERID} & \textit{BOSS} fibre-ID number  \\
57&\texttt{MJD} & \textit{BOSS} MJD \\
58&\texttt{Z\_BOSS} & Redshift from the automatic \textit{BOSS} pipeline\\
59&\texttt{ZERR\_BOSS} & Error for \texttt{Z\_BOSS}\\
60&\texttt{ZWARNING} & Redshift warning flag from the automatic \textit{BOSS} pipeline\\
61&\texttt{SN\_MEDIAN\_ALL} & Average S/N ratio of all five \textit{SDSS} bands from the automatic \textit{BOSS} pipeline \\
\hline
62&\texttt{Z} & Redshift after visual inspection and refit of spectrum with \texttt{spreduce1d.pro} \\
63&\texttt{ZERR} & Error for \texttt{Z}\\
64&\texttt{Z\_CONF} & Redshift confidence flag from visual inspection:  \texttt{3} - reliable redshift,\\
&& \texttt{2} - not robust redshift, \texttt{1} - bad spectrum,   \texttt{30}  - reliable visual redshift and pipeline failure,  \\
&&  \texttt{20}  - not robust visual redshift and pipeline failure\\
65&\texttt{CLASS}  & Classification for $\texttt{Z\_CONF}=3$ spectra: \texttt{BLAGN1}, \texttt{NLAGN2},  \texttt{NLAGN2cand}, \texttt{eAGN-ALG},\\  
&& \texttt{eAGN-SFG}, \texttt{NOC} (not classified)  \\
66&\texttt{STAR/BLLAC}& Star: \texttt{star}, BL LAC: \texttt{bllac} \\ 
\hline
  \end{tabular}
\end{scriptsize}

 \label{tab:catalogueparam}
 \end{table*}

   \begin{table*}
    \caption{Catalogue example of X-ray selected AGN in the northern XMM-XXL field}
\begin{scriptsize}
\begin{tabular}[width=20mm]{ccccccccccccccccc} 

\hline
\texttt{UXID} & ... & \texttt{LUM\_FULL} & ...  & \texttt{LR\_XMMSDSS}  & ... & \texttt{PQSOSUM}  & ... &  \texttt{LR\_XMMWISE} & .... & \texttt{CLEAN} \\
(1) & (2)-(7) & (8) & (9)-(15) & (16) & (17)-(37) & (38) & (39)-(44) &  (45) & (46)-(53) &(54) \\
\hline
N$\_14\_52$	&	...	&	9.837E+44	&...	&	4.057	&	...	&	0.2487	&	...	& 	5.1328 	&	...	&1		\\
N$\_17\_36$	&	...	&	1.782E+42	&...	&	8.391	&	...	&	 -9999		&	...	&	32.3115	&	...	&1		\\
N$\_12\_26$	&	...	&	1.059E+45	&...	&	35.170	&	...	&	0.9957	&	...	&	49.2858	&	...&1	\\
N$\_13\_30$	&	...	&	2.259E+45	&...	&	11.880	&	...	&	0.5135	&	...	&	9.3445&	...		&1	\\
N$\_12\_35$	&	...	&	7.265E+44	&...	&	3.615	&	...	&	0.9971	&	...	&	4.8066	&	...&1		\\
N$\_16\_5$	&	...	&	8.770E+42	&...	&	3.050	&	...	&	 -9999		&	...	&	2.9940&	...	&1		\\
  \end{tabular}
  \begin{tabular}[width=20mm]{ccccccccccccccccc} 
 \hline
\texttt{PLATE} & \texttt{FIBERID} & \texttt{MJD} & \texttt{Z\_BOSS} & \texttt{ZERR\_BOSS}& \texttt{ZWARNING} & \texttt{SN\_MEDIAN\_ALL} & \texttt{Z} & \texttt{ZERR} & \texttt{Z\_CONF} & \texttt{CLASS} & \texttt{STAR/BLLAC}\\
(55) & (56) & (57) & (58) &(59) & (60) & (61) & (62) & (64) & (64) & (65) & (66)\\
 \hline
4344		&	376	&	 55557	&	2.7035	& 6.31E-04 &	0	&	3.4623	&	2.7035	&	6.31E-04	&	3	&	BLAGN1	&	 -9999		\\
4344		&	424	&	55557	&	0.10823	&8.89E-06 &	0	&	38.1556	&	0.10823	&	8.89E-06	&	3	&	NLAGN2	& -9999			\\
4344 	&	448	&	55557 	&	2.47068	& 7.54E-04 &	0	&	5.2425	&	2.47068	&	7.54E-04	&	3	&	BLAGN1	&	 -9999		\\
4344		&	456 &	55557	&	2.40546	& 5.71E-04 &	0	&	1.5699	&	2.40546	&	5.71E-04	&	3	&	BLAGN1	& -9999			\\
4344		&	458	&	55557 	&	2.20862	&3.60E-04 &	0	&	12.7257	&	2.20862	&	3.60E-04	&	3	&	BLAGN1	&	 -9999		\\
4344		&	488 &	55557	&	0.28738	& 5.40E-05 &	0	&	9.9052	&	0.28738	&	5.40E-05	&	3	&	eAGN-ALG	& -9999		\\
  \end{tabular}
  
\end{scriptsize}

 \label{tab:catalog}
 \end{table*}
 
 We publish the catalogue of X-ray, optical, infrared, and spectroscopic properties of the 8445 X-ray selected sources in the northern XMM-XXL field, which is accessible in this website: \texttt{http://www.mpe.mpg.de/XraySurveys} under [Surveys] $>$  [XMM-XXL].
In Table \ref{tab:catalogueparam}, we list the catalogue parameters with information on:
\begin{compactenum}[(i)]
\item the X-ray imaging with \textit{XMM-Newton}; 
\item the optical imaging with \textit{SDSS} and XDQSO selection probabilities ;
\item the infrared imaging with \textit{WISE}; 
\item the \textit{BOSS} pipeline information; 
\item the redshift information of the visual inspection and the classification.
\end{compactenum}
The full X-ray catalogue of the X-ray selected sources in the XMM-XXL with complete information about their X-ray properties will be provided in the paper of Liu et al. (2015, in prep.). Table \ref{tab:catalog} in the appendix, shows the first 7 sources of the catalogue.

\label{lastpage}

\end{document}